\documentclass{JFM-FLM_Au}

\makeatletter

\def\pagelimitfooter{%
  \hbox to \textwidth{%
    {\cppagefont
    \ifx\@volume\undefined\else\textbf{\@volume}\fi\
    \ifx\@issue\undefined
      \ifpaper A\else\ifrapid R\else\ifpersp P\else\iffof F\else X\fi\fi\fi\fi1-
    \else
      \ifpaper A\else\ifrapid R\else\ifpersp P\else\iffof F\else X\fi\fi\fi\fi\@issue-
    \fi
    \thepage}%
    \hfill
  }%
}

\def\oddabsfooterflag{%
  \hbox to \textwidth{\hfill{\cppagefont\thepage}}%
}
\def\evenabsfooterflag{%
  \hbox to \textwidth{{\cppagefont\thepage}\hfill}%
}

\def\ps@titlepage{%
  \leftskip\z@
  \let\@mkboth\@gobbletwo
  \vfuzz=5\p@
  \def\@oddhead{%
    \vbox{\vspace*{-4pt}
    \hbox to \textwidth{\@j@urnal\hfill}}%
  }%
  \def\@evenhead{%
    \vbox{\vspace*{-4pt}
    \hbox to \textwidth{\@j@urnal\hfill}}%
  }%
  \def\@oddfoot{\hbox to \textwidth{\hfill{\cppagefont\thepage}}}%
  \def\@evenfoot{\hbox to \textwidth{{\cppagefont\thepage}\hfill}}%
  \def\sectionmark##1{}%
  \def\subsectionmark##1{}%
}

\makeatletter

\patchcmd{\@maketitle}
  {(Received xx; revised xx; accepted xx)\hfill}
  {}
  {}
  {\PackageWarning{JFM}{Could not remove received/revised/accepted line}}

\patchcmd{\@maketitle}
  {\vspace*{10\p@}\addvspace{4.6pc}}
  {\vspace*{5\p@}\addvspace{0pc}}
  {}
  {\PackageWarning{JFM}{Could not reduce space above title}}

\def\pagelimitfooter{%
  \hbox to \textwidth{%
    \ifodd\c@page
      \hfill{\cppagefont\thepage}%
    \else
      {\cppagefont\thepage}\hfill
    \fi
  }%
}
\makeatother

\makeatletter

\def\ps@headings{%
  \let\@mkboth\markboth
  \def\@oddhead{\hfill{\itshape\@righttitle}\hfill}%
  \def\@evenhead{\hfill{\itshape\@lefttitle}\hfill}%
  \def\@oddfoot{\hbox to \textwidth{\hfill{\cppagefont\thepage}}}%
  \def\@evenfoot{\hbox to \textwidth{{\cppagefont\thepage}\hfill}}%
  \def\sectionmark##1{\markboth{##1}{}}%
  \def\subsectionmark##1{\markright{##1}}%
}

\AtBeginDocument{\pagestyle{headings}}

\makeatother

\usepackage{bm}
\usepackage{xcolor}
\usepackage{tikz}
\usepackage{subcaption}

\graphicspath{{./}{Figures/}}

\lefttitle{Taborda \& van Wachem}
\righttitle{Orientation-dependent drag, lift, and torque correlations for regular Platonic polyhedral particles}

\title{Orientation-dependent drag, lift, and torque correlations for regular Platonic polyhedral particles}

\author{
Manuel A. Taborda\aff{1}
\and
Berend van Wachem\aff{1}
}

\affiliation{
\aff{1}
Lehrstuhl für Mechanische Verfahrenstechnik,
Fakultät für Verfahrens- und Systemtechnik,
Otto-von-Guericke-Universität,
39106 Magdeburg, Germany
}

\corresau{Manuel A. Taborda, \email{manuel.taborda@ovgu.de}}

\begin{document}
\maketitle

\begin{abstract}
In this work, particle-resolved direct numerical simulations are performed to investigate flow past the five Platonic solids, which represent a progression in particle sphericity with an increasing number of faces. The simulations cover particle Reynolds numbers in the range $0.1 \leq \mathrm{Re_p} \leq 300$ and multiple particle orientations relative to the incoming flow. Based on the numerical data, new correlations are developed for the drag, lift, and torque coefficients. The proposed drag correlation explicitly accounts for both Reynolds number and particle orientation, whereas the lift and torque coefficients are represented by orientation-dependent trigonometric and exponential basis functions whose coefficients vary with Reynolds number. The simulations are conducted using the immersed boundary method, and the resulting drag correlation accurately reproduces the numerical data. The lift and torque correlations capture the principal trends observed in the numerical simulations, including the strong dependence on particle orientation. The proposed correlations provide a computationally efficient framework for incorporating orientation-dependent hydrodynamic forces and torques into Euler--Lagrange and point-particle simulations, enabling a more realistic representation and predictions of non-spherical particle transport in multiphase flows.
\end{abstract}





\section{Introduction}
\label{sec:Introduction}

In many industrial and natural processes of particle transport in gas or liquid systems, such as sediment transport, particle separators, conveying in channels and pipes, dune formation, paper manufacturing, and insulation material processing, particles appear in a wide range of sizes and shapes (e.g., rods, plates, fibres, ground particles, and agglomerates). 
For instance, particle shape is known to affect particle transport in injection processes, sediment transport rates, and the stability of hydraulic structures~\citep{Kidanemariam2017}. These processes are often numerically predicted using the Euler–Lagrange approach; however, accurate expressions for flow resistance coefficients, wall effects, and non-ideal wall–particle interactions are still lacking. Traditionally, non-spherical particles are approximated as spheres, but such simplifications yield unrealistic predictions, especially in dense regions where non-spherical particles interact with other particles or boundaries~\citep{vanWachem2015}. In turbulent open-channel flows, most numerical studies investigating the transport of heavy particles model the dispersed phase as spherical particles, both in terms of the hydrodynamic force models and particle--wall collision dynamics. In natural environments, however, sediment particles are rarely spherical. Several experimental studies have therefore reported significant differences in the transport behaviour of spherical and non-spherical particles~\citep{Kussin2002, Lain2007, Taborda2025}. For instance,~\citet{Schmeeckle2001} experimentally analysed the collision of natural sediment particles with an inclined glass wall in water. They observed, among others, that particle orientation prior to impact is strongly influenced the normal restitution coefficient, a finding also confirmed in dry collision experiments using Platonic solids in~\citet{Shi2024}.

A widely used reference in the modelling of particle drag is the~\citet{Schiller1933} correlation, which expresses the drag coefficient of a spherical particle as a function of the particle Reynolds number, $\mathrm{Re_p}$. This correlation is simple, computationally efficient, and valid across a broad range of $\mathrm{Re_p}$. However, it is limited to spheres and cannot capture any influence of particle shape or orientation. The more general correlation of~\citet{Haider1989} extends drag predictions to non-spherical particles by introducing the sphericity $\phi$ as a shape parameter. 
Their correlation is expressed as $C_\mathrm{D} = \frac{24}{\mathrm{Re_p}}\left(1 + A \mathrm{Re_p^B}\right) + \frac{C}{1 + \frac{D}{\mathrm{Re_p}}},$ where $A$, $B$, $C$, and $D$ are empirical coefficients, which depends on $\phi$. This formulation effectively generalises the Schiller–Naumann approach by incorporating both inertial and shape effects, while maintaining a globally isotropic description of the drag. Its main advantage lies in its wide applicability to irregular particles with known sphericity. However, because it reduces shape effects to a single scalar parameter, it cannot account for orientation-dependent variations in drag, which are crucial for predicting the dynamics of anisotropic particles. A similar philosophy was adopted by~\citet{Ganser1993}, who proposed a drag correlation based on dynamic shape factors to improve predictions for irregular particles over a wide range of Reynolds numbers. More recently,~\citet{Bagheri2016} demonstrated that shape descriptors beyond sphericity are often required to accurately characterise the drag of non-spherical particles, highlighting the limitations of isotropic drag formulations for complex particle shapes. Building on this idea,~\citet{Holzer2008} proposed correlations that further refined drag predictions for non-spherical particles. Their model accounts not only for particle Reynolds number and sphericity, but also incorporates additional geometric descriptors to better represent elongated or flattened particles. Compared with~\citet{Haider1989}, the~\citet{Holzer2008} correlation provides improved accuracy in regimes where aspect ratio effects are important. Nevertheless, similar to the earlier correlations, it assumes isotropy in particle–fluid interactions and does not capture directly the role of instantaneous particle orientation. 

To overcome these limitations, direct numerical simulations (DNS) have increasingly been employed to derive drag, lift, and torque correlations for non-spherical particles~\citep{Zastawny2012c, Vergara2024, Cheron2024, vanWachem2024, Lain2024, Lain2025}. For example,~\citet{Zastawny2012c} performed PR-DNS of several axisymmetric non-spherical particles using the immersed boundary method (IBM) and proposed correlations for drag and lift forces, as well as pitching and rotational torques, explicitly depending on $\mathrm{Re_p}$ and the angle of incidence between particle and flow. Similarly,~\citet{Sanjeevi2017} investigated the orientational dependence of hydrodynamic forces acting on spheroidal particles and demonstrated the importance of accounting for particle orientation when modelling particle-fluid interactions. More recently, the scope of particle-resolved research has been extended towards increasingly complex particle shapes and flow configuration. \citet{Vergara2024} investigated the rotational dependence of the drag coefficient of irregularly shaped grains over a range of Reynolds numbers, further demonstrating the strong coupling between particle morphology, orientation, and hydrodynamic resistance. Using IBM, \citet{Wang2025c} developed a PR-DNS framework for general non-spherical particles, including some superellipsoids as well as cylinders, and analysed the associated drag, lift, and torque coefficients. Furthermore, \citet{Cheng2026} examined wall-bounded prolate spheroids using an extensive DNS database and demonstrated the coupled influence of Reynolds number, particle orientation, and wall distance on drag, lift, and pitching torque. These studies demonstrate the importance of orientation in determining hydrodynamic forces and torques, and provide a foundation for extending such correlations to new particle shapes.

For non-spherical particles, hydrodynamic forces and torques are intrinsically coupled through the particle orientation. The drag, lift, and torque coefficients depend on the instantaneous alignment of the particle relative to the incoming flow, while the hydrodynamic torque simultaneously governs the evolution of that orientation. Consequently, an accurate prediction of particle trajectory requires reliable models for all three quantities. The particle orientation directly influences its projected area, wake structure, lift force, and collision behaviour, thereby affecting its overall transport dynamics. Hydrodynamic torque is therefore of particular importance, as it determines particle rotation and preferential alignment, which in turn modify the resulting force coefficients. Although considerable progress has been made in developing drag correlations for non-spherical particles, significantly fewer studies have addressed orientation-dependent lift and torque effects, particularly for angular polyhedral particles. This limitation is especially relevant in Euler--Lagrange simulations, where an accurate representation of rotational dynamics is required to predict particle motion and orientation evolution.

Platonic-shaped particles provide a unique family of symmetric convex polyhedra that enables an analysis of shape effects in particle-laden flows. Unlike irregular particles, whose geometric complexity makes it difficult to describe the particle shape accurately, Platonic solids allow angularity, symmetry, and sphericity to be varied in a controlled manner while preserving geometric regularity. The five classical Platonic solids, i.e, tetrahedron, hexahedron, octahedron, dodecahedron, and icosahedron, span a broad range of sphericities and geometric complexities, making them idealized yet insightful shapes for studying fluid--particle interactions~\citep{Zhao2019a,Gai2023a,Gai2024}. Moving from the tetrahedron to the icosahedron progressively increases the number of faces and the particle sphericity, thereby providing a convenient framework for quantifying the influence of shape on hydrodynamic forces, torques, and wake dynamics. Their geometric regularity also allows precise evaluation of surface area, volume, and moments of inertia, facilitating both numerical and theoretical research. Furthermore, studying the complete family of Platonic solids enables the identification of trends associated with increasing sphericity and reduced angularity, which would be difficult to infer from studies restricted to a single particle shape.

In Euler–Lagrange simulations, spheres are still the most common assumption, but this approximation neglects the angularity and anisotropy inherent in many realistic particles~\citep{vanWachem2022, Taborda2023a, Cheron2024, Taborda2025}. Incorporating Platonic solids as reference particle shapes allows for the development of orientation-dependent hydrodynamic correlations that better reflect the dynamics of regular non-spherical particles. While drag predominantly governs particle transport and settling, lift forces can significantly influence lateral migration, near-wall dynamics, and preferential concentration, particularly when particle orientation evolves in response to the surrounding flow~\citep{taborda2026,taborda2026a}. Similarly, hydrodynamic torque determines particle rotation and preferential alignment, thereby influencing the projected area exposed to the flow and modifying the resulting drag and lift forces during its transport. Such correlations can bridge the gap between idealized spherical models and the complex behaviour of natural or industrial particles with faceted shapes. Recently,~\citet{Gai2023a} investigated flow regime transitions around stationary Platonic solids, showing that particle angularity and orientation strongly influence wake development. However, no orientation-dependent drag, lift, or torque correlations are reported. To the authors' knowledge, no comprehensive set of orientation-dependent drag, lift, and torque correlations currently exists for the complete family of Platonic solids. Consequently, Euler--Lagrange simulations involving angular polyhedral particles must rely on spherical approximations or correlations developed for fundamentally different particle shapes. This highlights the need for new correlations that explicitly account for both Reynolds number and orientation effects. In this context, Platonic solids serve as representative angular particle shapes, providing a framework for quantifying orientation-dependent hydrodynamic behaviour.

In this study, we simulate and analyse the flow past Platonic solids using particle-resolved direct numerical simulations (PR-DNS), with particular emphasis on hydrodynamic force and torque coefficients. We propose orientation-dependent correlations for drag, lift and torque coefficients that quantify the effect of particle alignment relative to the flow direction. These PR-DNS and correlations are validated against existing PR-DNS data~\citep{Gai2023a} and compared with classical drag models such as those of~\citet{Schiller1933} and~\citet{Haider1989}. The proposed framework naturally extends the predictive capabilities of Euler--Lagrange simulations to include Platonic solids, thereby enabling more realistic modelling of particulate flows involving angular, non-spherical particles over the range $0.1 \leq \mathrm{Re_p} \leq 300$. The paper is organized as follows. Section~\ref{sec:Numerical Methods} provides an overview of the numerical methodology, including the PR-DNS framework and the representation of Platonic solids. Section~\ref{sec:Simulation set-up} describes the simulation set-up and numerical discretization. Section~\ref{sec:Results and Discussion} presents the results and discussion for each Platonic shape, including the proposed drag, lift and torque coefficient correlations. Finally, Section~\ref{sec:Conclusions} summarizes the conclusions and outlook of the study.

\section{Numerical methods}
\label{sec:Numerical Methods}

PR-DNS simulations of the interaction between fluid and different Platonic-shaped particles in a uniform flow, varying particle orientation and fluid flow regime are carefully performed. The numerical simulations are achieved using the in-house code MultiFlow~\citep{vanWachem2023}, where the Navier-Stokes equation system is solved within a fully implicit pressure-velocity coupling, and the source terms are discretised ensuring the numerical balance with the flow pressure gradient. In order to account for the discretization of the particle surface, the smooth immersed boundary method is used~\citep{Uhlmann2005, Peskin1972}, which is based on the direct-forcing formulation of~\citet{AbdolAzis2019}.
The solution of the flow field is given by solving the continuity and momentum equations for an unsteady, incompressible flow, with an additional source term as follows:

\begin{equation}
    \nabla \cdot \bm{u} = 0
\label{eq:conservationEquation}
\end{equation}

\begin{equation}
\rho \frac{\partial \bm{u}}{\partial t} + \rho \nabla \bm{u} \bm{u}  =
-\nabla p +  \mu \nabla^2 \bm{u} + \bm{s} \
    \label{eq:momentumEquation}
\end{equation}

where $\bm{u}$ corresponds to the fluid velocity field, $p$ is the pressure, $\bm{s}$
is the source term that represents the force due to the immersed boundary (IB),
and $\rho$ and $\mu$ are the density and dynamic viscosity of the fluid, respectively.
The solver is based on the finite-volume method, employing an implicit scheme to couple pressure and velocity. In this framework, the smooth-interface IBM discretizes the solid-fluid interface into a set of Lagrangian markers, $\bm{X}_j$, $j = 1,2,\dots,N_L$, where $N_L$ corresponds to the total number of markers, overlapping the Eulerian mesh. Since the Eulerian mesh does not coincide with the Lagrangian markers, information must be exchanged between the two discrete frameworks to ensure proper coupling of the immersed boundary with the surrounding fluid. Flow variables at Eulerian points, $\bm{u}_i$, are interpolated, or spread back, to the Lagrangian points using a discrete delta function, $\delta$, with weights, $\omega_{i,j}$. The velocity at the Lagrangian point, $j$, is expressed as

\begin{equation}
\bm{U}_{j} = \sum_i^{N_E} \bm{u}_i \, \omega_{i,j} \Delta v_i \ ,
\label{eq:lagInterpolation}
\end{equation}

where $\Delta v_i$ denotes the Eulerian cell volume, with compact support chosen to always contain several fluid cells. The Lagrangian momentum source term required to enforce the desired velocity, $\bm{U}_{\mathrm{IB},j}$, at the Lagrangian point is obtained from the discretized momentum equation at time $k$:

\begin{equation}
\bm{S}_j^{\;k} = \frac{\bm{U}_{\text{IB},j}^{\;k} - \bm{U}_j^{\;k-1}}{\Delta t} + \bm{C}_j^{\;k} + \bm{B}_j^{\;k} - \bm{D}_j^{\;k} \ ,
\label{eq:lagForce}
\end{equation}

where $\bm{C}_j$, $\bm{B}_j$, and $\bm{D}_j$ represent the Lagrangian interpolations of the convective, pressure-gradient, and viscous terms, respectively, obtained from the corresponding Eulerian quantities $\bm{c}_i$, $\bm{b}_i$, and $\bm{d}_i$:
\begin{align}
\bm{C}_{j} &= \sum_i^{N_E} \bm{c}_i \, \omega_{i,j} \Delta v_i \ , &
\bm{B}_{j} &= \sum_i^{N_E} \bm{b}_i \, \omega_{i,j} \Delta v_i \ , &
\bm{D}_{j} &= \sum_i^{N_E} \bm{d}_i \, \omega_{i,j} \Delta v_i \ .
\end{align}

The Lagrangian forces are subsequently spread back to the Eulerian mesh as source terms in the momentum equation (equation~\ref{eq:momentumEquation}):

\begin{equation}
\bm{s}_i = \sum_j^{N_L} \bm{S}_j \, \omega_{i,j} \Delta V_j \ ,
\label{eq:eulForcing}
\end{equation}

where $\Delta V_j$ are the Lagrangian volumes defined to ensure exact conservation of the total force between the Lagrangian and Eulerian representations:

\begin{equation}
\sum_j^{N_L} \bm{S}_j \Delta V_j = \sum_i^{N_E} \bm{s}_i \Delta v_i \ .
\end{equation}

The Lagrangian volumes are then obtained by solving the following linear system:

\begin{equation}
\sum_l^{N_L} \bm{q}_{j,l} \Delta V_l = \bm{S}_j \ ; \quad \text{in which }  
\bm{q}_{j,l} = \sum_i^{N_{E,j}} \omega_{i,j} \omega_{i,l} \Delta v_i ~ \bm{S}_l \
\end{equation}

Here, $\bm{S}_l$ denotes the Lagrangian IBM forcing associated with the $l$th Lagrangian marker, whose contribution to marker $j$ is determined by the overlap of their respective interpolation and spreading supports. This formulation ensures that the spreading and subsequent interpolation of the forces reproduce the intended Lagrangian forcing at each marker. The accurate computation of the face advecting velocity used to evaluate the convective fluxes requires inclusion of the source terms arising from the IBM into the momentum-weighted interpolation (MWI) scheme~\citep{Bartholomew2018}. This procedure guarantees that the discrete pressure gradient and the IBM momentum source term are numerically balanced, avoiding unphysical solutions. The source terms, $\bm{s}_i$, as computed in equation~\ref{eq:eulForcing}, are consistently included in the MWI, ensuring enforcement of the no-slip condition on the immersed boundary while maintaining the stability and accuracy of the pressure-velocity coupling. Moreover, the interpolation and spreading compact supports are constructed using a modified moving-least-squares (MLS) algorithm~\citep{Bale2021}. This approach reconstructs the velocity field from neighbouring Eulerian cells without requiring explicit information along the solid interface, thereby improving the enforcement of the no-slip condition and enhancing numerical stability in the vicinity of complex geometries. The size of the compact support determines both the number of fluid cells involved in the interpolation and the extent of the spreading of fluid variables. In this work, a five-point spline kernel function is employed~\citep{Bao2016}. The spreading of the IBM source terms towards the source terms of the fluid momentum equations is scaled by a relaxation factor~\citep{Zhou2021}. This relaxation factor is based on stability condition criterion, and controls the rate at which the no-slip condition is reached as well the magnitude of the no-slip error. For further details on the IBM implementation, the reader is referred to the original work of \citet{AbdolAzis2019} and \citet{Cheron2023a}.

\begin{figure}[htbp!]
\centering
\includegraphics[width=0.95\textwidth]{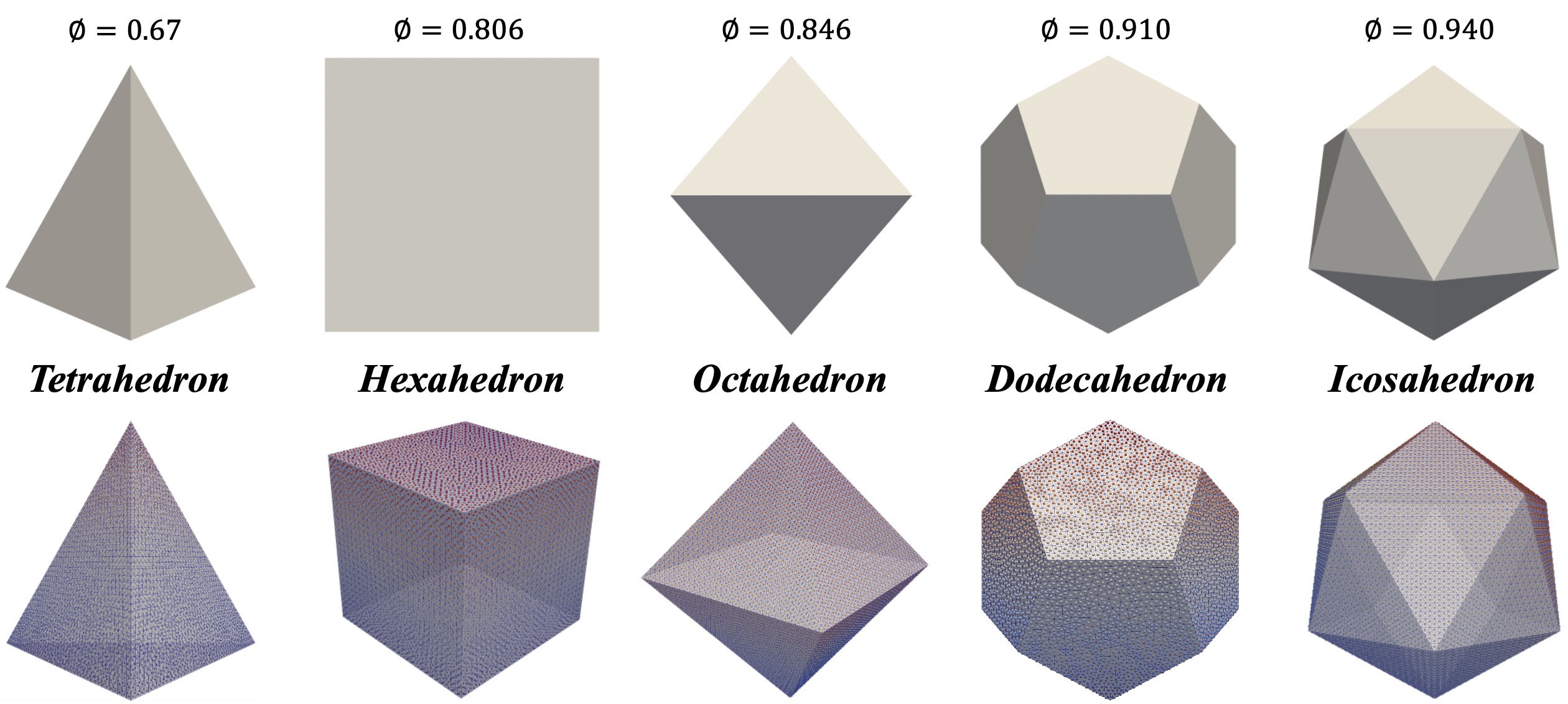}
\caption{Platonic polyhedra and Lagrangian markers superposed on the initial triangulated mesh Lagrangian markers are located in the centre of the triangle mesh (the number of markers is reduced to allow for visibility). The sphericity $\phi$ is also shown for each particle as reference.}
\label{fig:PlatonicShapeswitMarkers}
\end{figure}

Furthermore, the fluid forces acting on the non-spherical particle are obtained directly from the immersed boundary forcing applied on the Lagrangian markers. For each marker $j$, the IBM force $\bm{S}_j$ are computed from the discrete momentum equation. Since $\bm{S}_j$ represents the force exerted on the fluid, the corresponding hydrodynamic force acting on the particle is obtained with the opposite sign. The total hydrodynamic force acting on the particle is then evaluated as
\begin{equation}
\bm{F} = -\sum_{j=1}^{N_L} \bm{S}_j \, \Delta V_j ,
\label{eq:totalForce}
\end{equation}
where $\Delta V_j$ is the effective Lagrangian volume associated with marker $j$. The force components are calculated by projecting $\bm{F}$ onto the coordinate axes as
\begin{equation}
F_{x} = F_{\mathrm{D}} = \bm{F} \cdot \hat{e}_x, 
\qquad
F_{y} = \bm{F} \cdot \hat{e}_y,
\qquad
F_{z} = F_{\mathrm{L}} = \bm{F} \cdot \hat{e}_z ,
\label{eq:forceProjections}
\end{equation}
where $F_{\mathrm{D}}$ denotes the streamwise drag force, while $F_y$ and $F_z$ denote the spanwise and vertical components of the lift force, respectively. In the following, the term lift force refers specifically to the vertical component, $F_z$.

In addition, the total hydrodynamic torque acting on the particle is computed as
\begin{equation}
\bm{T}
=
-\sum_{j=1}^{N_L}
\left(
\bm{x}_j-\bm{x}_p
\right)
\times
\bm{S}_j \, \Delta V_j ,
\label{eq:totalTorque}
\end{equation}
where $\bm{x}_j$ denotes the position of Lagrangian marker $j$ and $\bm{x}_p$ is the particle-centre position. The torque component considered in the present work is obtained by projection onto the y-axis (perpendicular to the page),
\begin{equation}
T_y = \bm{T}\cdot\hat{e}_y ,
\label{eq:torqueProjection}
\end{equation}
where $\hat{e}_y$ denotes the unit vector along the y-axis.

For comparison along different flow conditions, the drag-, lateral- (spanwise), lift-forces and torque quantities are expressed in terms of dimensionless coefficients such as:
\begin{equation}
C_{x} = C_\mathrm{D} =
\frac{F_{x}}
{\tfrac{1}{2}\rho_F |\bm{\tilde{u}}|^2 A_\mathrm{p}},
\qquad
C_{y} =
\frac{F_{y}}
{\tfrac{1}{2}\rho_F |\bm{\tilde{u}}|^2 A_\mathrm{p}},
\qquad
C_{z} = C_\mathrm{L} =
\frac{F_{z}}
{\tfrac{1}{2}\rho_F |\bm{\tilde{u}}|^2 A_\mathrm{p}},
\label{eq:forceCoeffs}
\end{equation}

and

\begin{equation}
C_\mathrm{T} =
\frac{T_y}
{\tfrac{1}{2}\rho_F |\bm{\tilde{u}}|^2 A_\mathrm{p} D_\mathrm{eq}},
\label{eq:torqueCoeff}
\end{equation}
where $\rho_F$ is the fluid density, $\bm{\tilde{u}}$ is the relative velocity between the velocity of the locally undisturbed fluid at the particle centre, $\bm{u}_{\mathrm{f@p}}$, and the particle velocity, $\bm{u}_\mathrm{p}$, and $A_\mathrm{p}=\pi D_{\mathrm{eq}}^2/4$ is the fixed reference cross-sectional area of the volume equivalent sphere. The equivalent diameter $D_\mathrm{eq}$ is used as the characteristic length scale for torque normalization. These coefficients provide a normalized description of the hydrodynamic forces and torque that is independent of the specific flow conditions.

Figure~\ref{fig:PlatonicShapeswitMarkers} presents the five Platonic solids considered in this study, namely, the tetrahedron, hexahedron, octahedron, dodecahedron, and icosahedron. As the number of regular faces on a particle increases from 4 to 20, the particle sphericity also increases, ranging from $\phi=0.67$ for the tetrahedron to $\phi=0.94$ for the icosahedron. To discretize the particle surface within the immersed boundary framework, each geometry is represented by an STL file and subdivided into a collection of small triangular surface elements. Lagrangian markers are then placed at the centroids of the triangles to represent the particle surface in the numerical domain. For all five Platonic solids, a minimum of 21,000 Lagrangian markers is employed per particle to ensure an accurate surface representation, with the marker spacing chosen to be comparable to the local Eulerian mesh resolution. The lower panels of figure~\ref{fig:PlatonicShapeswitMarkers} show the resulting marker distributions.

\section{Simulation configuration}
\label{sec:Simulation set-up}

The numerical methods described in the previous section are employed to
simulate the flow around the stationary Platonic-shaped particles, wherefrom the hydrodynamic forces are determined. The computational domain is defined as a box of dimensions $20D_{\mathrm{eq}} \times 15D_{\mathrm{eq}} \times 15D_{\mathrm{eq}}$, corresponding to approximately 900,000 control volumes, where $D_{\mathrm{eq}}$ corresponds to the equivalent volume diameter. For low particle Reynolds numbers ($0.1 \leq \mathrm{Re_p} \leq 1$), a larger domain of
$30D_{\mathrm{eq}} \times 30D_\mathrm{eq} \times 30D_{\mathrm{eq}}$ is employed to account
for the long-range nature of viscous disturbances in the Stokes-flow
regime. This larger domain reduces confinement effects and ensures that
the imposed boundary conditions do not influence the computed drag,
lift, and torque acting on the particle. This configuration is carefully selected to minimize computational cost while guaranteeing domain-independent results for the range of Reynolds numbers considered, in agreement with the findings of~\citet{Cheron2024}. The particle centre is positioned at a distance of $5D_{\mathrm{eq}}$ from the inlet boundary of the domain. Subsequently, the range of particle Reynolds number, spanning from $0.1$ to $300$, and orientation with respect to the flow field are varied. The particle orientation is described by the rotation angle $\theta$, defined as the rotation of the particle about the $y$-axis, while the remaining rotational degrees of freedom are kept fixed. The orientation angle $\theta$ is presented in degrees throughout the figures and discussion, whereas it is expressed in radians when used in the trigonometric terms of the proposed correlations. Consequently, the present study considers a one-parameter family of orientations rather than the complete three-dimensional orientation space. This configuration was chosen to consistently investigate the influence of orientation while maintaining a tractable number of simulations. At higher particle Reynolds numbers, typically above $\mathrm{Re_p} \approx 300$, the hydrodynamic coefficients of non-spherical particles may exhibit temporal fluctuations \citep{Cheron2024,Lain2025}. To account for this behaviour, all simulations are performed over sufficiently long physical times to reach statistically converged flow conditions. The presented hydrodynamic coefficients for such cases are obtained by averaging the instantaneous values over the final 40\% of the simulation time, thereby minimizing the influence of the remaining fluctuations while maintaining the statistically stationary response.

\begin{figure}[htbp!]
\centering
\includegraphics[width=0.7\textwidth]{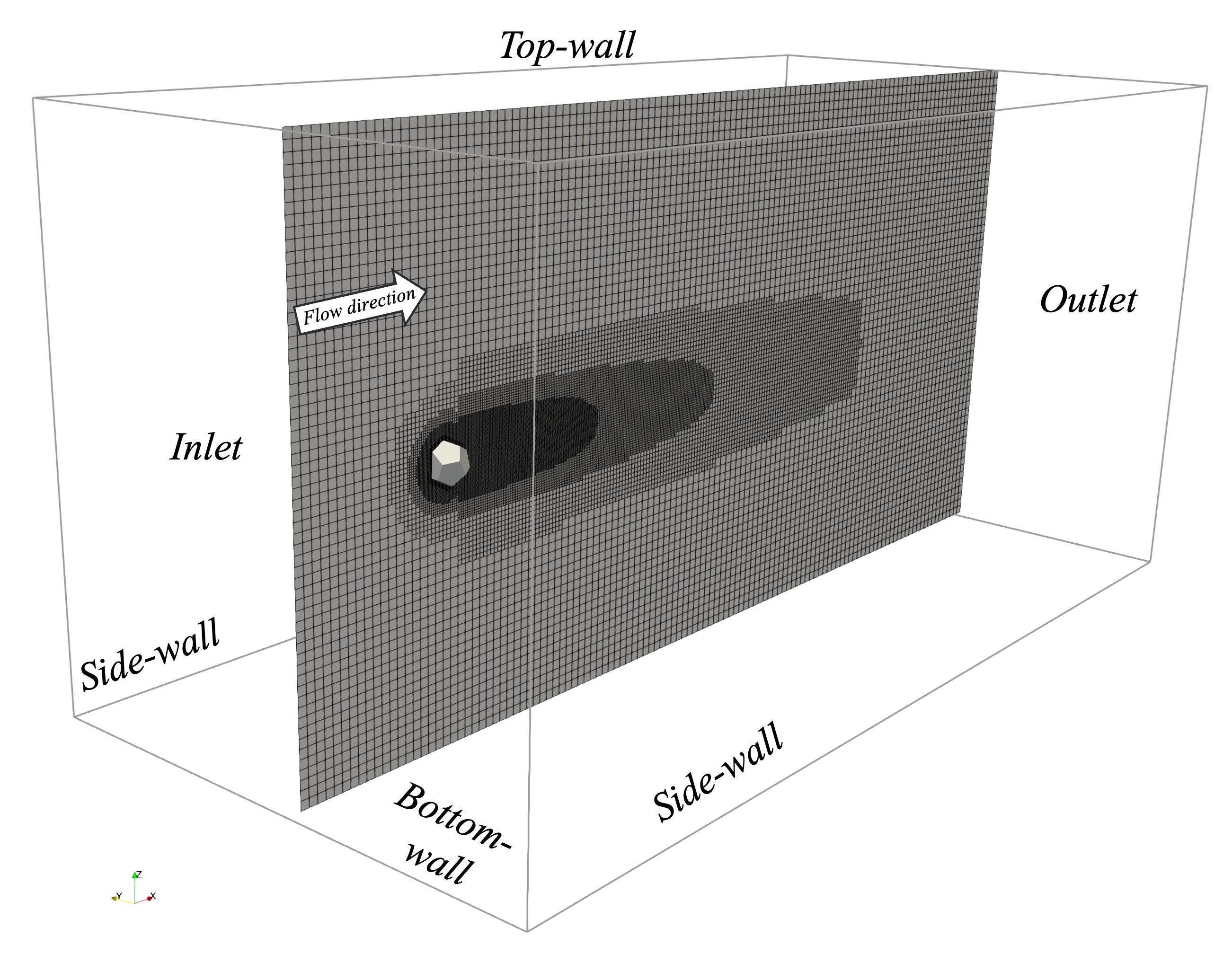}
\caption{Numerical domain showing a cross-section of the computational mesh with the corresponding boundary conditions and the refinement near the particle and behind the particle to reproduce any wake flow structures.}
\label{fig:NumericalDomain}
\end{figure}

The computational domain, a cross-section of the computational mesh, and the imposed boundary conditions are illustrated in figure~\ref{fig:NumericalDomain} and are described as follows. At the inlet, a uniform velocity profile is prescribed along with a zero pressure gradient. At the outlet, a zero-gradient condition is applied for the velocity field, while the pressure is fixed to a reference value of zero. A constant velocity boundary condition corresponding to the free-stream velocity is prescribed at the domain boundaries (top, bottom and side walls). To enhance the resolution near the particle surface while maintaining computational efficiency, the simulations employ adaptive mesh refinement (AMR) for the Eulerian fluid grid. The refinement strategy is based on a distance-based criterion, ensuring that the mesh is locally refined in the vicinity of the particle surface and within a prescribed downstream region, scaled with the particle equivalent diameter and elongated in the streamwise direction. Simulations are conducted over sufficiently long physical times to ensure the establishment of statistically stationary flow regimes.

To solve the discretized equations governing the fluid flow, the PR-DNS are carried out with second-order spatial and temporal accuracy. An implicit scheme is employed for the diffusion term, while the advection term is discretized using a central differencing scheme. The transient terms are integrated with a second-order backward Euler method. The time step is selected such that the Courant–Friedrichs–Lewy (CFL) number remains below 0.001 for all simulated cases. Such a small time step is required to ensure the accurate imposition of the no-slip condition on the Platonic particle surfaces within the immersed boundary method. This requirement is particularly important at low Reynolds numbers, where the flow is viscous-dominated and even small inaccuracies in the fluid--solid surface coupling can lead to noticeable errors in the computed hydrodynamic forces and torques. The chosen CFL threshold is found to effectively minimize numerical slip and ensure grid-independent force calculation. Across all particle orientations considered, the resulting penetration error at the immersed boundary remains below 4\%, confirming the accurate enforcement of the no-slip condition. Moreover, a minimum grid resolution of $D_{\mathrm{eq}}/\Delta x = 40$ is employed.

\section{Results and discussion}
\label{sec:Results and Discussion}

The drag, lift, and torque coefficients obtained from the PR-DNS simulations are presented and discussed in this section for the five Platonic particle shapes over a range of Reynolds numbers and particle orientations. First, the drag coefficient is analysed and compared with the PR-DNS data of \citet{Gai2023a}, the classical spherical correlation of \citet{Schiller1933}, and the non-spherical particle correlation of \citet{Haider1989}. It should be noted that the data of \citet{Gai2023a} are available only for three representative orientations, namely vertex-, edge-, and face-facing configurations. The PR-DNS data are subsequently used to derive correlations that account for particle shape, orientation relative to the flow direction, and particle Reynolds number. Thereafter, the effects of particle shape and orientation on the lift and torque coefficients are investigated. Finally, correlations for all hydrodynamic coefficients are developed based on the complete data, with parameters determined through a least-squares regression procedure.\\
As demonstrated in the following sections, for highly angular particles, the projected frontal area exposed to the incoming flow varies significantly with particle orientation. This effect is further analysed through the evolution of the surrounding flow field, illustrated by the velocity contours and streamlines for the tetrahedron, hexahedron, and octahedron in Figures~\ref{fig:TetrahedronContourRe},~\ref{fig:HexahedronContourRe}, and~\ref{fig:OctahedronContourRe}, respectively, for different Reynolds numbers and particle orientations. However, variations in projected frontal area alone do not fully explain the orientation dependence of the hydrodynamic forces. For faceted particles, the sharp edges and vertices introduce geometric discontinuities that strongly influence the flow-separation process. Once inertial effects become sufficiently important for separation to occur, these edges act as preferred, or ``pinned'', separation locations, in contrast to smooth particles, for which the separation point is determined primarily by the evolution of the boundary layer under an adverse pressure gradient. Rotating a faceted particle therefore changes not only the projected frontal area, but also the orientation of the edges and faces relative to the incoming flow. This modifies the upstream pressure distribution, the location of the pinned separation lines, and the subsequent wake development. The resulting differences in wake size and pressure recovery lead to orientation-dependent pressure drag, an effect that becomes increasingly pronounced as the Reynolds number increases and inertial forces dominate over viscous diffusion.

On the other hand, as the number of faces on the particle surface increases, the sphericity $\phi$ increases and the geometry approaches a more isotropic, sphere-like shape. Consequently, the dependence of both the projected frontal area and the flow-separation process on particle orientation becomes progressively weaker. The increasing number of smaller faces reduces the influence of individual edges as pinned separation locations, causing the separation behaviour to approach that of a smooth sphere, where the separation point is governed primarily by the pressure distribution rather than by geometric constraints. This behaviour is well represented by the more refined polyhedral shapes, such as the dodecahedron and icosahedron, which exhibit reduced angularity compared to the tetrahedron, hexahedron, and octahedron. Due to the inherent shape anisotropy of the Platonic solids, at least one component of the transverse force generally becomes non-zero for certain orientations of the tetrahedron and octahedron, even under steady flow conditions. While the magnitude of the lift coefficient ($C_\mathrm{L} = C_z$) may remain relatively small, particularly for highly spherical particles such as the dodecahedron and icosahedron, it is important to note that $C_\mathrm{L}$ is, in general, non-zero except for symmetry-aligned configurations. This behaviour arises from asymmetries in the pressure distribution induced by the instantaneous orientation of the particle relative to the incoming flow. In contrast, the spanwise force component $C_y$ is typically one order of magnitude smaller than $C_z$ and remains close to zero for all orientations. This behaviour is a direct consequence of the symmetry of the particle in the spanwise direction, which leads to a near-cancellation of pressure and viscous contributions when calculated over the particle surface. As a result, $C_y$ can be considered negligible compared to the lift force $C_z$.

\subsection{Drag coefficient}

This section examines the influence of particle shape, orientation, and Reynolds number on the drag coefficient. First, the flow surrounding the platonic-shaped particles is given, and later the PR-DNS results are compared with the PR-DNS data of~\citet{Gai2023a} and the classical Schiller--Naumann correlation for spherical particles~\citep{Schiller1933}, together with the Haider--Levenspiel correlation for non-spherical particles~\cite{Haider1989}, both of which provide a basis for the development of orientation-dependent drag correlations.

\subsubsection{Tetrahedron}
\label{sec:tetrahedron}

\begin{figure}[htbp!]
\centering
\includegraphics[width=0.95\textwidth]{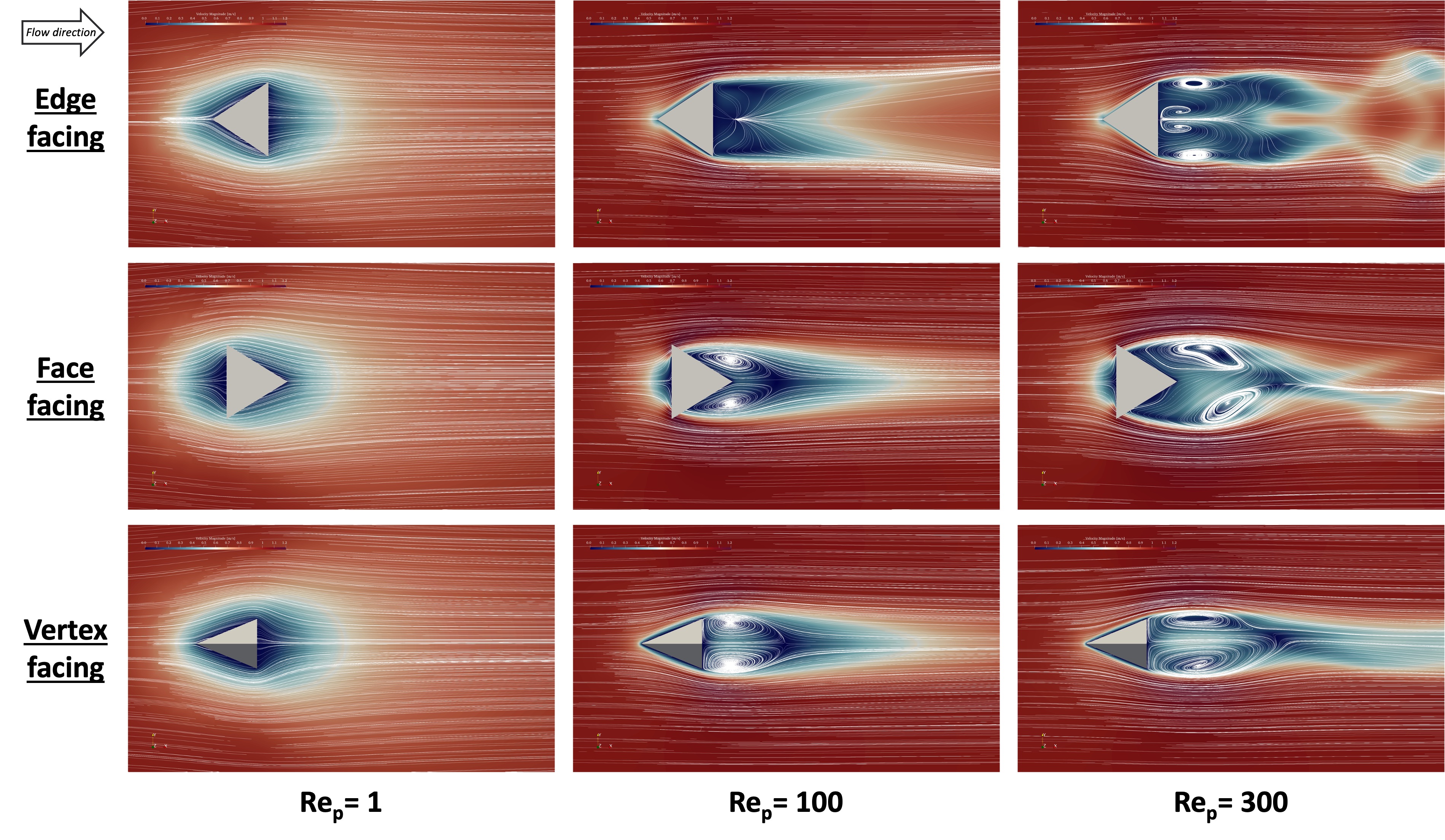}
\caption{Simulation results with tetrahedral particle. Fluid speed in a cross section through the middle of the particle, and flow streamlines surrounding the particle at different Reynolds number and three particle orientations, namely, edge-facing (top figures), face-facing (middle figures), and vertex-facing (bottom figures). The contours are shown on the central plane passing through the particle centre.}
\label{fig:TetrahedronContourRe}
\end{figure}

In order to show the evolution of the flow around the considered particles, figure~\ref{fig:TetrahedronContourRe} presents the velocity field around a tetrahedral particle at three distinct orientations, namely, edge-facing, face-facing, and vertex-facing, under varying particle Reynolds numbers, specifically $\mathrm{Re_p} = 1$, $100$ and $300$. These particular orientation are also presented in~\citet{Gai2023a}. The results are visualized through velocity magnitude contours, with superimposed streamlines to highlight the local flow structure. The velocity scale ranges from blue (low velocity) to red (high velocity), allowing clear identification of wake developing regions, with some separation zones, followed by the recirculation behaviour depending on the flow regime. At low Reynolds number, namely at $\mathrm{Re_p} = 1$, the flow remains viscous-dominated and is close to the Stokes-flow regime. As expected, the influence of particle orientation on the flow field is considerably weaker than at higher Reynolds numbers. For all three orientations, the flow is steady and remains fully attached around the tetrahedron, with no signs of flow separation. The wake development is limited and restricted to a small low-velocity region immediately downstream of the particle. Streamlines bend smoothly around the geometry and converge gradually behind the particle, which shows that viscous effects dominate over inertial effects. Under these conditions, the drag force is determined primarily by viscous stresses, resulting in only modest variations with particle orientation.

As the particle Reynolds number increases to $\mathrm{Re_p} = 100$, inertial effects become significant, and the influence of orientation is more pronounced. For the edge-facing orientation, flow separation occurs near the rear edges of the tetrahedron, forming a symmetric but narrow wake characterized by two recirculating vortices. In this orientation, the projected frontal area exposed to the incoming flow is comparatively small, allowing the flow to remain attached over a larger portion of the particle surface before separating. Consequently, the recirculation region is relatively compact, promoting a faster pressure recovery in the wake and resulting in a moderate pressure drag. The face-facing orientation leads to a much broader wake structure, with stronger and more prominent recirculation zones. This occurs because the frontal triangular face imposes a sudden blockage to the incoming flow, generating a large stagnation region upstream and a strong pressure gradient along the lateral edges. The abrupt turning of the flow around the sharp edges promotes early separation, causing the separated shear layer to move away from the particle surface and increasing the size of the low-pressure wake region. Here, one triangular face is oriented normal to the incoming flow, producing the largest projected frontal area among the considered configurations. However, the increase in drag is not caused directly by the larger projected area, but rather by its effect on the pressure distribution around the particle. The increased blockage generates a stronger stagnation region upstream and a more pronounced adverse pressure gradient near the sharp edges, promoting earlier flow separation and enlarging the low-pressure wake region. The vertex-facing configuration yields a wake structure that is intermediate between the two previous cases. Because the incoming flow first encounters a single vertex rather than an entire face, the fluid is progressively guided around the inclined surfaces before separating. This reduces the size of the separated region compared with the face-facing orientation while still producing a larger wake than the edge-facing configuration. This shows that even at moderate particle Reynolds numbers, particle orientation can significantly alter the flow topology, and consequently, the hydrodynamic forces acting on the non-spherical particle. At $\mathrm{Re_p} = 300$, the effects of particle orientation are even more pronounced. The flow becomes unsteady, with signs of vortex shedding and asymmetric wake development, particularly in the face-facing orientation. In this case, the wide and elongated wake means that the separated shear layers remain detached over a longer downstream distance before recombining. This delays pressure recovery behind the particle and promotes the formation of large coherent vortical structures. This behaviour implies a significant relative increase in drag due to both pressure and viscous contributions. For the edge-facing orientation at the same particle Reynolds number, the wake remains relatively narrow and well-organized, although unsteadiness begins to develop. The smaller and more coherent vortical structures show that the separated shear layers remain closer to the particle centreline, allowing the wake to recover more rapidly. As a result, the low-pressure region behind the particle is weakened, leading to a lower drag than for the face-facing orientation. The vertex-facing orientation again exhibits intermediate behaviour, with a more organized but broader wake than in the edge-facing configuration. The symmetry of the flow is slightly disrupted, but less so than in the face-facing case. In general, the influence of particle orientation becomes increasingly pronounced with Reynolds number. At low particle Reynolds numbers, the flow remains laminar and only weakly dependent on orientation. At higher particle Reynolds numbers, the face-facing orientation produces the largest wakes and the highest drag, whereas the edge-facing orientation generates narrower wakes and lower drag. The vertex-facing configuration shows intermediate behaviour.\\

\begin{figure}[htbp!]
\centering
\includegraphics[width=0.6\textwidth]{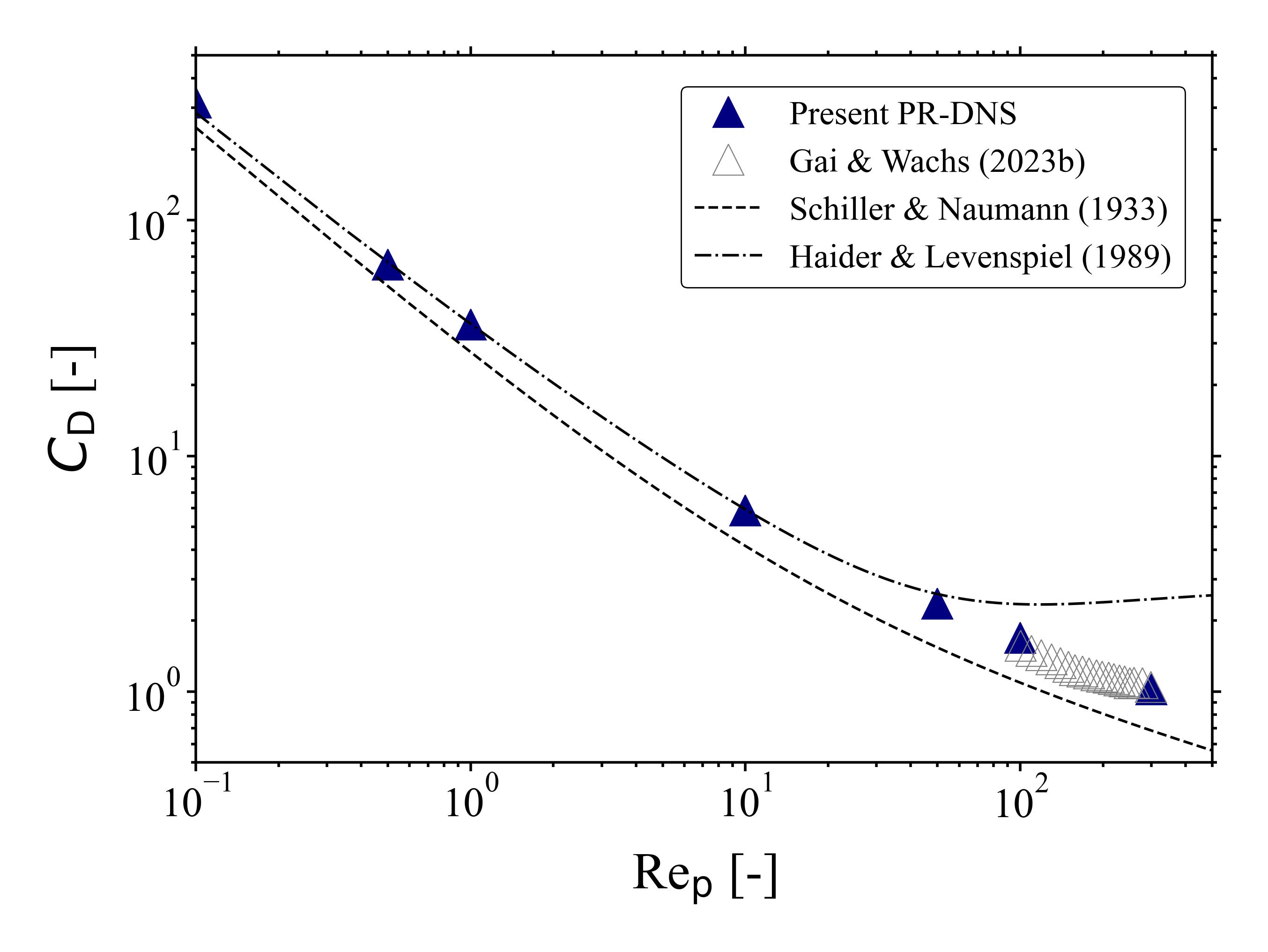}
\caption{Drag coefficient as a function of the Reynolds number for tetrahedral particle when the particle is oriented with the vertex against the flow. Lines correspond to the~\citet{Schiller1933} (dashed) and the~\citet{Haider1989} (dash-dotted) correlations, open symbol corresponds to the PR-DNS of~\citet{Gai2023a}, and the filled blue symbols represent the present PR-DNS.}
\label{fig:Tetra_Cd_Re}
\end{figure}

Figure~\ref{fig:Tetra_Cd_Re} presents the drag coefficient $C_\mathrm{D}$ as a function of the particle Reynolds number $\mathrm{Re_p}$ for a tetrahedral particle oriented with one of its vertex facing the flow. The vertical axis is plotted on a logarithmic scale to highlight variations across several orders of magnitude. The results from the present PR-DNS simulations (solid blue triangles) are compared with the PR-DNS data from~\citet{Gai2023a}, the~\citet{Schiller1933} correlation for spherical particles, and the~\citet{Haider1989} correlation for non-spherical particles. The drag coefficient obtained in this study consistently exceeds the prediction of the~\citet{Schiller1933} correlation along the entire range of particle Reynolds numbers, with deviations reaching over 50\% at intermediate $\mathrm{Re_p} \sim 50$. This shows the strong influence of particle shape and orientation, particularly outside the Stokes regime. The~\citet{Haider1989} correlations shows better agreement, especially in the intermediate regime ($\mathrm{Re_p} \sim 20$), where the predicted drag closely matches the PR-DNS values. 
However, it under-predicts the drag at lower Reynolds numbers and over-predicts it at higher ones. The results obtained from our simulations are in very good agreement with those reported by~\citet{Gai2023a}, showing close agreement in both the drag coefficient values and their variation with particle Reynolds number. Additionally, a decrease in the drag coefficient with increasing particle Reynolds number is observed in both cases, validating our PR-DNS simulations. Furthermore, this comparison confirms that classical drag correlations, even those accounting for sphericity~\citep{Haider1989}, may not accurately capture the behaviour of the sharply faceted tetrahedral particle when its orientation is fixed.\\

\begin{figure}[htbp!]
\centering
\includegraphics[width=1.0\textwidth]{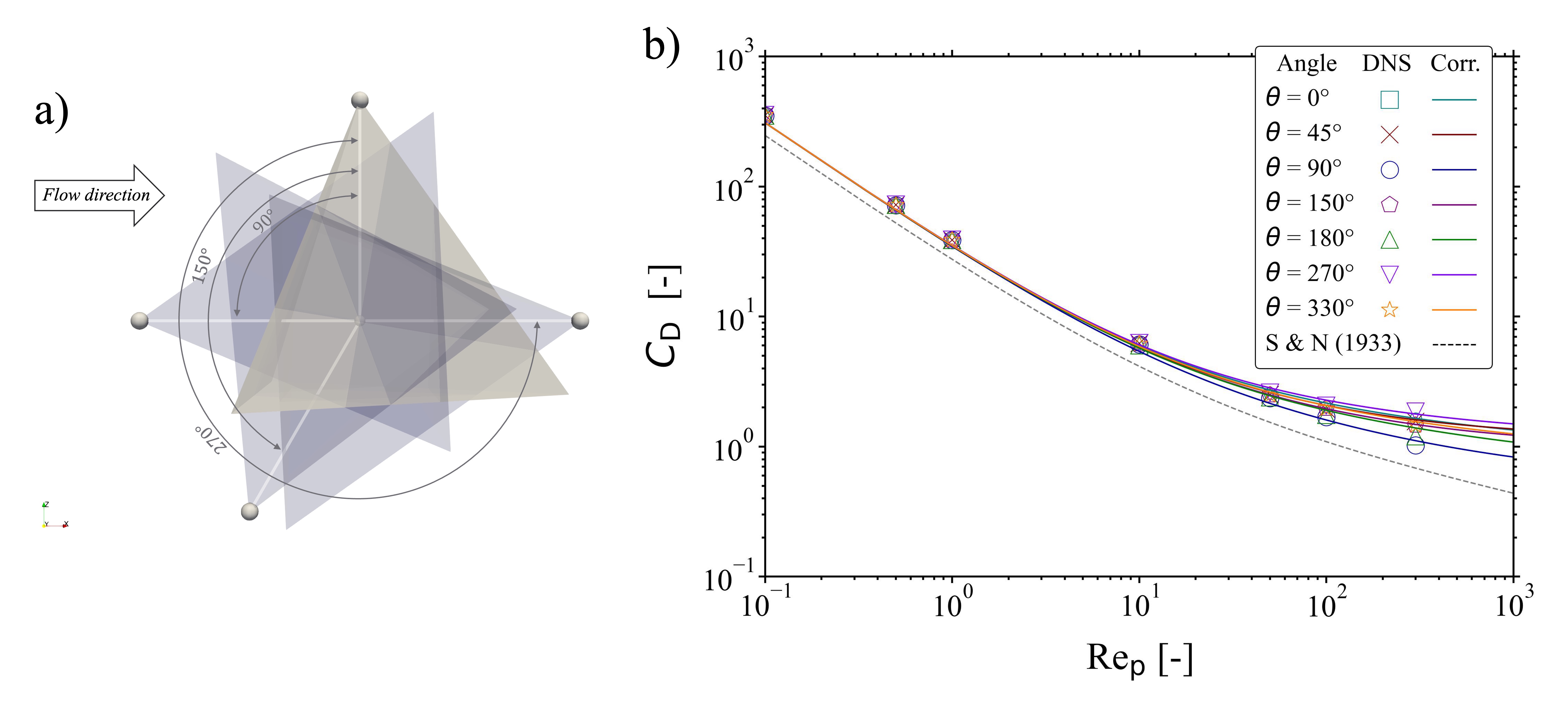}
\caption{a) Tetrahedron orientation angle with respect to the flow. The initial orientation is represented by the gray-colour tetrahedron and the small spheres are presented to guide the initial orientation, b) Drag coefficient as a function of the Reynolds number for tetrahedral particle. Symbols correspond to the PR-DNS data and the solid lines correspond to the proposed correlation.}
\label{fig:TetraCorrelation}
\end{figure}

Following the comparison with the PR-DNS data of \citet{Gai2023a}, additional simulations are carried out to quantify the effect of particle orientation on the drag coefficient. Beginning from a reference orientation, the particle is consistently rotated to analyse different orientations with respect to the main flow direction. The adopted orientation convention and rotation angles are illustrated in figure~\ref{fig:TetraCorrelation}a, where the initial orientation is represented by the gray-colour tetrahedron. The small spheres are presented to guide the initial orientation. Moreover, the corresponding drag coefficients are presented in figure~\ref{fig:TetraCorrelation}b as a function of the particle Reynolds number for the full set of orientations considered, obtained from our PR-DNS simulations. Both the particle Reynolds number and drag coefficient are displayed on logarithmic scales, to better highlight the trends for the investigated $\mathrm{Re_p}$-range. In the figure, the PR-DNS data are represented by symbols, while the classical correlation proposed by~\citet{Schiller1933} is shown as a dashed line. At low particle Reynolds numbers, the simulation results exhibit a trend similar to that predicted by~\citet{Schiller1933}; however, the drag coefficients are higher, primarily due to the reduced sphericity of the tetrahedron, as previously discussed by~\citet{Haider1989}. A pronounced deviation from the spherical correlation emerges as the particle is rotated with respect to the flow direction, particularly for Reynolds numbers exceeding $10$. Below this threshold, particle orientation has a negligible influence on the drag coefficient. Beyond it, orientation-dependent effects become increasingly significant, with differences in drag values surpassing 40\% at higher Reynolds numbers.

Based on the PR-DNS data, a correlation is derived to predict the drag coefficient of an isolated tetrahedron-shaped particle subjected to uniform flow. This formulation explicitly accounts for both the particle Reynolds number $\mathrm{Re_p}$ and its orientation relative to the flow direction, $\theta$. The proposed expression for the drag coefficient is given in equation~\ref{eq:tetraCorre} and presented with solid lines in figure~\ref{fig:TetraCorrelation}b.\\

\begin{equation}
\begin{aligned}
C_\mathrm{D} =\;& \frac{30}{\mathrm{Re_p}}\left(1 + 0.15\,\mathrm{Re_p}^{0.687}\right) \\
&+ \bigl(1 - H(\theta-\pi)\bigr)
\frac{0.45\left(1.275 + 0.25\cos \theta - \sin^2 \theta\right)}
{-0.8 - 0.15\,|\mathrm{Re_p}\sin \theta|^{0.125}
+ 1.65\,|\mathrm{Re_p}\cos \theta|^{0.0012}} \\
&+ H(\theta-\pi)
\frac{0.45\left(1.275 - 0.25\sin \theta + \cos^2 \theta\right)}
{-0.2 - 0.1\,|\mathrm{Re_p}\sin \theta|^{0.125}
+ 1.65\,|\mathrm{Re_p}\cos \theta|^{0.012}}
\end{aligned}
\label{eq:tetraCorre}
\end{equation}

where the Heaviside function is defined as:
\begin{equation*}
H(\theta-\pi)=
\begin{cases}
0, & \theta \le \pi,\\
1, & \theta > \pi.
\end{cases}
\end{equation*}

The drag coefficient expression proposed in equation~\ref{eq:tetraCorre} represents an extension of the classical correlation introduced by~\citet{Schiller1933}. While the original model provides a widely used empirical formulation for spherical particles, the modified expression incorporates additional terms to account for orientation-dependent effects, thereby extending its applicability to tetrahedral particles. The classical~\citet{Schiller1933} correlation, $C_\mathrm{D} = \frac{24}{\mathrm{Re_p}} \left(1 + 0.15 \mathrm{Re_p}^{0.687}\right)$, is valid only for spherical particles in the intermediate particle Reynolds-number regime and its formulation naturally assumes isotropic drag. 
Here, the proposed modified drag correlation preserves the general structure of the Schiller--Naumann model in the low-to-intermediate Reynolds-number regime through the first term, $\frac{30}{\mathrm{Re_p}} \left(1 + 0.15 \mathrm{Re_p}^{0.687} \right)$. This expression closely follow the original correlation but with an increased prefactor of $30$ instead of $24$, which reflects the higher drag associated with the angular shape of the tetrahedron. It also extends the widely used~\citet{Haider1989} correlation by explicitly incorporating orientation effects. 
A second term is introduced that accounts for the orientation angle $\theta$, as well as on trigonometric and weakly nonlinear power-law functions of the particle Reynolds number. These additional terms capture variations in the drag coefficient, arising from variations primarily in the facing area, wake dynamics, and flow separation, which are strongly influenced by particle orientation (see also figure~\ref{fig:TetrahedronContourRe}). Unlike the correlations of~\citet{Schiller1933} and~\citet{Haider1989}, the proposed model explicitly captures anisotropic, orientation-dependent drag behaviour. The angular correction is defined in a piecewise manner through the Heaviside function over the intervals $0^\circ \leq \theta \leq 180^\circ$ and $180^\circ< \theta \leq 360^\circ$. This formulation captures the asymmetric dependence of the drag coefficient on particle orientation,
which arises from the low geometric symmetry of the tetrahedron. For example, beyond $180^\circ$, the particle predominantly faces the flow with an edge rather than a flat face as in the $0^\circ$-$90^\circ$ range (figure~\ref{fig:TetraCorrelation}a), leading to substantially different wake development (see also figure~\ref{fig:TetrahedronContourRe}). The proposed correlation for the tetrahedral particles provides an accurate prediction of the drag coefficient over the entire range of particle Reynolds numbers from $\mathrm{Re_p} = 0.1$ up to $\mathrm{Re_p} = 300$, and for particle orientation angles ranging from $\theta = 0^\circ$ to $\theta = 360^\circ$. The lowest coefficient of determination obtained when applying this correlation for each orientation and $\mathrm{Re_p}$ is $R^2 = 0.9796$, confirming its high accuracy.

\begin{figure}[htbp!]
\centering
\includegraphics[width=1.0\textwidth]{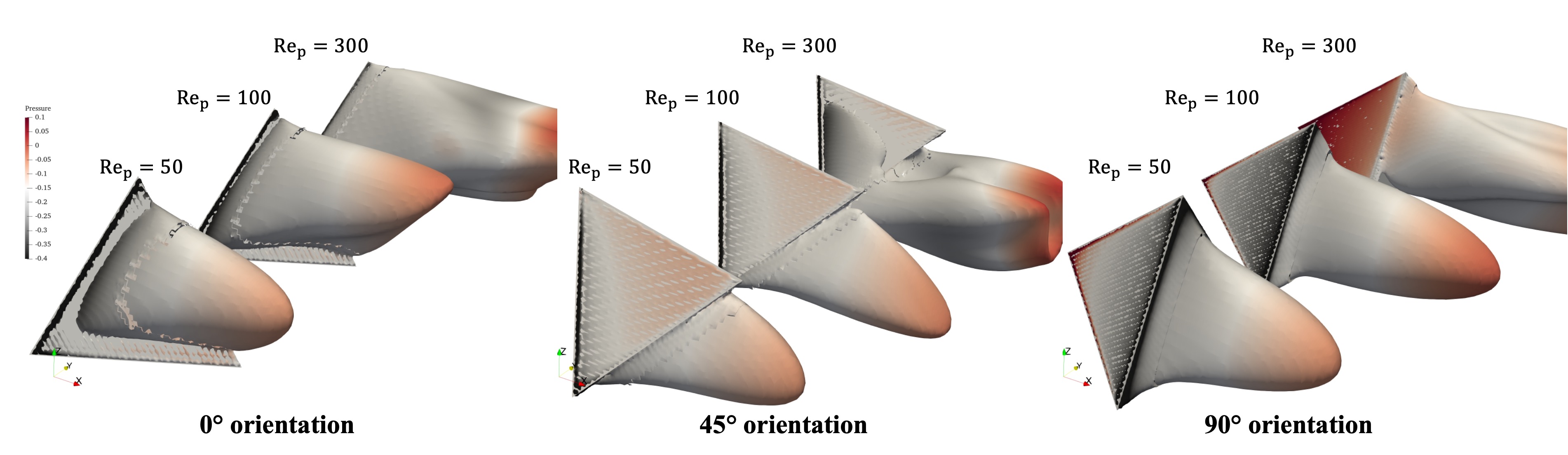}
\caption{Iso-surfaces of zero velocity for an tetrahedron particle at $\mathrm{Re_p}=50$, $100$, and $300$ and orientations $\theta = 0^\circ$, $45^\circ$, and $90^\circ$.}
\label{fig:TetrahedronIsoContour}
\end{figure}

A complementary perspective on the flow behaviour around the tetrahedral particle is obtained from the iso-surface of zero streamwise velocity for three investigated particle Reynolds numbers, namely, $\mathrm{Re_p}=50$, $100$, and $300$, as well as for three orientations $\theta = 0^\circ$, $45^\circ$, and $90^\circ$ in figure~\ref{fig:TetrahedronIsoContour}. Whereas figure~\ref{fig:TetrahedronContourRe} illustrates the velocity magnitude contours and streamlines on the central plane, the present three-dimensional iso-surface visualizations provide direct insight into the onset and evolution of flow separation and the structure of the recirculation region as both inertia and orientation are varied. Since the $u_x=0$ iso-surface approximately defines the boundary of the recirculation bubble, it provides a direct measure of the wake extent and how the separated flow evolves with particle Reynolds number and orientation. For $\theta = 0^\circ$, the tetrahedron adopts a pyramid-like orientation in which a triangular face is tilted away from the incoming flow and the downstream edge lies behind the particle. In this configuration, the flow encounters a moderately inclined surface rather than a frontal face, resulting in a reduced effective frontal area and a smoother redistribution of the incoming flow. At $\mathrm{Re_p}=50$, the iso-surface of $u_x=0$ encloses a compact and nearly symmetric recirculation region, showing that separation occurs relatively late along the rear edge. As the Reynolds number increases to $\mathrm{Re_p}=100$, the recirculation region extends farther upstream, which indicates that separation occurs earlier and the separated shear layers remain detached over a longer distance. At $\mathrm{Re_p}=300$, the recirculation region becomes larger and more elongated, still the wake remains relatively narrow because the inclined faces guide the flow smoothly toward the rear edges before separation, limiting the lateral growth of the recirculation region. This behaviour is similar to the face-facing configuration discussed earlier, where the flow remains attached over a substantial portion of the particle surface before separating. At $\theta = 45^\circ$, the particle presents its most front-facing orientation, with a triangular face directly normal to the incoming flow. This configuration produces the largest effective frontal blockage among the three orientations and generates a strong stagnation region in the bottom face. At $\mathrm{Re_p}=50$, the iso-surface of zero velocity shows a broader separated region, with the recirculation bubble extending farther downstream than in the $\theta = 0^\circ$ case. Increasing the particle Reynolds number to $\mathrm{Re_p}=100$ intensifies this effect, where the separated shear layers detach in a similar location, the recirculation region widens, and the wake becomes more asymmetric due to the strong adverse pressure gradients generated along the lateral edges of the frontal face. At $\mathrm{Re_p}=300$, the wake is substantially larger and more complex, with pronounced lateral spreading and with indications of instability. This behaviour reflects the face-facing orientation in figure~\ref{fig:TetrahedronContourRe}, where the frontal face induces strong stagnation and early separation, leading to the largest wakes and highest drag. For $\theta = 90^\circ$, the particle presents a vertex or sharp apex to the incoming flow. At $\mathrm{Re_p}=50$, separation occurs earlier than in the $\theta = 0^\circ$ case but later than in the $\theta = 45^\circ$ configuration, producing a recirculation region of intermediate size. At $\mathrm{Re_p}=100$, the wake becomes more three-dimensional, with the separated shear layers detaching along the inclined edges. At $\mathrm{Re_p}=300$, the recirculation region expands significantly, and the iso-surface reveals a wake that is both elongated and laterally broadened, however less asymmetric than in the $\theta = 45^\circ$ case. This intermediate behaviour is consistent with the vertex-facing configuration previously discussed, where the flow accelerates around the converging faces before separating. The iso-surfaces further demonstrate the separation-pinning mechanism introduced above. Although the recirculation region grows substantially with particle Reynolds number, the separation lines remain attached to the sharp edges of the tetrahedron. Consequently, rotating the particle changes which edges control the separation process, producing different wake topology and pressure-drag characteristic.

\subsubsection{Hexahedron}
\label{sec:hexahedron}

\begin{figure}[htbp!]
\centering
\includegraphics[width=0.95\textwidth]{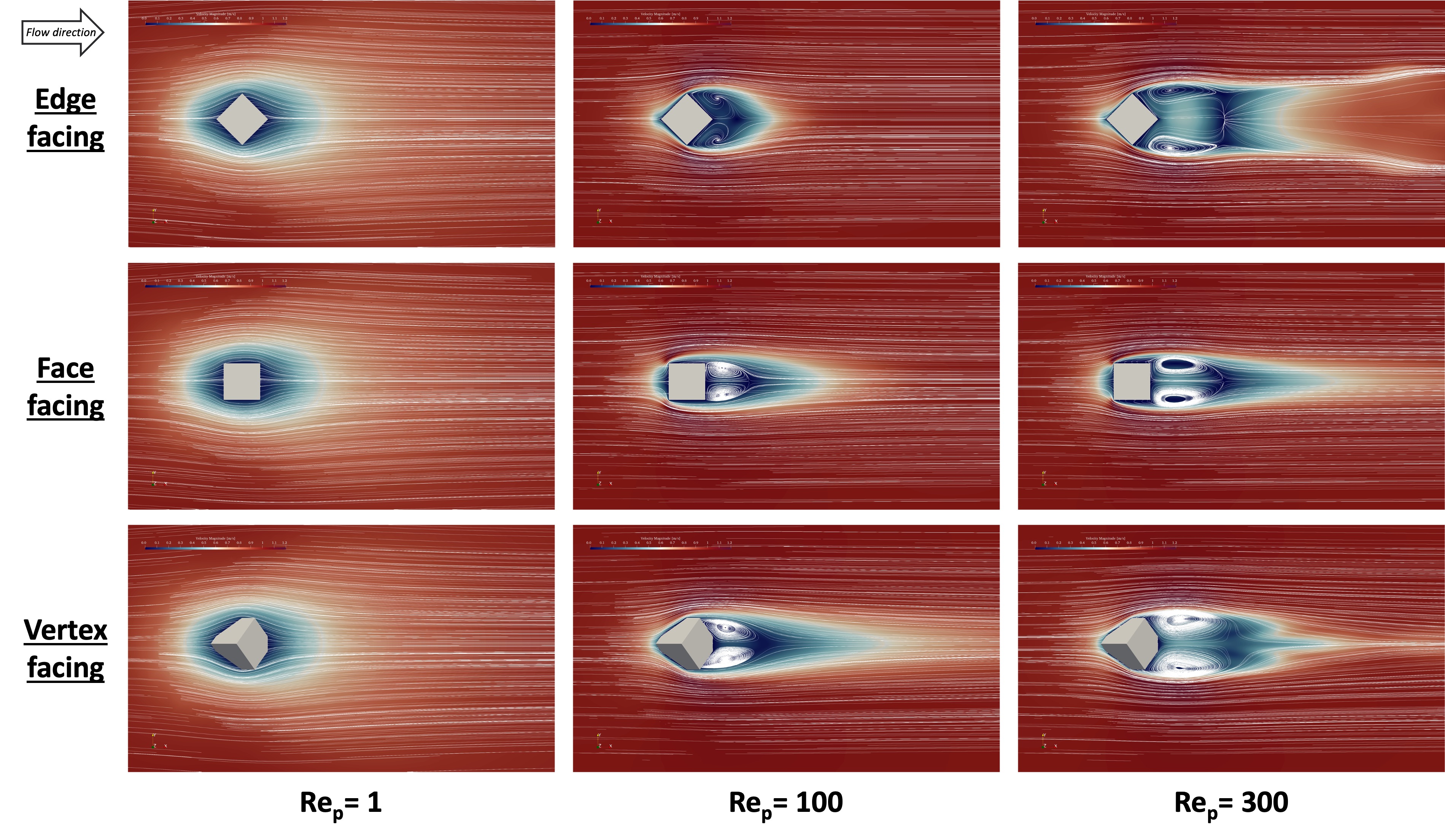}
\caption{Simulation results with hexahedral particle. Fluid speed in a cross section through the middle of the particle, and flow streamlines surrounding the particle at different Reynolds number and three particle orientations, namely, edge-facing (top figures), face-facing (middle figures), and vertex-facing (bottom figures). The contours are shown on the central plane passing through the particle centre.}
\label{fig:HexahedronContourRe}
\end{figure}

Figure~\ref{fig:HexahedronContourRe} presents PR-DNS results for the flow around a hexahedral particle subjected to a uniform flow at three Reynolds numbers, $\mathrm{Re_p} = 1$, $100$, and $300$. The flow fields are shown for three distinct particle orientations with respect to the incoming flow, namely, edge-, face-, and vertex-facing. Each panel shows contours of velocity magnitude, with superposed streamlines to visualize flow structure, wake development, and recirculation behaviour. The velocity field is colour-coded, with red indicating high-speed flow and blue for low-speed regions, typically associated with separation and vortex formation. At the lowest Reynolds number, $\mathrm{Re_p} = 1$, the flow regime is dominated by viscous effects and remains symmetric, laminar, and steady for all orientations. The streamlines lie smoothly around the particle, with no evidence of separation or wake formation. The velocity contours exhibit a low-velocity ``enclosure" surrounding the hexahedral particle’s surface, without visible recirculation zones. These flow characteristics are consistent with the creeping-regime behaviour, where inertia is negligible and particle orientation has minimal influence on the drag. The wake patterns are nearly identical across all three orientations, confirming that the hydrodynamic response at this Reynolds number is largely insensitive to how the hexahedron is aligned relative to the flow direction.

As the Reynolds number increases to $\mathrm{Re_p} = 100$, inertial effects become more pronounced, and the impact of particle orientation begins to be more evident. In the edge-facing orientation, the flow separates smoothly at the rear edges of the particle, forming two compact, symmetric recirculation zones downstream. Since the incoming flow first encounters an edge rather than an entire face, the fluid is divided more gradually around the particle, reducing the effective frontal blockage. Here, the wake remains narrow and aligned with the flow direction, and the streamlines reattach relatively quickly compared with the other two orientations. Consequently, the separated shear layers remain relatively close to the particle centreline, resulting in a narrow wake, and a moderate pressure drag. For the face-facing orientation, one square face is oriented normal to the incoming flow, producing the largest projected frontal area of the three configurations. The sharp leading edges generate a strong adverse pressure gradient that causes the boundary layer to separate almost immediately after the flow passes the front face. As a result, the separated shear layers move farther away from the particle, forming a broader recirculation region with delayed pressure recovery. This flow configuration leads to increased drag due to the larger separation region and the altered wake structure that prevents flow reattachment. In the vertex-facing case, the wake structure is intermediate between the edge- and face-facing orientations. The flow initially impinges on a single vertex and is subsequently guided along several inclined faces before separating. This gradual redistribution of the flow reduces the size of the separated region compared with the face-facing configuration while still producing a larger wake than in the edge-facing orientation. These differences will show that, at moderate Reynolds numbers, the particle orientation becomes a key factor in determining both wake dynamics and hydrodynamic coefficients. At $\mathrm{Re_p}=300$, inertial effects dominate over viscous diffusion, allowing the separated shear layers to persist farther downstream before dissipating. Consequently, the wakes become longer, increasingly asymmetric, and more susceptible to vortex shedding, which reflects the transition towards an unsteady flow regime. In the edge-facing case, the wake elongates and shows signs of unsteady behaviour. Although the wake becomes longer, the separated shear layers remain relatively coherent and close to the wake centreline. This limits the lateral expansion of the recirculation region and allows a more efficient pressure recovery than in the other orientations. Furthermore, the face-facing orientation produces two large-scale vortices downstream of the particle. Although the separation region becomes longer, the wake remains relatively confined laterally because the separated shear layers originate from opposite sharp edges of the square face and remain approximately parallel during their downstream evolution. Compared to the lower $\mathrm{Re_p}$, the separation zones grow in length, but the flow remains more organized with respect to the other orientations.  The vertex-facing orientation again represents a compromise between the other two configurations. Although the inclined faces reduce the abruptness of flow separation compared with the face-facing case, the wake still grows sufficiently to sustain larger vortical structures than in the edge-facing configuration. Notably, the evolution of the wake is governed by the interaction between the particle shape and the increasing importance of inertial effects. Orientations exposing a larger projected frontal area promote earlier flow separation, leading to broader wakes. Conversely, orientations that guide the flow around edges or inclined faces produce more compact recirculation regions. These mechanisms explain the increasing sensitivity of the hydrodynamic coefficients to particle orientation as the particle Reynolds number increases.

\begin{figure}[htbp!]
\centering
\includegraphics[width=0.6\textwidth]{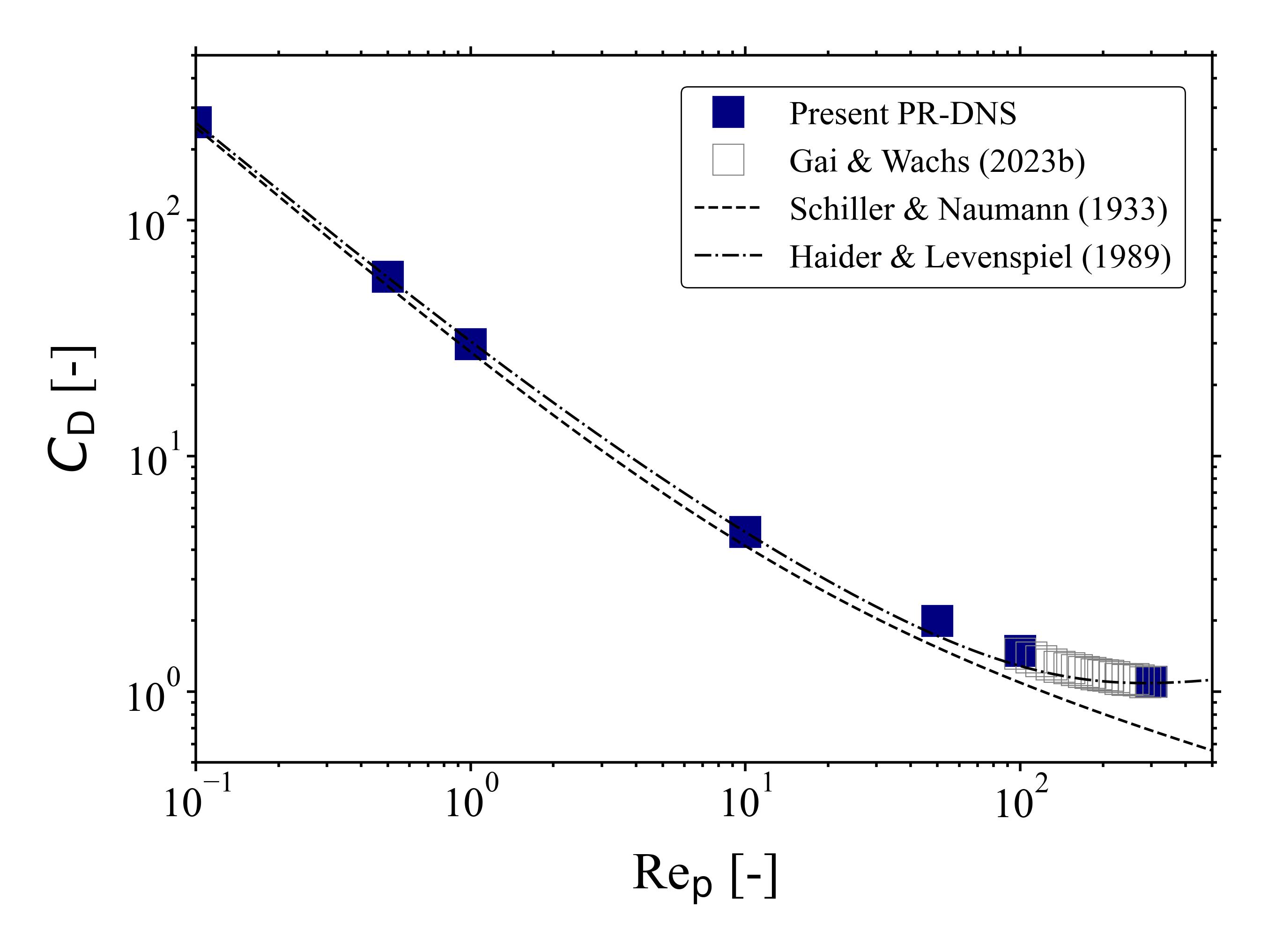}
\caption{Drag coefficient as a function of the Reynolds number for hexahedral particle when the particle is oriented with the vertex against the flow. Lines correspond to the~\citet{Schiller1933} (dashed) and the~\citet{Haider1989} (dash-dotted) correlations, open symbol corresponds to the PR-DNS of~\citet{Gai2023a}, and the filled blue symbols represent the present PR-DNS.}
\label{fig:Hexa_Cd_Re}
\end{figure}

To evaluate quantitatively the numerical results, figure~\ref{fig:Hexa_Cd_Re} illustrates the evolution of the drag coefficient $C_\mathrm{D}$ as a function of the particle Reynolds number $\mathrm{Re_p}$ for a hexahedral particle oriented such that one of its vertices points in the direction of the flow.  The results from the present PR-DNS (solid blue squares) are compared against the classical~\citet{Schiller1933} spherical correlation, the~\citet{Haider1989} correlation for non-spherical particles, and the PR-DNS data presented by~\citet{Gai2023a}.  It can be seen that at low Reynolds numbers ($\mathrm{Re_p} < 10$), the drag on the hexahedral particle is higher than that predicted by either empirical correlation, with deviations exceeding 20\% relative to, for instance, the~\citet{Schiller1933} correlation. This discrepancy shows the expected deficiency of spherical-based correlations for capturing the drag behaviour of faceted particles, particularly in the Stokes and transitional regimes. 
The~\citet{Haider1989} correlation provides an improved agreement, especially for $\mathrm{Re_p} \sim 50-100$, although it underestimates the drag coefficient at the lowest Reynolds numbers. At higher Reynolds numbers, $\mathrm{Re_p} > 100$, the present PR-DNS results show a slow decay in $C_\mathrm{D}$, with values remaining consistently above those predicted by both models. Compared to the PR-DNS data from~\cite{Gai2023a}, the present results are in good agreement, reproducing both the trends and the drag coefficient values for the hexahedral particle across their investigated particle Reynolds number range ($\mathrm{Re_p} \ge 100$). Moreover, it can be seen that the drag coefficient on a hexahedron is significantly higher than that of a sphere, especially in low to moderate particle Reynolds-number regimes. While the~\citet{Haider1989} model approximates the trend more closely than~\citet{Schiller1933}, it still fails to capture the full extent of the orientation-dependent drag increase observed with the PR-DNS results.

\begin{figure}[htbp!]
\centering
\includegraphics[width=1.0\textwidth]{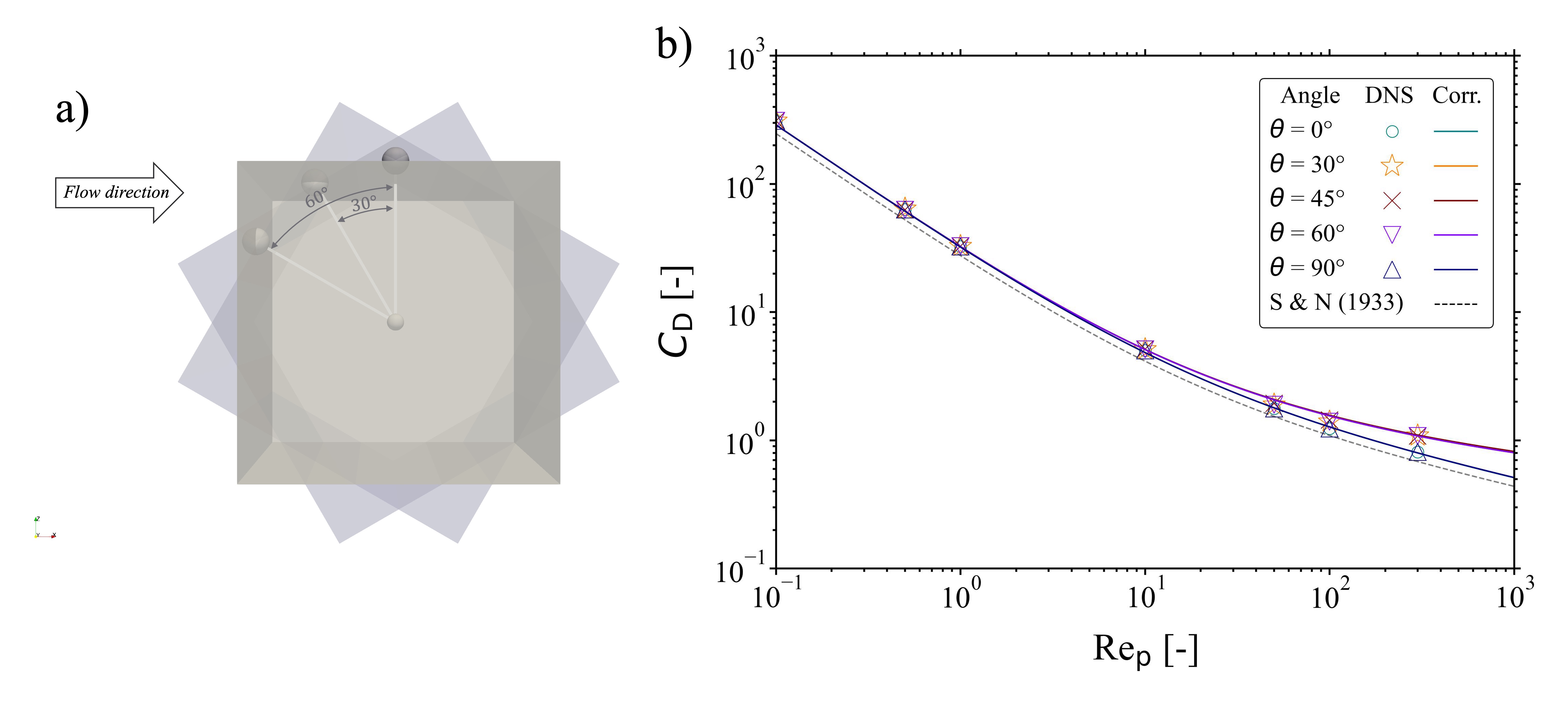}
\caption{a) Hexahedron orientation angle with respect to the flow, b) Drag coefficient as a function of the Reynolds number for hexahedral particle. Symbols correspond to the PR-DNS data and the solid lines correspond to the proposed correlation.}
\label{fig:HexaCorrelation}
\end{figure}

After the comparison against the PR-DNS data of \citet{Gai2023a}, additional simulations are performed to investigate the influence of particle orientation on the drag coefficient of the hexahedron. The adopted orientation convention and rotation angles are illustrated in figure~\ref{fig:HexaCorrelation}a, while the corresponding drag coefficients obtained from the PR-DNS simulations are shown in figure~\ref{fig:HexaCorrelation}b as a function of the particle Reynolds number. As observed for the tetrahedron, the drag coefficient follows a trend similar to the spherical correlation of \citet{Schiller1933} at low particle Reynolds numbers, although consistently higher values are obtained due to the non-spherical particle shape. The influence of particle orientation is relatively weak at low Reynolds numbers but becomes increasingly pronounced as $\mathrm{Re_p}$ increases, leading to variations in the drag coefficient among the different orientations considered. Based on the PR-DNS data, a correlation is proposed for the drag coefficient of an isolated hexahedral particle in uniform flow. The proposed correlation accounts explicitly for the combined influence of particle Reynolds number and particle orientation with respect to the incoming flow. The correlation is presented in equation~\ref{eq:hexaCorre}, while its predictions are shown as solid lines in figure~\ref{fig:HexaCorrelation}b.

\begin{equation}
C_\mathrm{D} = \frac{28}{\mathrm{Re_p}} \left(1.0 + 0.15 \mathrm{Re_p}^{0.687} \right)
+ \frac{0.45(1.2 + \left|\sin(2\theta)\right|)}{3.0 + 0.8 \left( \mathrm{Re_p} \left|\sin(2\theta)\right| \right)^{-0.2}}
\label{eq:hexaCorre}
\end{equation}

The first term in the correlation preserves the structural form of the expression proposed by~\citet{Schiller1933}, with a slightly increased prefactor, $28$ instead of $24$, obtained from a regression against the PR-DNS data for the hexahedral particle. The second term introduces an orientation-dependent correction via a sinusoidal function, which is symmetric about $\theta = 45^\circ$ and periodic over $90^\circ$. This formulation ensures that the drag force remains invariant under reflections about the diagonal plane, while maintaining a continuous and symmetric angular correction. Furthermore, the denominator of the orientation-dependent term incorporates a weakly negative exponent on the Reynolds number, enhancing the angular correction at low $\mathrm{Re_p}$ and progressively reducing its influence at higher Reynolds numbers. The proposed correlation for hexahedral particles accurately predicts the drag coefficient over the full range of particle Reynolds numbers, from $\mathrm{Re_p} = 0.1$ to $\mathrm{Re_p} = 300$, and for orientation angles ranging from $\theta = 0^\circ$ to $\theta = 360^\circ$. When applied to all orientations and Reynolds numbers, the minimum coefficient of determination is $R^2 = 0.9757$, confirming the high accuracy of the correlation.

\begin{figure}[htbp!]
\centering
\includegraphics[width=1.0\textwidth]{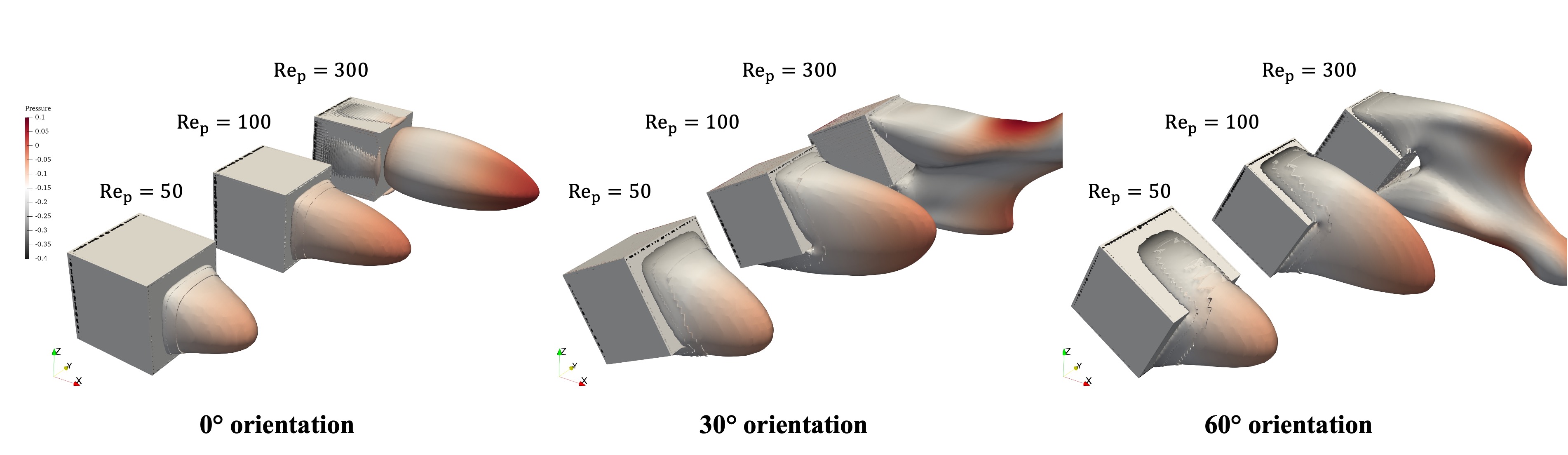}
\caption{Iso-surfaces of zero velocity for an hexahedral particle at $\mathrm{Re_p}=50$, $100$, and $300$ and orientations $\theta = 0^\circ$, $30^\circ$, and $60^\circ$.}
\label{fig:HexahedronIsoContour}
\end{figure}

To further clarify how these orientation-dependent flow characteristics translate into three-dimensional wake development, figure~\ref{fig:HexahedronIsoContour} complements the planar velocity–contour visualizations by illustrating the evolution of the recirculation region through iso-surfaces of zero streamwise velocity for the three investigated orientations and Reynolds numbers. For an orientation of $\theta = 0^\circ$, the particle presents a square face directly normal to the incoming flow, producing the largest projected frontal area among the three orientations. At $\mathrm{Re_p}=50$, the iso-surface of $u_x=0$ encloses a compact but clearly defined recirculation region, with separation occurring almost immediately at the sharp trailing edges of the front face. As the Reynolds number increases to $\mathrm{Re_p}=100$, the separated shear layers detach farther from the particle surface, and the wake widens due to the strong adverse pressure gradients generated along the lateral edges. At $\mathrm{Re_p}=300$, the recirculation region becomes substantially larger and more elongated, with coherent separated shear layers persisting over long downstream distances. The wake remains relatively symmetric because the separated flow originates from opposite edges of the square face, but its size and coherence show the strong influence of frontal blockage on separation behaviour and drag. At $\theta = 30^\circ$, the particle is rotated such that the incoming flow encounters a partially inclined face and a more complex arrangement of edges. This reduces the effective frontal area compared with the $\theta = 0^\circ$ case and modifies the distribution of stagnation pressure. At $\mathrm{Re_p}=50$, the iso-surface of zero velocity presents a recirculation region that is narrower and slightly skewed, with separation occurring later along the downstream edges. Increasing the particle Reynolds number to $\mathrm{Re_p}=100$ leads to earlier separation along the exposed edges and a more three-dimensional wake structure. The recirculation region expands laterally, and the wake becomes increasingly asymmetric due to the uneven exposure of edges and faces. At $\mathrm{Re_p}=300$, the wake grows substantially in both length and width, and the iso-surface shows pronounced lateral spreading and signs of instability. This behaviour shows the sensitivity of the separation process to moderate changes in particle inclination, consistent with the orientation-dependent wake development observed in figure~\ref{fig:HexahedronContourRe}. For $\theta = 60^\circ$, the particle is strongly inclined relative to the incoming flow, exposing a combination of edges and vertices rather than a full face. At $\mathrm{Re_p}=50$, although separation still occurs at the downstream edges, the inclined surfaces guide the flow more smoothly, reducing the extent of the separated region. As the Reynolds number increases to $\mathrm{Re_p}=100$, the wake becomes more three-dimensional, with the separated shear layers detaching along the inclined faces and forming a moderately skewed recirculation region. At $\mathrm{Re_p}=300$, the iso-surface shows a wake that is elongated and laterally broadened, though still relatively more organized than in the $\theta = 30^\circ$ case. Although the separation is pinned to the exposed edges, the orientation modifies the direction of the separated shear layers, resulting in a narrower wake than in the in the $\theta = 0^\circ$  (face-facing) configuration.

\subsubsection{Octahedron}
\label{sec:octahedron}

\begin{figure}[htbp!]
\centering
\includegraphics[width=0.95\textwidth]{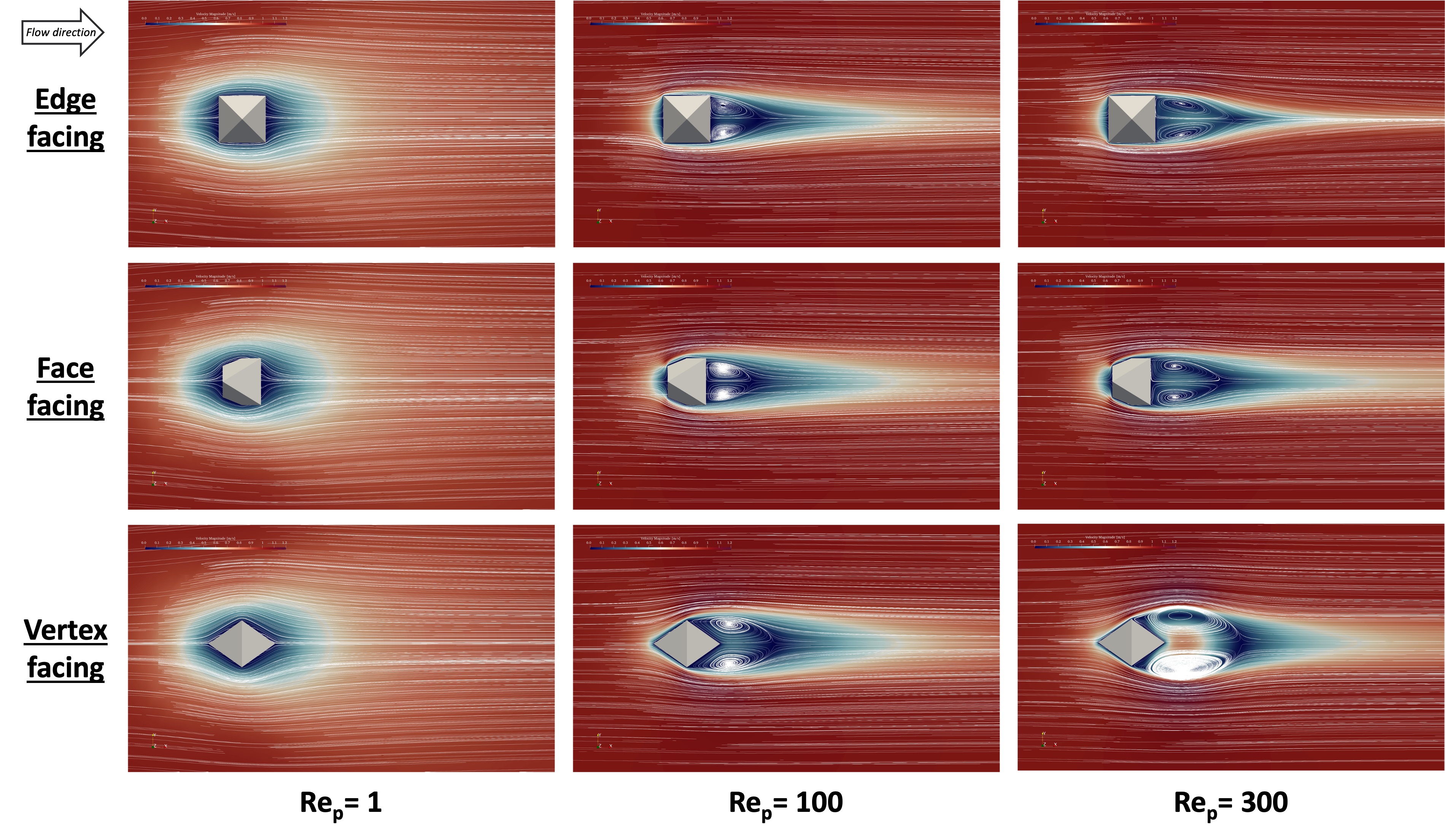}
\caption{Simulation results with octahedral particle, Fluid speed in a cross section through the middle of the particle, and flow streamlines surrounding a single particle at different Reynolds number and three particle orientations, namely, edge-facing (top figures), face-facing (middle figures), and vertex-facing (bottom figures). The contours are shown on the central plane passing through the particle centre.}
\label{fig:OctahedronContourRe}
\end{figure}

Figure~\ref{fig:OctahedronContourRe} shows the PR-DNS simulations of the flow around an octahedral particle subjected to uniform incoming flow. The flow fields are visualized via velocity magnitude contours with superposed streamlines, spanning three particle Reynolds numbers ($\mathrm{Re_p} = 1$, $100$, and $300$) and three particle orientations relative to the flow direction, namely, edge-facing, face-facing, and vertex-facing. Red colours correspond to high velocities, while blue indicates low-velocity regions. At the lowest Reynolds number ($\mathrm{Re_p} = 1$), the flow remains steady, symmetric, and entirely dominated by viscous forces. For all three orientations, streamlines wrap smoothly around the particle, and no separation or recirculation is observed. The octahedron generates a narrow, symmetric low-velocity wake that extends only a short distance downstream. The near-identical appearance of all three configurations confirms that at very low particle Reynolds numbers, orientation has negligible influence on the flow field or drag.

Moreover, as the Reynolds number increases to $\mathrm{Re_p} = 100$, inertial effects become evident, and orientation-dependent differences emerge. In the edge-facing configuration, flow separates at the trailing edges of the octahedron, creating two compact, symmetric recirculation regions. Since the incoming flow encounters an edge first, it is smoothly redistributed over the adjacent inclined faces before reaching the rear of the particle. This delays the growth of the separated region, producing a relatively narrow wake, which suggest a moderate increase in drag compared to the creeping flow regime ($\mathrm{Re_p} < 1$). In the face-facing orientation, one triangular face is directly exposed to the incoming flow, increasing the projected frontal area and modifying the local pressure distribution around the particle. Consequently, the separated shear layers move farther away from the particle centreline, forming a broader wake with larger recirculation zones and a corresponding increase in drag. In the vertex-facing orientation, the incoming flow is forced to accelerate around the converging inclined faces before separating along the rear surfaces. 
Although the flow initially follows the particle shape, the rapid expansion behind the particle promotes a larger separated region, resulting in a longer wake compared with the edge-facing configuration. These results show that, in this regime, the vertex-facing configuration is associated with slightly higher drag than the edge-facing and face-facing cases. At $\mathrm{Re_p}=300$, inertial effects dominate the flow dynamics, causing the separated shear layers to remain coherent over longer downstream distances. As a result, the wakes become substantially larger and increasingly susceptible to wake instabilities, making the influence of particle orientation much more pronounced. For the edge-facing configuration, the wake elongates considerably. The inclined faces continue to guide the separated flow toward the wake centreline, limiting the lateral growth of the recirculation region despite the increased Reynolds number. In the face-facing orientation, the larger frontal blockage promotes an earlier separation of the boundary layer and delays pressure recovery, allowing the wake to maintain its coherence over a longer downstream distance. The vertex-facing configuration displays the most complex behaviour because the flow accelerates around the inclined faces before undergoing an abrupt expansion downstream of the particle. This rapid change in flow direction produces strong shear layers with enhanced velocity gradients, which are more susceptible to instability when inertia dominates. Consequently, the wake expands laterally and exhibits stronger vortex activity compared with the other orientations. This shows that the vertex-facing orientation may yield the highest drag at $\mathrm{Re_p} = 300$, consistent with the extensive wake development and strong separation effects. These orientation-dependent differences are primarily a consequence of the octahedral geometry and its distinct flow-separation mechanism compared with the previous particles. Unlike the hexahedron, whose flat faces produce abrupt flow separation at sharp edges, the inclined triangular faces of the octahedron guide the incoming flow around the particle before separation occurs. Consequently, differences between orientations are governed primarily by how the inclined faces redistribute the flow and modify the downstream velocity field. Although octahedral particles exhibit only a weak orientation dependence at low Reynolds numbers, orientation effects become increasingly significant with increasing flow inertia, where particle orientation modifies the projected frontal area and the direction in which the inclined faces guide the flow, thereby controlling the separation process and wake development. This again highlights that particle orientation should be accounted for in the hydrodynamic force correlations employed, for instance, in Euler--Lagrange simulations.\\

\begin{figure}[htbp!]
\centering
\includegraphics[width=0.6\textwidth]{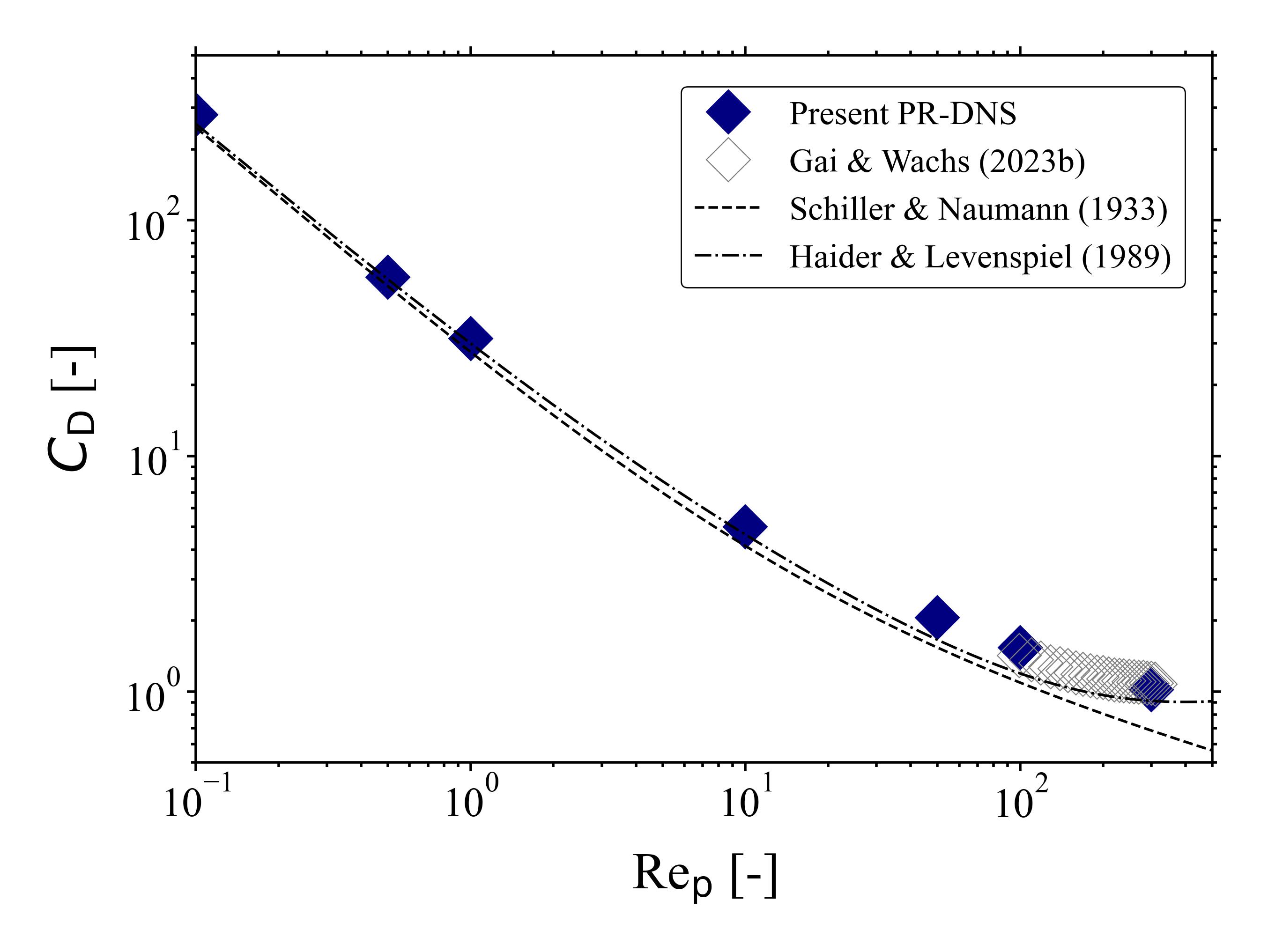}
\caption{Drag coefficient as a function of the Reynolds number for octahedral particle when the particle is oriented with the vertex against the flow. Lines correspond to the~\citet{Schiller1933} (dashed) and the~\citet{Haider1989} (dash-dotted) correlations, open symbol corresponds to the PR-DNS of~\citet{Gai2023a}, and the filled blue symbols represent the present PR-DNS.}
\label{fig:Octa_Cd_Re}
\end{figure}

A comparison between the present PR-DNS results and existing numerical data and drag correlations is presented in figure~\ref{fig:Octa_Cd_Re}. The figure shows the variation of the drag coefficient $C_\mathrm{D}$ with the particle Reynolds number $\mathrm{Re_p}$ for an octahedral particle oriented with one vertex facing the incoming flow. The present PR-DNS results (solid blue diamonds) are compared with the spherical correlation of~\citet{Schiller1933}, the non-spherical particle correlation of~\citet{Haider1989}, and the PR-DNS data reported by~\citet{Gai2023a}. It can be seen that at low Reynolds numbers ($\mathrm{Re_p} < 10$), the drag on the octahedrons is significantly higher than that predicted by either~\citet{Schiller1933} or the~\citet{Haider1989} correlations, with deviations exceeding 20\% relative to the~\citet{Schiller1933} correlation. On the other hand, the~\citet{Haider1989} correlation provides lower values for all considered particle Reynolds number. At higher Reynolds numbers ($\mathrm{Re_p} > 50$), the present PR-DNS results show a continuous decay in $C_\mathrm{D}$, with values remaining above those predicted by the empirical models. Compared to the PR-DNS data from~\citet{Gai2023a}, the present results are slightly higher for particle Reynolds number of 100, but slightly lower for particle $\mathrm{Re_p} = 300$, with differences of about 2--5\%. In general the drag coefficient on an octahedron is significantly higher than that of a sphere, especially in low to high particle Reynolds-number regimes.

\begin{figure}[htbp!]
\centering
\includegraphics[width=1.0\textwidth]{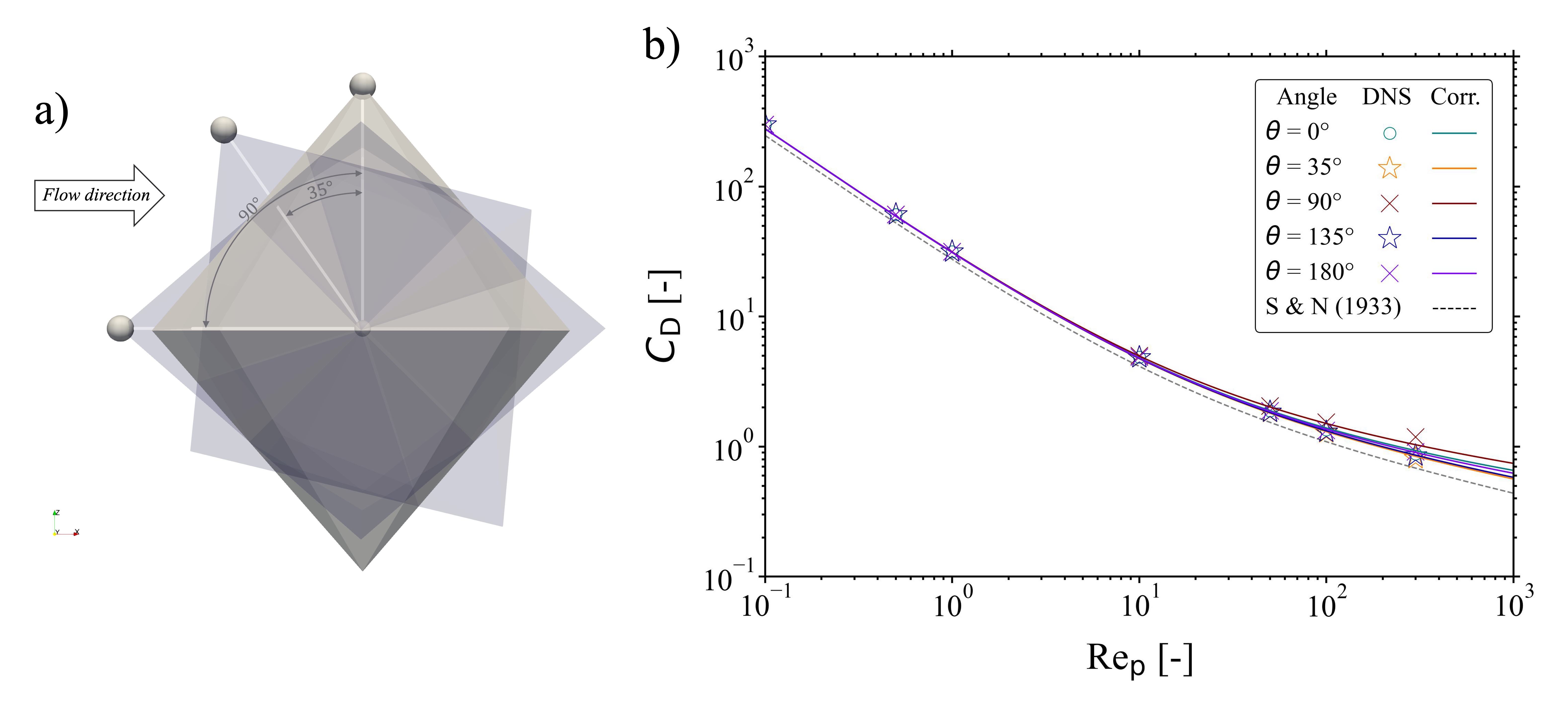}
\caption{a) Octahedron orientation angle with respect to the flow, b) Drag coefficient as a function of the Reynolds number for octahedral particle. Symbols correspond to the PR-DNS data and the solid lines correspond to the proposed correlation.}
\label{fig:OctaCorrelation}
\end{figure}

After the comparison against the PR-DNS data of \citet{Gai2023a}, additional simulations are performed to investigate the influence of particle orientation on the drag of the octahedron. Figure~\ref{fig:OctaCorrelation} summarizes the results. The adopted orientation convention is illustrated in figure~\ref{fig:OctaCorrelation}a, where the gray octahedron denotes the reference configuration and the auxiliary spheres are included to aid in visualizing the particle rotation. The corresponding PR-DNS results are shown in figure~\ref{fig:OctaCorrelation}b. Consistent with the trends observed for the tetrahedral and hexahedral particles, the influence of orientation is weak at low Reynolds numbers but becomes increasingly pronounced as $\mathrm{Re_p}$ increases. The PR-DNS data is subsequently employed to develop the orientation-dependent correlation given in equation~\ref{eq:octaCorre} and presented with solid lines in figure~\ref{fig:OctaCorrelation}b. The proposed model extends the previous orientation-dependent formulations for the octahedral particle by incorporating symmetric angular contribution and is defined as:

\begin{equation}
C_\mathrm{D} = \frac{27}{\mathrm{Re_p}} \left(1.0 + 0.15 \mathrm{Re_p}^{0.687} \right) +
\frac{0.45\left( 1.0 + \cos^2\theta + 4.0\left|\sin\theta\right|^5 \right)}
{0.5 + 5.0\cos^2\theta + 6.0\left(\mathrm{Re_p}\left|\sin\theta\right|\right)^{0.05}}
\label{eq:octaCorre}
\end{equation}

As in the previous Platonic shapes, the first term is based on the correlation proposed by~\citet{Schiller1933}, with the prefactor increased to $27$ to reproduce the hydrodynamic response of the octahedron observed in the PR-DNS. The second term introduces an orientation-dependent correction based on the symmetric functions $\cos^2\theta$ and $|\sin\theta|$. This formulation ensures that the drag coefficient is identical for orientations related by the shape symmetry of the octahedron (for instance at $\theta=0^\circ$ and $180^\circ$), thereby providing a continuous and physically consistent description over the entire orientation range. The numerator combines a quadratic cosine contribution with a strongly non-linear $|\sin\theta|^5$ term, increasing the sensitivity of the drag coefficient at intermediate orientations. The denominator is similarly modulated by $\cos^2\theta$ and incorporates a weak Reynolds number dependence through the term $(\mathrm{Re_p} |\sin\theta|)^{0.05}$, allowing the influence of particle orientation to vary smoothly across the investigated particle Reynolds number range. This correlation structure presents directional symmetry as observed in the drag behaviour in figure~\ref{fig:Octa_Cd_Re}. The smooth transition enabled by the used trigonometric functions allows the model to capture gradual changes in projected area and flow separation as the orientation varies (see also figure~\ref{fig:OctahedronContourRe}). Comparing the correlation to both classical models and previous symmetric orientation-dependent correlations, equation~\ref{eq:octaCorre} offers improved flexibility for capturing anisotropic hydrodynamic behaviour with a continuous angular representation for the octahedral particles. The proposed drag correlation for octahedral particles accurately predicts its coefficient over the full range of particle Reynolds numbers, from $\mathrm{Re_p} = 0.1$ to $\mathrm{Re_p} = 300$, and for orientation angles ranging from $\theta = 0^\circ$ to $\theta = 360^\circ$. When applied to all orientations and Reynolds numbers, the obtained minimum coefficient of determination corresponds to $R^2 = 0.9855$, which present a high accuracy.

\begin{figure}[htbp!]
\centering
\includegraphics[width=1.0\textwidth]{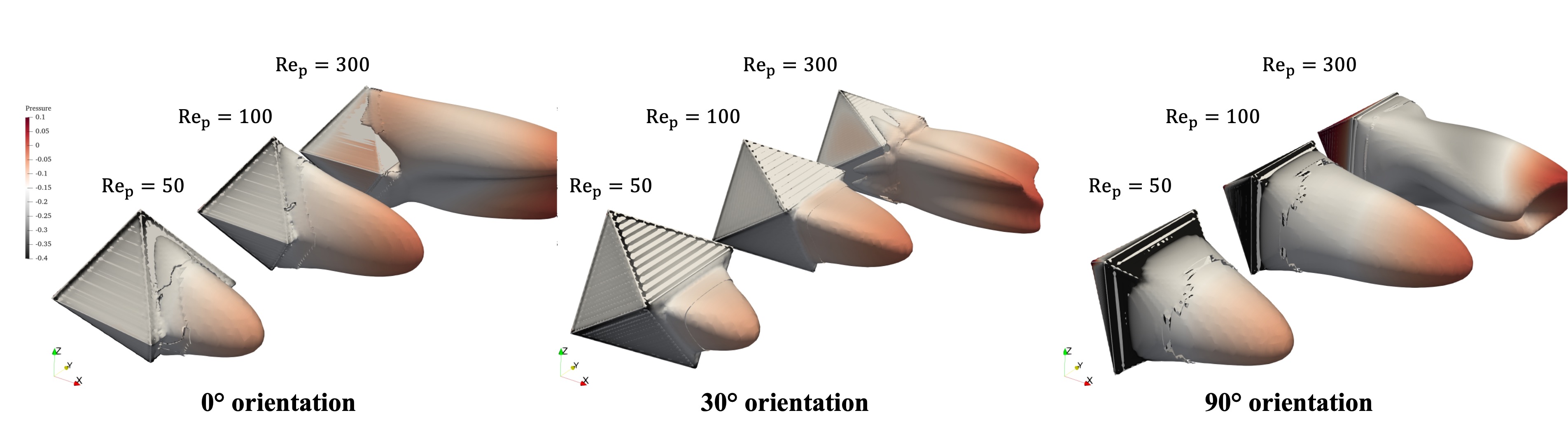}
\caption{Iso-surfaces of zero velocity for an octahedral particle at $\mathrm{Re_p}=50$, $100$, and $300$ and orientations $\theta = 0^\circ$, $30^\circ$, and $90^\circ$.}
\label{fig:OctahedronIsoContour}
\end{figure}

The physical basis of this correlation becomes more apparent when examining the evolution of the separation point and wake topology across both Reynolds number and orientation. A complementary view of the flow behaviour (figure~\ref{fig:OctahedronContourRe}) is obtained from the iso-surface of zero streamwise velocity ($u_x=0$) for three of the investigated particle Reynolds numbers, $\mathrm{Re_p}=50$, $100$, and $300$, and the three octahedron orientations, $\theta = 0^\circ$, $30^\circ$, and $90^\circ$ in figure~\ref{fig:OctahedronIsoContour}. These iso-surfaces provide direct insight into how the recirculation region develops as both inertia and orientation are varied, thereby explaining the orientation-dependent terms introduced in the proposed drag correlation. For $\theta = 0^\circ$, the particle presents, with two triangular faces, a large area normal to the incoming flow, resulting in an extended stagnation region and separation occurring predominantly along the downstream edges of this frontal facet. At $\mathrm{Re_p}=50$, the wake remains compact and nearly symmetric, with the separation line located close to the trailing edges. As the Reynolds number increases to $\mathrm{Re_p}=100$ and $\mathrm{Re_p}=300$, the separation remains pinned to the downstream edges, while the separated shear layers remain coherent over longer downstream distances, producing a larger recirculation region. Despite this growth, the wake maintains a largely symmetric structure due to the geometric alignment of the particle with the flow. At an orientation of $\theta = 30^\circ$, the particle is inclined relative to the flow direction, modifying both the projected area and the exposure of sharp edges. When a $\mathrm{Re_p}=50$ is considered, the iso-surface of $u_x=0$ already attaches along oblique edges and partially inclined facets, producing a wake that is more three-dimensional and moderately skewed. Increasing the Reynolds number to $\mathrm{Re_p}=100$ and $\mathrm{Re_p}=300$ intensifies this behaviour, here the separation remains pinned to the exposed inclined edges, but the direction of the separated shear layers changes, producing a larger and increasingly asymmetric wake. This behaviour explains the strong orientation dependence captured by the proposed correlation, whose non-linear angular terms reproduce the rapid increase in wake size and drag observed at intermediate orientations. For $\theta = 90^\circ$, the inclination is more pronounced and the flow encounters a combination of inclined facets and strongly exposed edges. At $\mathrm{Re_p}=50$, separation already occurs earlier along these edges, and the wake shows a clear three-dimensional character. As the Reynolds number increases to $\mathrm{Re_p}=100$ and $\mathrm{Re_p}=300$, the iso-surface of zero velocity encloses a substantially larger and more asymmetric recirculation region. The separation points migrate further upstream toward the most exposed edges, and the wake becomes strongly skewed with respect to the main flow direction. This pronounced wake asymmetry explains the strong orientation dependence of the drag coefficient and motivates the non-linear angular terms incorporated into the proposed correlation.

In summary, the velocity contours and streamlines obtained from the PR-DNS for the highly angular (low-sphericity) particles, namely the tetrahedron, hexahedron, and octahedron, demonstrate the coupled influence of particle shape, orientation, and particle Reynolds number on the surrounding flow field. In the low-Reynolds-number regime, the flow is dominated by viscous effects and remains largely attached to the particle surface for all considered orientations. Consequently, the wake remains nearly symmetric, and only minor differences are observed between the various particle alignments, resulting in a weak orientation dependence of the hydrodynamic forces. As the flow enters the intermediate Reynolds-number regime, inertial effects become increasingly important and orientation-dependent differences begin to emerge. Flow separation occurs behind the particles, leading to the formation of recirculation regions whose size and topology depend strongly on the exposed particle surfaces and the resulting pressure distribution. The hexahedron generally produces the largest recirculation regions, the tetrahedron exhibits the strongest wake asymmetry, and the octahedron displays an intermediate behaviour. At high Reynolds numbers, inertial effects dominate the flow dynamics and the influence of particle geometry becomes considerably more pronounced. The wakes become substantially larger and more complex, producing significant orientation-dependent variations in the drag coefficient. Among the considered particles, the tetrahedron exhibits the greatest degree of wake asymmetry, while the octahedron maintains a comparatively more balanced, although still strongly orientation-dependent, flow field.

The observed orientation dependence of faceted particles can be related to their distinct flow-separation mechanism compared with smooth spherical particles. For particles composed of planar faces, such as the tetrahedron, hexahedron, and octahedron, sharp edges and vertices introduce geometric discontinuities that strongly constrain the location at which the flow separates from the particle surface. Unlike a smooth sphere, where separation develops as a consequence of the evolving adverse pressure gradient along the curved surface, faceted particles impose preferred separation locations associated with their edges and corners, where the abrupt change in surface orientation prevents the flow from continuously following the particle geometry once inertial effects become significant. Therefore, rotating the particle modifies the position, inclination, and number of these separation sites exposed to the incoming flow, leading to changes in the pressure distribution, wake topology, and pressure drag contribution. At low particle Reynolds numbers, viscous stresses dominate the force balance and rapidly diffuse velocity gradients generated by the geometric discontinuities, reducing differences in wake structure between orientations. However, as particle Reynolds number increases, the separated shear layers become less affected by viscous diffusion and preserve the influence of the geometrically constrained separation points over longer downstream distances. This amplifies the differences in wake development and pressure recovery between orientations, resulting in stronger variations in hydrodynamic forces. Conversely, particles with higher sphericity, such as the dodecahedron and icosahedron, which are presented below in detail, possess a larger number of smaller faces and a geometry that more closely approximates a smooth spherical surface. In these cases, the influence of individual edges on the separation process is reduced, and the flow experiences a more gradual variation in surface direction (see below figure~\ref{fig:DodecahedronIsoContour} and figure~\ref{fig:IcosahedronIsoContour}). Consequently, the wake becomes less sensitive to particle orientation, approaching the behaviour of a sphere, where separation is primarily governed by the pressure distribution rather than by sharp geometric constraints.\\

In the following subsections, an analysis is presented for particles with higher sphericity, namely, dodecahedron, and icosahedron. As shown below, the results increasingly resemble those of a perfect sphere. Consequently, despite the orientation-dependent nature of the analysis, only minor differences are observed between the various orientations.

\subsubsection{Dodecahedron}
\label{sec:dodecahedron}

\begin{figure}[htbp!]
\centering
\includegraphics[width=0.95\textwidth]{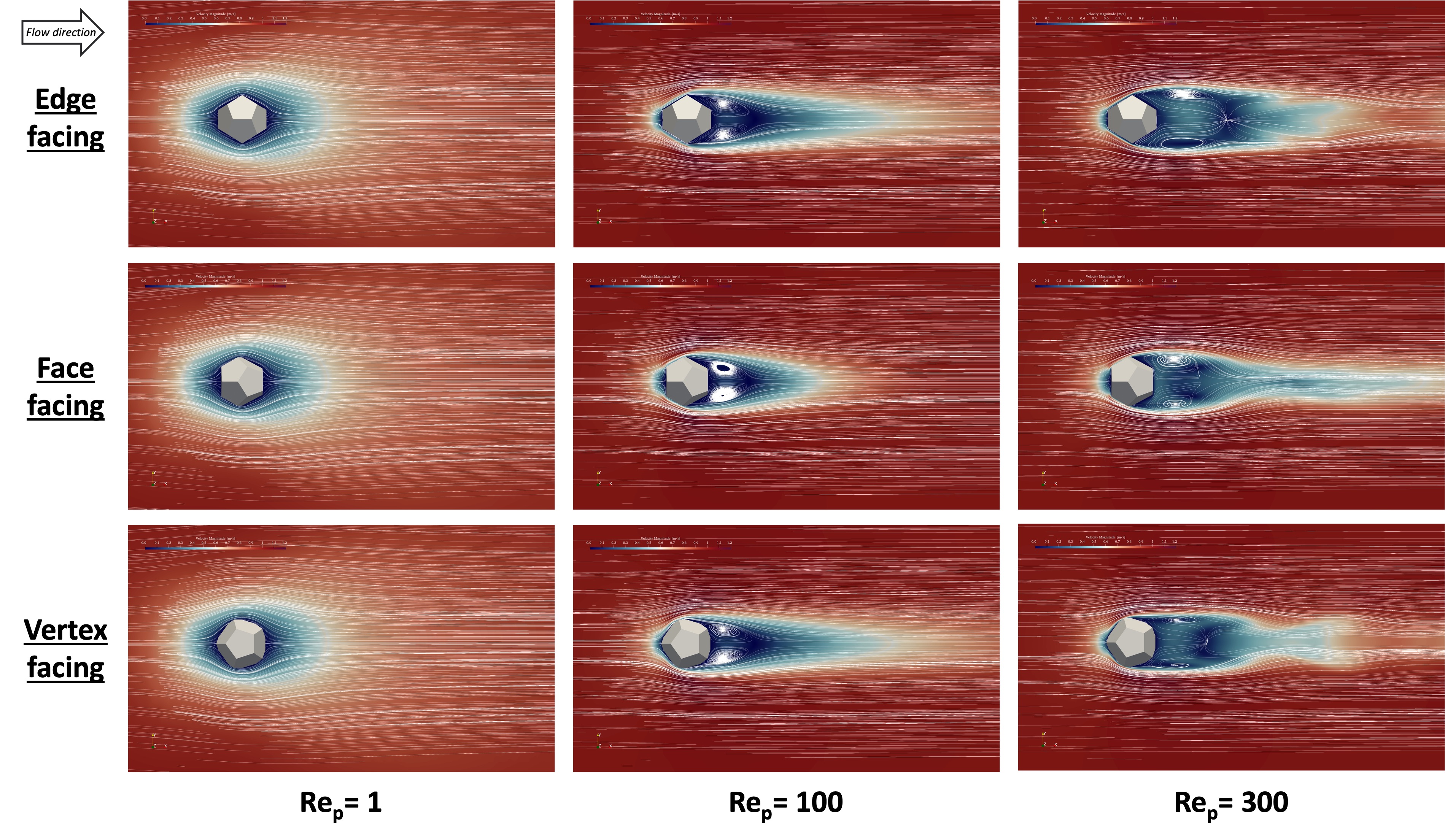}
\caption{Simulation results with dodecahedral particle, Fluid speed in a cross section through the middle of the particle, and flow streamlines surrounding a single particle at different Reynolds number and three particle orientations, namely, edge-facing (top figures), face-facing (middle figures), and vertex-facing (bottom figures). The contours are shown on the central plane passing through the particle centre.}
\label{fig:DodecahedronContourRe}
\end{figure}

Figure~\ref{fig:DodecahedronContourRe} presents velocity magnitude contours with streamlines around a dodecahedron immersed in a uniform flow. Three distinct particle orientations, edge-, face-, and vertex-facing, are shown across three particle Reynolds numbers, $\mathrm{Re_p} = 1$, $\mathrm{Re_p} = 100$, and $\mathrm{Re_p} = 300$. The velocity field is colour-coded, with red indicating high velocities in the free stream and dark blue representing low velocities, near or within the wake region. At low particle Reynolds number ($\mathrm{Re_p} = 1$), the flow remains steady and symmetrical for all orientations. The streamlines smoothly wrap around the particle, and only a weak velocity variation is observed downstream. The wake regions for all orientations are compact and nearly indistinguishable, which unsurprisingly means that viscous effects dominate and the influence of the dodecahedron's orientation on the flow field is minimal at this regime. At an intermediate particle Reynolds number ($\mathrm{Re_p} = 100$), a marked separation in wake structures is observed depending on the orientation. The edge-facing configuration produces a more elongated and symmetric wake, characterized by two symmetric vortices. The face-facing orientation results in a slightly broader wake, with vortical structures showing earlier separation and asymmetric elongation. For the vertex-facing orientation, the wake appears wider and slightly more asymmetric compared to the other two orientations as inertial effects become relevant. These differences arise from the way each orientation modifies the pressure distribution and the location of flow separation. The edge-facing configuration presents a relatively gradual expansion of the body to the incoming flow, allowing the boundary layer to remain attached over a larger portion of the particle before separating. Consequently, the separated region is comparatively narrow and the pressure reduction in the wake remains limited. On the other hand, the face-facing orientation exposes a larger projected area normal to the flow and introduces an abrupt geometric discontinuity at the rear of the particle. This promotes an adverse pressure gradient, causing earlier boundary-layer separation and producing a broader wake with stronger recirculation. The vertex-facing orientation exhibits an intermediate behaviour, where the inclined faces partially guide the flow around the particle while still generating sufficiently strong adverse pressure gradients to produce a wider wake than in the edge-facing configuration. 
At the highest analysed particle Reynolds number ($\mathrm{Re_p}=300$), inertial forces dominate and the separated shear layers become increasingly unstable. As a result, the wakes behind the particle exhibit significant elongation and more complex vortical structures. The larger momentum carried by the incoming flow delays viscous diffusion of vorticity, allowing the separated shear layers to persist further downstream before dissipating. This increases both the size of the recirculation region and the pressure deficit behind the particle. In all three orientations, well-defined recirculation zones are visible. The edge-facing case continues to display a relatively streamlined and symmetric wake. However, for the face-facing and vertex-facing orientations, the wake becomes more asymmetric and broader, which shows enhanced separation and increased drag. Especially in the vertex-facing orientation, the wake shows strong lateral expansion.\\

These qualitative differences emphasize the sensitivity of the flow structures around the dodecahedral particles to both particle Reynolds number, and more weakly, to orientation. While the influence of orientation is negligible in the creeping flow regime, it becomes progressively significant as the particle Reynolds number increases. The relative increase of wake complexity with Reynolds number demonstrates that, despite its high sphericity, both particle geometry and orientation should be accounted for in drag force models for regular non-spherical particles such as the dodecahedron.\\

\begin{figure}[htbp!]
\centering
\includegraphics[width=0.6\textwidth]{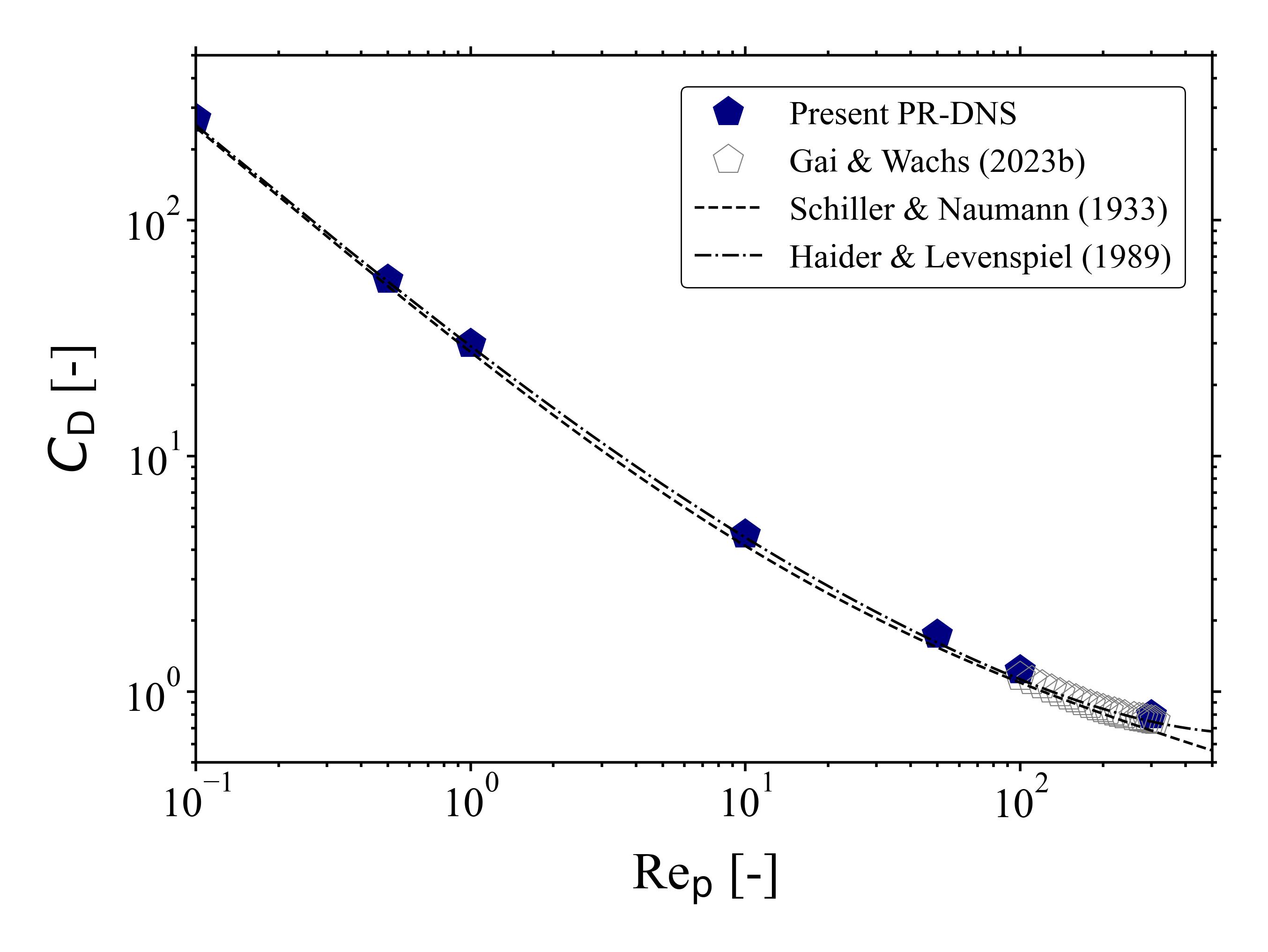}
\caption{Drag coefficient as a function of the Reynolds number for dodecahedral particle when the particle is oriented with the vertex against the flow. Lines correspond to the~\citet{Schiller1933} (dashed) and the~\citet{Haider1989} (dash-dotted) correlations, open symbol corresponds to the PR-DNS of~\citet{Gai2023a}, and the filled blue symbols represent the present PR-DNS.}
\label{fig:Dode_Cd_Re}
\end{figure}

The hydrodynamic behaviour of a dodecahedral particle placed with a vertex aligned in the direction of flow is shown in figure~\ref{fig:Dode_Cd_Re}, where the drag coefficient $C_\mathrm{D}$ is plotted against the Reynolds number $\mathrm{Re_p}$. This comparison includes the present PR-DNS results, as well as the known spherical and non-spherical correlations~\citep{Schiller1933, Haider1989}, and high-resolution PR-DNS data from~\citet{Gai2023a}. At low Reynolds numbers, where viscous effects dominate, the drag observed for the dodecahedron significantly exceeds predictions from both the~\citet{Schiller1933} and the~\citet{Haider1989} correlations. The~\citet{Haider1989} correlation, proposed to account for non-sphericity, better captures the overall drag trend, particularly in the transitional regime around $\mathrm{Re_p} = 50-100$. Nevertheless, even this model consistently underestimates the PR-DNS values at lower particle Reynolds numbers. On the other hand, at higher Reynolds numbers ($\mathrm{Re_p} > 150$), all models, including the PR-DNS from~\citet{Gai2023a}, begin to align more closely with our PR-DNS data, with deviations narrowing to below 5\%. Although the dodecahedron possesses a faceted geometry characterized by sharp edges and flat faces, rather than a smooth, continuous surface typical of a sphere, its hydrodynamic response, here quantified by the drag coefficient, presents a high similarity to that of a spherical particle across a wide range of particle Reynolds numbers. This means that, despite its geometric deviation from perfect sphericity, the dodecahedron behaves in a manner closely approximating that of a sphere, particularly in regimes where inertial effects dominate. The relatively small discrepancy in drag may be attributed to the high degree of geometric symmetry and the compactness of the dodecahedral shape, which together minimize the perturbations to the surrounding flow field typically induced by its angular form.

\begin{figure}[htbp!]
\centering
\includegraphics[width=1.0\textwidth]{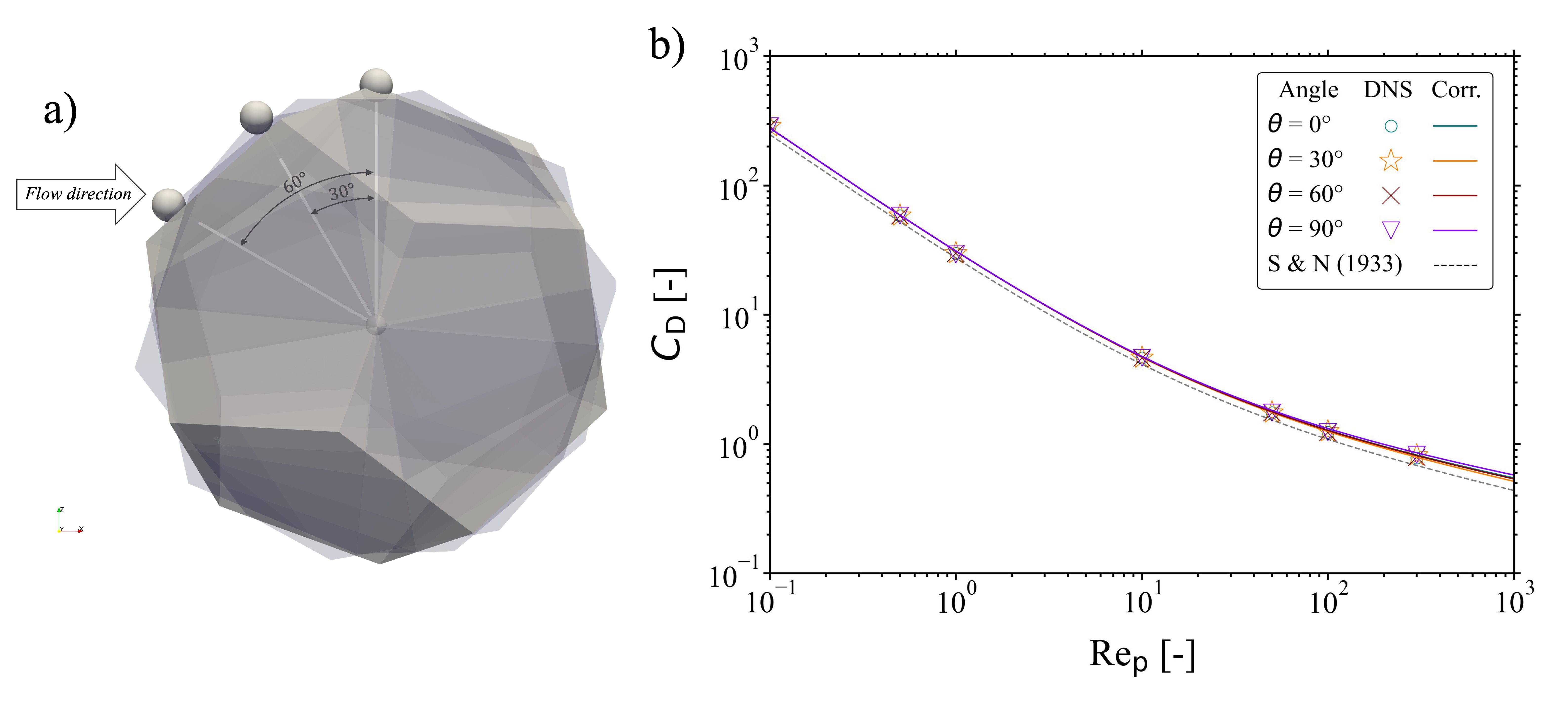}
\caption{a) Dodecahedron orientation angle with respect to the flow, b) Drag coefficient as a function of the Reynolds number for dodecahedral particle. Symbols correspond to the PR-DNS data and the solid lines correspond to the proposed correlation.}
\label{fig:DodeCorrelation}
\end{figure}

Following the previous comparison against the PR-DNS data of \citet{Gai2023a}, the influence of particle orientation on the drag of the dodecahedron is further investigated through additional PR-DNS simulations. Figure~\ref{fig:DodeCorrelation} summarizes the results. The adopted orientation convention is illustrated in figure~\ref{fig:DodeCorrelation}a, where the gray dodecahedron denotes the reference configuration and the auxiliary spheres are included to facilitate the visualization of the particle rotation. The corresponding drag coefficients obtained from the PR-DNS simulations are shown in figure~\ref{fig:DodeCorrelation}b. In contrast to the previously discussed Platonic particles, the influence of orientation on the drag coefficient remains relatively limited over the entire $\mathrm{Re_p}$-range considered. Owing to its
high sphericity, the dodecahedron exhibits only modest variations in drag among the different orientations, even at higher Reynolds numbers. 

The resulting PR-DNS results are subsequently used to derive the orientation-dependent drag correlation given in equation~\ref{eq:dodeCorre} and presented with solid lines in figure~\ref{fig:DodeCorrelation}b. The correlation proposed is presented in equation~\ref{eq:dodeCorre} and describes the drag coefficient for the dodecahedral particle with a moderate orientation sensitivity. The correlation is given as:

\begin{equation}
C_\mathrm{D} = \frac{27}{\mathrm{Re_p}} \left(1.0 + 0.15 \mathrm{Re_p}^{0.687} \right) +
\frac{0.15\left( 1.0 + \cos^2\theta + 4.0\left|\sin\theta\right|^5 \right)}
{0.5 + 5.0\cos^2\theta + 6.0\left(\mathrm{Re_p}\left|\sin\theta\right|\right)^{0.05}}
\label{eq:dodeCorre}
\end{equation}

where the first term maintains the structure of the modified~\citet{Schiller1933} correlation, with a geometry-dependent prefactor of $27$ calibrated from the PR-DNS data. The second term introduces a relatively weak orientation-dependent correction, reflected by its lower coefficient of $0.15$. The angular dependence is governed by the symmetric functions $\cos^2\theta$ and $|\sin\theta|$, ensuring a consistent response for geometrically equivalent orientations while preserving the mild influence of particle alignment on the drag coefficient. The denominator incorporates a weak Reynolds number dependence through the term $(\mathrm{Re_p} |\sin\theta|)^{0.05}$, which allows orientation effects to persist across the investigated particle Reynolds number range while gradually diminishing as inertial effects become dominant. With the proposed drag correlation for dodecahedral particles, accurate predictions of the drag coefficient over the entire range of particle Reynolds numbers, from $\mathrm{Re_p} = 0.1$ to $\mathrm{Re_p} = 300$, and for orientation angles spanning $\theta = 0^\circ$ to $\theta = 360^\circ$ are obtained. When applied to all considered orientations and Reynolds numbers, the minimum coefficient of determination obtained is $R^2 = 0.9987$, which demonstrates good agreement with the PR-DNS data.

\begin{figure}[htbp!]
\centering
\includegraphics[width=1.0\textwidth]{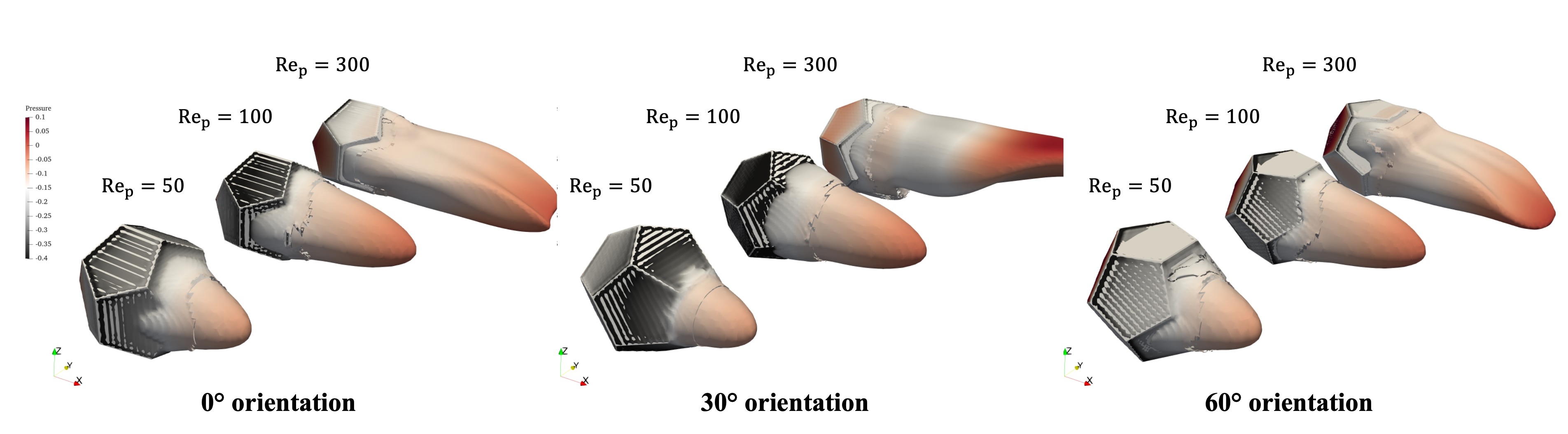}
\caption{Iso-surfaces of zero velocity for an dodecahedral particle at $\mathrm{Re_p}=50$, $100$, and $300$ and orientations $\theta = 0^\circ$, $30^\circ$, and $60^\circ$.}
\label{fig:DodecahedronIsoContour}
\end{figure}

A more complete understanding of the flow behaviour around the dodecahedral particle is provided by the iso-surfaces of zero streamwise velocity shown for $\mathrm{Re_p}=50$, $100$, and $300$ and orientations $\theta = 0^\circ$, $30^\circ$, and $60^\circ$ in figure~\ref{fig:DodecahedronIsoContour}. While figure~\ref{fig:DodecahedronContourRe} illustrates the planar velocity field and streamline patterns, the three-dimensional iso-surface representation provides further insight into the evolution of the separated region and the influence of particle orientation on the wake structure. Compared with the highly angular particles discussed previously, the dodecahedron exhibits a weaker dependence of the separation process on individual edges, as the larger number of smaller faces produces a smoother approximation of a spherical surface. For $\theta = 0^\circ$, where the particle is oriented with an edge facing the incoming flow, the fluid is progressively redistributed along the adjacent inclined faces before reaching the downstream region. This configuration produces the smallest effective frontal area and limits the lateral expansion of the separated wake. At $\mathrm{Re_p}=50$, the recirculation region is compact and relatively symmetric, with separation occurring near the downstream edges of the exposed facets. As $\mathrm{Re_p}$ increases to $100$ and $300$, the wake elongates due to the increased influence of inertia, but it remains comparatively narrow and coherent. This shows that, although separation becomes stronger at higher particle Reynolds numbers, the orientation-induced modifications of the wake remain limited. At an orientation of $\theta = 30^\circ$, a pentagonal face is oriented normal to the incoming flow, increasing the frontal blockage and modifying the upstream pressure distribution. The larger stagnation region promotes stronger adverse pressure gradients near the surrounding edges, resulting in an earlier detachment of the flow compared with the edge-facing case. At $\mathrm{Re_p}=50$, the recirculation region is already broader than for $\theta=0^\circ$. As $\mathrm{Re_p}$ increases to $100$ and $300$, the wake expands in both length and width, with the separated shear layers remaining coherent over longer downstream distances. However, compared with the tetrahedron and hexahedron, the increase in wake size remains moderate because the individual edges have a weaker influence on fixing the separation location. For $\theta = 60^\circ$, the incoming flow first encounters a vertex before being redistributed along several inclined faces. At $\mathrm{Re_p}=50$, the recirculation region exhibits an intermediate size between the previous orientations. Increasing the Reynolds number to $100$ and $300$ produces a more three-dimensional wake, with separation occurring along several inclined facets and edges, resulting in an elongated but relatively organized recirculation region. The wake remains narrower than in the $\theta = 30^\circ$ orientation because the inclined surfaces reduce the abruptness of the flow deflection and limit the lateral growth of the separated region. The progressive reduction of orientation effects observed for the dodecahedron is therefore not only associated with the decrease in projected-area changes, but also with the weakening of the shape effect imposed by individual edges on the flow separation.

\subsubsection{Icosahedron}
\label{sec:icosahedron}

\begin{figure}[htbp!]
\centering
\includegraphics[width=0.9\textwidth]{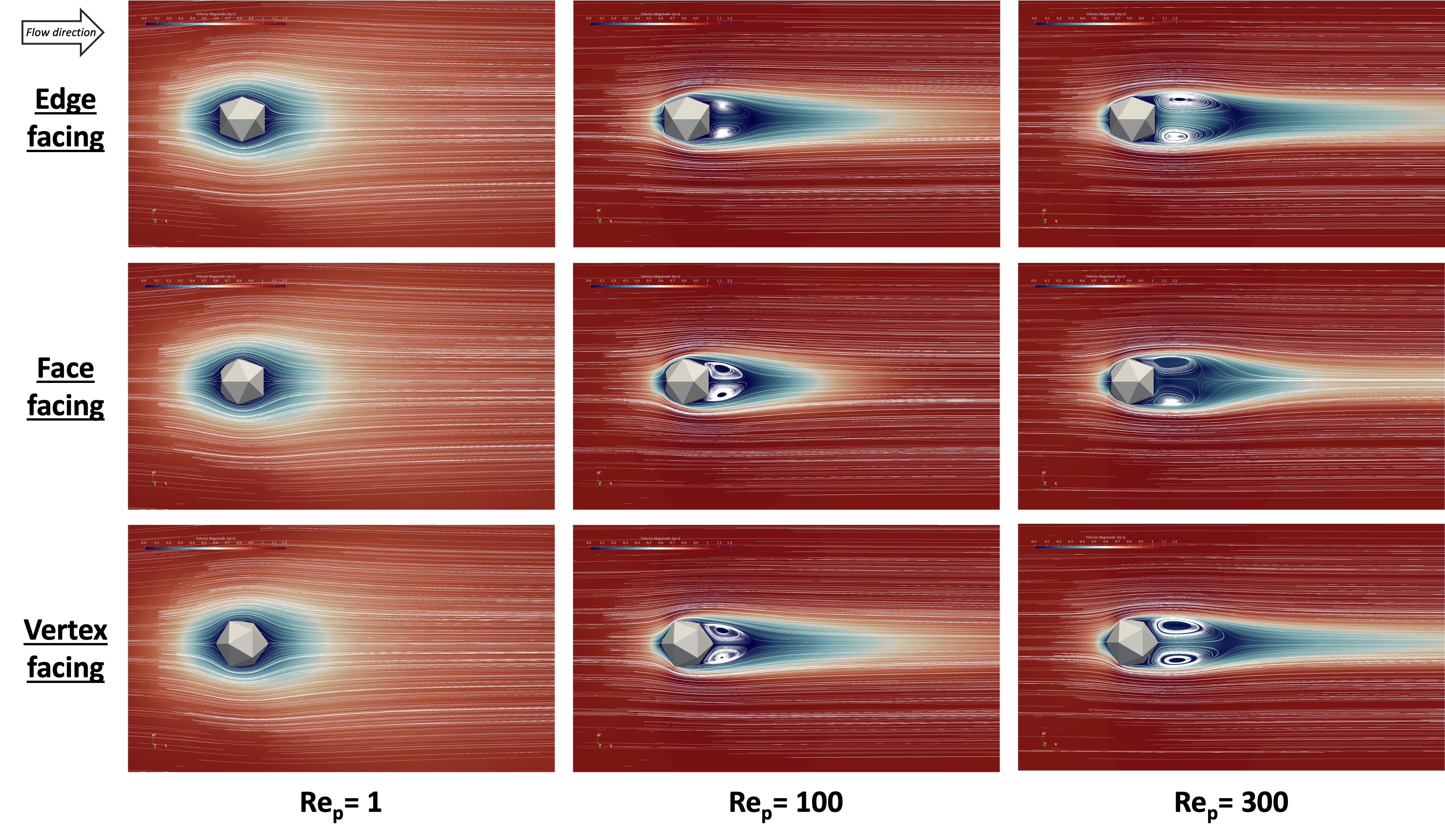}
\caption{Simulation results with icosahedral particle, Fluid speed in a cross section through the middle of the particle, and flow streamlines surrounding a single particle at different Reynolds number and three particle orientations, namely, edge-facing (top figures), face-facing (middle figures), and vertex-facing (bottom figures). The contours are shown on the central plane passing through the particle centre.}
\label{fig:IcosahedronContourRe}
\end{figure}

Lastly, figure~\ref{fig:IcosahedronContourRe} presents the velocity contours and streamlines around the icosahedron at particle Reynolds number of $\mathrm{Re_p}=1$, $100$, and $300$ for the edge-, face-, and vertex-facing configurations. As observed for the other Platonic particles, the flow remains attached and nearly symmetric at $\mathrm{Re_p}=1$, resulting in only minor differences among the considered orientations. With increasing particle Reynolds number, wake development and flow separation become more pronounced. However, compared to the tetrahedron, hexahedron, and octahedron, the influence of orientation on the wake structure remains relatively weak. This behaviour can be attributed to the geometrical characteristics of the icosahedron. Among the considered Platonic particles, the icosahedron has the highest number of faces and the largest sphericity, resulting in a surface geometry that closely approaches a sphere. Consequently, changes in orientation produce only small variations in the projected frontal area and in the inclination of the surfaces exposed to the incoming flow. This naturally reduces differences in the pressure distribution around the particle and promotes similar separation locations for the different orientations. 
In addition, the multiple inclined faces provide a smoother transition for the surrounding flow compared with particles with fewer and sharper faces, such as the tetrahedron or hexahedron. Therefore, the separated shear layers remain comparable between orientations, limiting variations in wake topology and hydrodynamic forces. Even at $\mathrm{Re_p}=300$, the differences between the considered orientations are modest, reflecting the high sphericity of the icosahedron and the resulting similarity in projected area. The edge-facing orientation produces a slightly narrower wake due to the reduced separated region behind the particle, whereas the face-facing and vertex-facing configurations generate somewhat broader recirculation regions as a consequence of small variations in frontal blockage and surface inclination. However, these geometric differences are substantially weaker than those observed for particles with fewer faces, resulting in only small changes in pressure distribution, separation location, and therefore hydrodynamic drag. Nevertheless, the overall flow field exhibits only limited sensitivity to orientation across the investigated particle Reynolds-number range.

\begin{figure}[htbp!]
\centering
\includegraphics[width=0.6\textwidth]{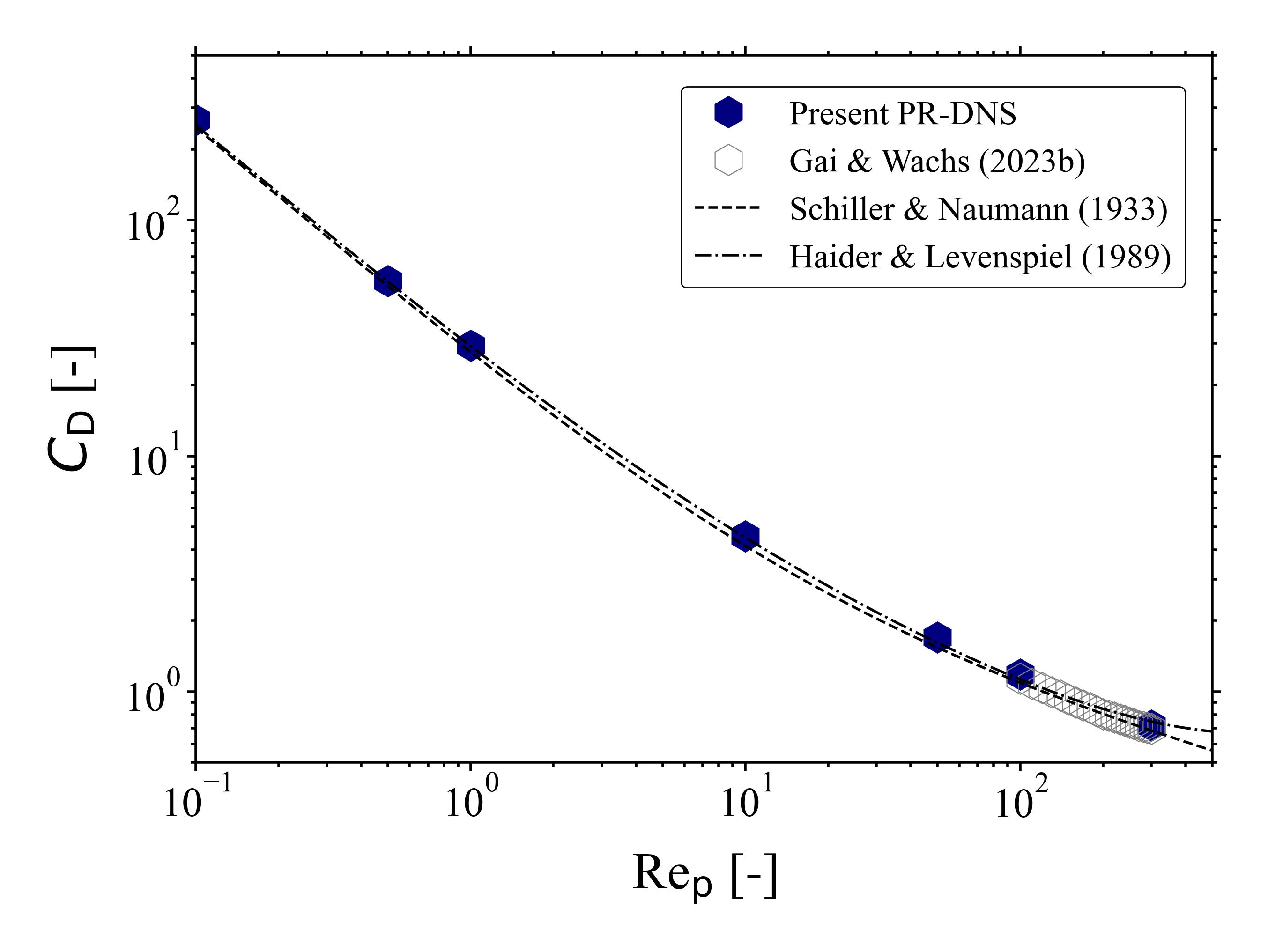}
\caption{Drag coefficient as a function of the Reynolds number for icosahedral particle when the particle is oriented with the vertex against the flow. Lines correspond to the~\citet{Schiller1933} (dashed) and the~\citet{Haider1989} (dash-dotted) correlations, open symbol corresponds to the PR-DNS of~\citet{Gai2023a}, and the filled blue symbols represent the present PR-DNS.}
\label{fig:Ico_Cd_Re}
\end{figure}

To further validate the numerical results, the drag coefficient is compared with the PR-DNS data of \citet{Gai2023a}, the classical spherical correlation of \citet{Schiller1933}, and the widely used non-spherical correlation of \citet{Haider1989}, as shown in figure~\ref{fig:Ico_Cd_Re}, over a range of particle Reynolds numbers for the specific vertex-facing orientation, in which a particle vertex is aligned with the flow direction. The correlations of \citet{Schiller1933} and \citet{Haider1989} predict smooth, decreasing drag coefficient curves with increasing particle Reynolds number. Both correlations slightly underestimate the drag coefficient compared to the present PR-DNS results and the data reported by \citet{Gai2023a} over the entire investigated Reynolds number range. The PR-DNS results exhibit consistently higher drag coefficients, which highlight the influence of particle shape on the flow resistance, despite the relatively high sphericity of the icosahedron. Furthermore, the close agreement among the PR-DNS data and the limited spread in $C_\mathrm{D}$ show that the drag of the icosahedron is only weakly affected by orientation. This behaviour can be attributed to its high sphericity, which leads to only minor variations in projected area and wake structure across the considered orientations.

\begin{figure}[htbp!]
\centering
\includegraphics[width=1.0\textwidth]{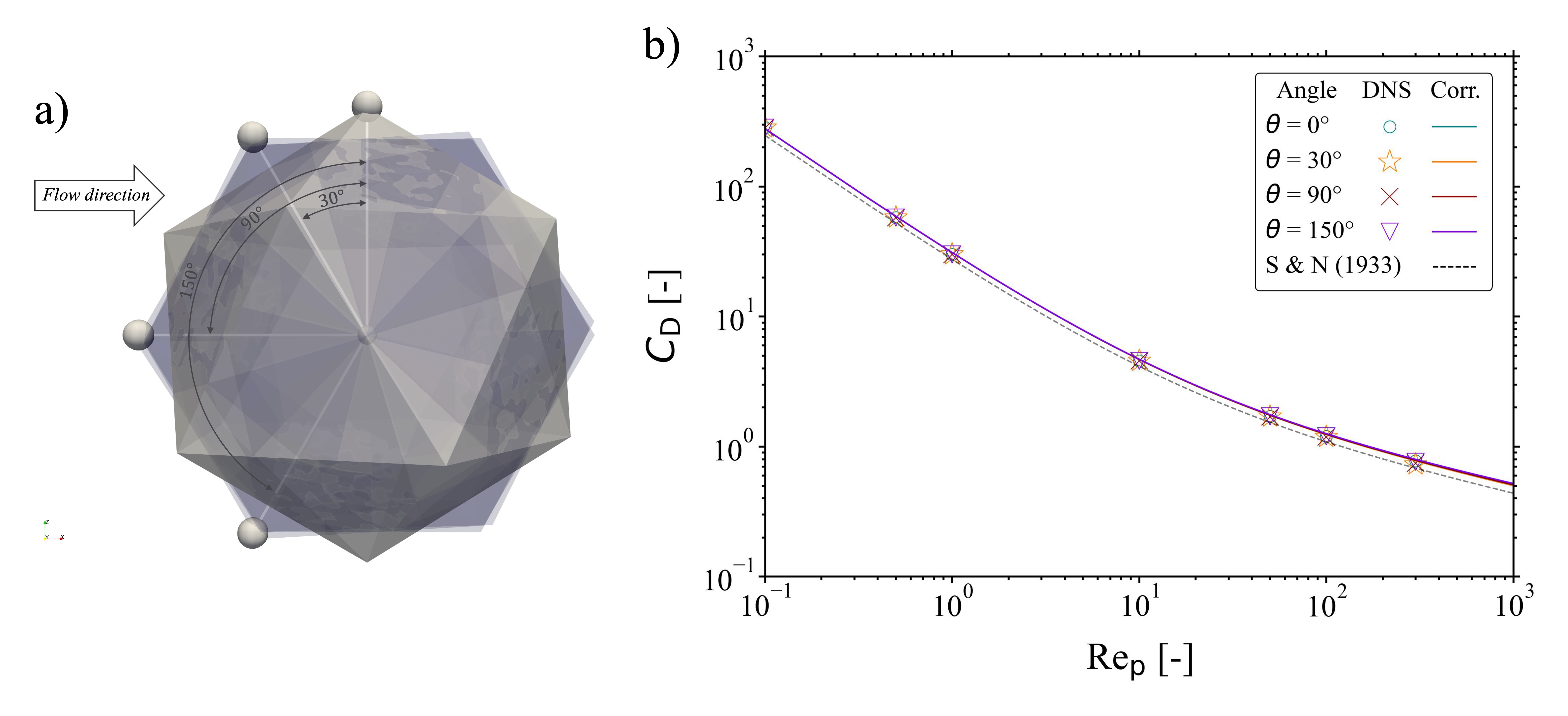}
\caption{a) Icosahedron orientation angle with respect to the flow, b) Drag coefficient as a function of the Reynolds number for icosahedral particle. Symbols correspond to the PR-DNS data and the solid lines correspond to the proposed correlation}
\label{fig:IcoCorrelation}
\end{figure}

To evaluate and extend a correlation capable of capturing the dependence of the drag force on both particle orientation and Reynolds number, we propose a formulation following a similar coefficient structure as used for the previously considered particle shapes. The correlation is given in equation~\ref{eq:icoCorre} and shown as solid lines in figure~\ref{fig:IcoCorrelation}b. This correlation is inherently suitable for particles with high geometric symmetry, but presenting some angular structures, such as the icosahedron, and naturally reproduces the nearly isotropic drag behaviour observed in this particle shape. At the same time, it incorporates a weak orientation-dependent correction, allowing for the subtle variations in drag induced by particle rotation relative to the flow as follows:

\begin{equation}
C_\mathrm{D} = \frac{27}{\mathrm{Re_p}} \left(1.0 + 0.15 \mathrm{Re_p}^{0.687} \right) +
\frac{0.05\left( 1.0 + \cos^2\theta + 4.0\left|\sin\theta\right|^5 \right)}
{0.5 + 5.0\cos^2\theta + 6.0\left(\mathrm{Re_p}\left|\sin\theta\right|\right)^{0.05}}\label{eq:icoCorre}
\end{equation}

where the first term is based on the correlation of
\citet{Schiller1933} and accounts for the Reynolds number dependence of
the drag coefficient. The second term introduces a relatively small
angular correction, reflecting the high sphericity of the icosahedron
and the resulting weak sensitivity of the drag to particle orientation.
The proposed correlation accurately reproduces the PR-DNS results for
$0.1 \leq \mathrm{Re_p} \leq 300$ and $0^\circ \leq \theta \leq 360^\circ$,
yielding a minimum coefficient of determination of $R^2 = 0.9994$ over
the entire dataset.

\begin{figure}[htbp!]
\centering
\includegraphics[width=1.0\textwidth]{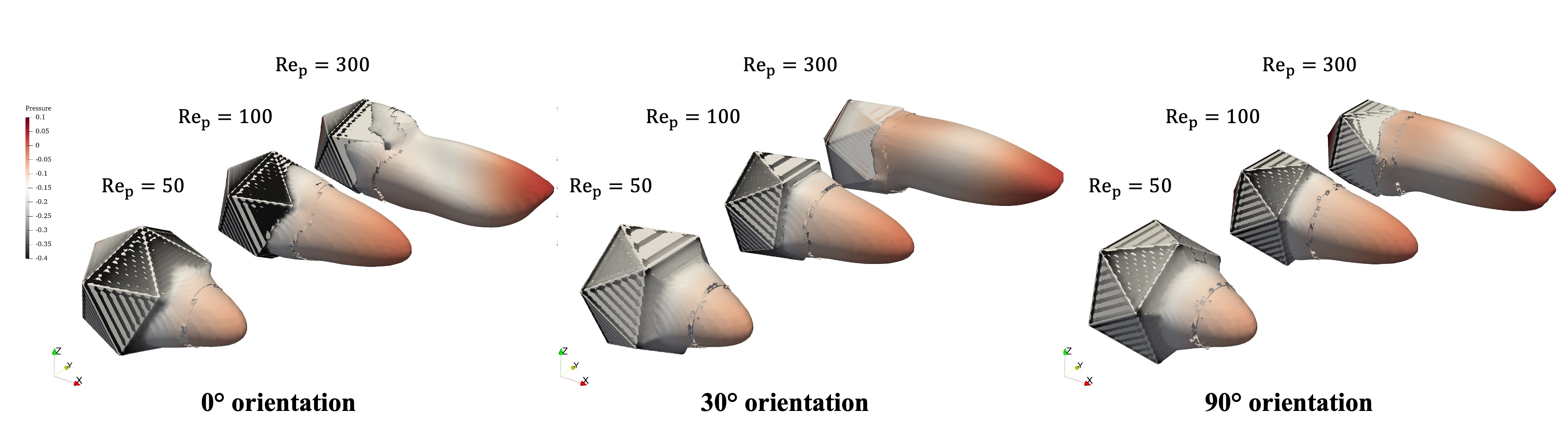}
\caption{Iso-surfaces of zero velocity for an icosahedral particle at $\mathrm{Re_p}=50$, $100$, and $300$ and orientations $\theta = 0^\circ$, $30^\circ$, and $90^\circ$.}
\label{fig:IcosahedronIsoContour}
\end{figure}

Although the icosahedron is still a polyhedral particle characterized by flat faces and sharp edges, its high sphericity and large number of faces result in a geometry that closely approaches that of a sphere. Consequently, the influence of individual edges on the flow-separation process is strongly reduced compared with the lower-sphericity particles discussed previously. The separation locations become less constrained by specific geometric features, and variations in particle orientation produce only minor changes in the wake topology and pressure distribution. This explains why the drag coefficient of the icosahedron exhibits behaviour close to that of a spherical particle over a broad range of particle Reynolds numbers. To further illustrate this behaviour, the iso-surfaces of zero streamwise velocity shown in figure~\ref{fig:IcosahedronIsoContour} provide a three-dimensional representation of the wake development around the icosahedron for $\mathrm{Re_p}=50$, $100$, and $300$ and orientations $\theta = 0^\circ$, $30^\circ$, and $90^\circ$. These visualizations show the comparatively weak influence of particle orientation on the separation process and recirculation region as the particle geometry approaches a spherical shape. For $\theta = 0^\circ$, the flow encounters an inclined face of the icosahedron, resulting in a gradual redistribution of the incoming fluid along the neighbouring facets and a relatively small projected frontal area. At $\mathrm{Re_p}=50$, the recirculation region is compact and nearly symmetric. As $\mathrm{Re_p}$ increases to $100$ and $300$, the wake elongates due to the increased influence of inertia, but it remains narrow and coherent, showing that the orientation-induced modifications of the separated flow are limited. 
At an orientation of $\theta = 30^\circ$, the flow impinges on a vertex of the icosahedron before being redistributed over the surrounding inclined facets. Compared with the $\theta=0^\circ$ configuration, this orientation modifies the local stagnation region and the arrangement of the downstream edges exposed to the flow, resulting in a slightly different pressure distribution. However, due to the high sphericity of the particle and the large number of neighbouring facets, no dominant separation location is imposed by a single geometric part. Consequently, the wake remains similar in size and topology to that observed for $\theta=0^\circ$. At $\mathrm{Re_p}=100$ and $300$, the recirculation region becomes progressively elongated, but only minor differences in lateral expansion and wake asymmetry are observed. Finally, at $\theta = 90^\circ$, the flow also encounters a vertex-facing configuration, although the exposed vertex and surrounding facets differ from those in the $\theta=30^\circ$ case. The incoming flow is again smoothly redistributed along multiple inclined surfaces, resulting in only small changes in the separation process and wake development. The resulting recirculation region remains comparable in size to the other orientations, with slight differences related to the local arrangement of facets around the exposed vertex. Increasing $\mathrm{Re_p}$ produces a larger and more elongated wake; however, the multiple inclined faces distribute the flow smoothly around the particle and limit large variations in the wake development. The iso-surfaces confirm that the high sphericity of the icosahedron weakens the influence of particle orientation by reducing the dominance of individual shape characteristics on the separation process, resulting in flow structures that increasingly resemble those observed for a sphere.

In general, the dodecahedron and icosahedron, which possess the highest sphericities among the Platonic solids, exhibit qualitatively similar hydrodynamic behaviour over the entire range of particle Reynolds numbers investigated. In the low Reynolds-number regime, the flow remains attached to the particle surface for all considered orientations, producing compact, nearly symmetric wake structures with only minor orientation-dependent variations. Consequently, the hydrodynamic response is only weakly affected by particle alignment. As the flow transitions to the intermediate Reynolds-number regime, inertial effects become more pronounced and flow separation gives rise to recirculation regions behind the particles. However, compared with the tetrahedron, hexahedron, and octahedron, the influence of orientation on the wake topology is considerably weaker. Owing to its higher sphericity, the icosahedron consistently generates slightly smaller and more compact wake structures than the dodecahedron, reflecting its closer similarity to a sphere. In the high particle Reynolds-number regime, the wakes become increasingly elongated and orientation-dependent flow features become more apparent. Nevertheless, the variation in wake size and structure with particle orientation remains substantially smaller than for the lower-sphericity Platonic solids. Face-facing and vertex-facing configurations generally produce broader wake regions than edge-facing orientations, although the differences are comparatively modest.\\

The proposed drag correlations for the five Platonic solids extend the classical correlation of \citet{Schiller1933} by incorporating angular correction terms that account for orientation-dependent drag. All formulations keep the particle Reynolds number dependence of the original model while introducing shape-specific trigonometric functions whose magnitude reflects the degree of particle anisotropy. The tetrahedron exhibits the strongest orientation dependence and therefore requires the most complex angular representation, whereas the hexahedron follows a simpler periodic behaviour associated with its symmetry. The octahedron, dodecahedron, and icosahedron share a common angular structure, with the magnitude of the correction progressively decreasing as particle sphericity increases. Consequently, the influence of orientation is largest for the tetrahedron and smallest for the icosahedron, which approaches nearly spherical behaviour. Together, these correlations provide a consistent structure for predicting the drag of regular, faceted particles over a wide range of particle Reynolds numbers and orientations with respect to the fluid flow. Having analysed the flow structure and drag coefficient, the following sections examine the lift and torque coefficients for the considered Platonic particle shapes.

\subsection{Lift coefficient}

The lift force acting on a particle immersed in a fluid flow originates from asymmetries in the pressure and viscous stress distributions over the particle surface. For non-spherical particles, these asymmetries depend not only on particle shape, but naturally on the particle Reynolds number and its orientation relative to the incoming flow. As the particle rotates, the projected area, flow separation locations, wake topology, and pressure distribution around the particle are naturally modified (see for instance, figure~\ref{fig:TetrahedronContourRe}), leading to significant variations in both the magnitude and sign of the lift force. Consequently, the lift coefficient cannot be represented solely as a function of particle Reynolds number, but must also incorporate the angular dependence associated with the particle shape. Owing to the rotational symmetry of the hexahedron, octahedron, dodecahedron, and icosahedron, the lift coefficient exhibits a periodicity of $180^\circ$, and the proposed correlations are therefore valid over the angular range $0^\circ \leq \theta \leq 180^\circ$. In contrast, the tetrahedron does not possess this rotational symmetry, requiring the correlation to be defined over the full range $0^\circ \leq \theta \leq 360^\circ$ of orientation.

The PR-DNS simulations performed for the five Platonic solids, show that the orientation dependence of the lift coefficient, $C_\mathrm{L}$, differs among the five Platonic solids owing to their different geometric symmetries and wake characteristics. Although the individual functional forms are shape dependent, all proposed correlations share the same general strategy. The orientation dependence is represented through combinations of trigonometric and exponential functions, while the amplitudes of the individual terms vary smoothly with particle Reynolds number through quadratic polynomials in $x=\log_{10}(\mathrm{Re_p})$. This formulation provides sufficient flexibility to reproduce the different orientation-dependent behaviour observed for each particle while maintaining compact analytical expressions suitable for engineering applications. Based on the resulting PR-DNS simulations, the proposed lift coefficient is correlated with the following form:

\begin{equation}
C_L=f_{\mathrm{shape}}(\theta,\mathrm{Re_p})
\end{equation}

where the functional form $f_{\mathrm{shape}}$ depends on the particle geometry. The analytical expressions proposed for each Platonic solid are given in table~\ref{tab:lift_correlations}. In every case, the particle Reynolds-number dependence (A to F) is expressed as quadratic polynomials such as:

\begin{equation}
\Phi(x)=\phi_0+\phi_1x+\phi_2x^2,
\qquad
\Phi\in\{A,B,C,D,E,F\}
\label{eq:CL_poly}
\end{equation}

where $x=\log_{10}(\mathrm{Re_p})$, and the corresponding coefficients $\phi_i$ are provided in table~\ref{tab:lift_coefficients}. For instance, the coefficients for $B$ are $b_0$, $b_1$, and $b_2$.

\begin{table}
\centering
\begin{tabular}{lc}
\toprule
\textbf{Particle shape} & \textbf{Correlation form for $C_\mathrm{L}$} \\
\midrule
Tetrahedron &
$A+B\cos(4\theta)+C\sin(6\theta)+C\cos(2\theta)+D\cos(3\theta)+E\sin(9\theta)$
\\
Hexahedron &
$A+B\sin(\theta)+C\sin(4\theta)$
\\
Octahedron &
$A+B\sin(\theta)+C\cos(\theta)+De^{\theta}+Ee^{2\theta}$
\\
Dodecahedron &
$A+B\sin(\theta)+C\cos(\theta)+D\cos(3\theta)+Ee^{\theta}+Fe^{2\theta}$
\\
Icosahedron &
$A+B\sin(\theta)+C\cos(\theta)+D\sin(3\theta)+Ee^{\theta}+Fe^{2\theta}$
\\
\bottomrule
\end{tabular}
\caption{Functional forms used for the lift coefficient correlations of the Platonic solids.}
\label{tab:lift_correlations}
\end{table}

Although the analytical expressions differ among the Platonic solids, they all follow the same physical rationale. The coefficient $A$ represents the Reynolds-number-dependent baseline contribution to the lift coefficient, whereas the remaining terms describe the orientation-dependent modulation of the lift generated by the interaction between the particle shape and the surrounding flow. Depending on the particle symmetry, different combinations of trigonometric and exponential functions are required to reproduce the numerical data accurately. The trigonometric terms account for the periodic changes associated with repeated exposure of equivalent faces and edges during particle rotation, whereas the exponential terms provide additional flexibility for particles whose lift variation exhibits strongly asymmetric or rapidly varying the angular behaviour. The amplitudes of all terms evolve smoothly with particle Reynolds number through quadratic polynomial functions, allowing the correlations to reproduce the continuous transition from viscous- to inertia-dominated flow regimes.

\begin{table}
\centering
\small
\begin{tabular}{cccccc}
\toprule
\textbf{Coefficient} &
\textbf{Tetrahedron} &
\textbf{Hexahedron} &
\textbf{Octahedron} &
\textbf{Dodecahedron} &
\textbf{Icosahedron} \\
\midrule

$a_0$ & 6.5625E-2 & 6.5401E-3 & 1.4938E-1 & -2.0382E-1 & 4.5181E-3 \\
$a_1$ & -9.1801E-2 & -1.6910E-2 & 9.1229E-2 & -7.2518E-2 & -5.3738E-1 \\
$a_2$ & 5.5375E-2 & 5.9802E-3 & -6.0584E-2 & 1.8814E-1 & 3.7562E-1 \\
\midrule

$b_0$ & -3.7069E-2 & 2.8707E-3 & -2.2474E-2 & 1.2234E-1 & 2.4749E-2 \\
$b_1$ & 3.2735E-2 & -5.3407E-3 & -1.1840E-1 & -5.4768E-3 & 8.6838E-2 \\
$b_2$ & -4.5359E-2 & 1.4349E-3 & 4.4599E-2 & -8.3788E-2 & -7.9798E-2 \\
\midrule

$c_0$ & -7.5956E-3 & -3.1204E-2 & -1.0497E-1 & 1.5064E-1 & -6.7922E-3 \\
$c_1$ & 4.3643E-3 & -6.9725E-2 & -1.1116E-1 & 2.3686E-2 & 4.1170E-1 \\
$c_2$ & -1.0262E-2 & 2.2624E-2 & 5.3359E-2 & -1.1984E-1 & -2.8565E-1 \\
\midrule

$d_0$ & -1.1544E-1 & -- & -2.7979E-2 & 2.9724E-2 & 7.3621E-3 \\
$d_1$ & 2.0943E-1 & -- & 4.1376E-3 & 3.6669E-3 & -3.3315E-2 \\
$d_2$ & -1.9612E-1 & -- & 4.4885E-3 & -2.4788E-2 & 1.8001E-2 \\
\midrule

$e_0$ & -1.7553E-2 & -- & 7.6793E-4 & 2.0242E-2 & -1.7840E-3 \\
$e_1$ & 3.3091E-2 & -- & -5.8597E-4 & 1.7776E-2 & 9.7719E-2 \\
$e_2$ & -7.2439E-3 & -- & 1.1706E-5 & -2.4967E-2 & -6.7748E-2 \\
\midrule

$f_0$ & -- & -- & -- & -1.6321E-4 & 4.8485E-5 \\
$f_1$ & -- & -- & -- & -6.3416E-4 & -2.5074E-3 \\
$f_2$ & -- & -- & -- & 4.9287E-4 & 1.7375E-3 \\
\bottomrule

\end{tabular}
\caption{Fit coefficients for the lift coefficient $C_\mathrm{L}$ correlation.}
\label{tab:lift_coefficients}
\end{table}

\begin{figure}[htbp!]
\centering
\includegraphics[width=0.95\textwidth]{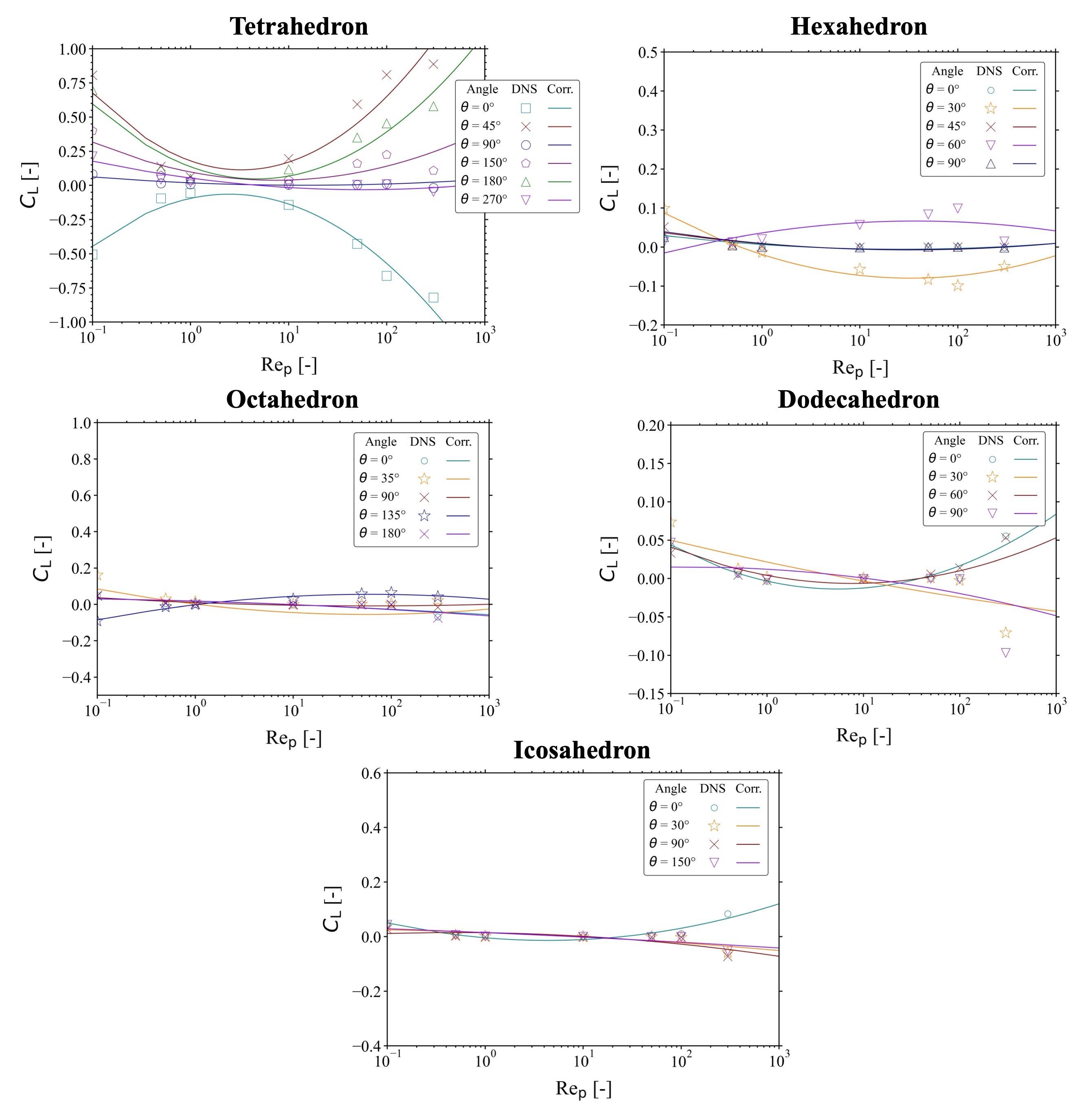}
\caption{Lift coefficient as a function of the Reynolds number and particle orientation for the considered Platonic particles. Symbols correspond to the PR-DNS data and the solid lines correspond to the proposed correlation.}
\label{fig:LiftCorrelation}
\end{figure}

Figure~\ref{fig:LiftCorrelation} compares the lift coefficient values predicted by the proposed correlation with those obtained from the PR-DNS simulations for each of the considered Platonic shapes. As can be seen, the correlation reproduces the trends observed in the numerical simulations, including both the particle Reynolds number dependence and the pronounced orientation-dependent variations of the different Platonic solids. The agreement between the correlation and the PR-DNS data is generally very good over the investigated particle Reynolds number. In particular, the model accurately captures the locations of maxima and minima in the lift coefficient. 
It is also worth noting that, compared to the drag coefficient, the lift coefficient is typically one to two orders of magnitude smaller. This reflects its physical origin as a consequence of flow asymmetries and orientation-dependent pressure imbalances, rather than the dominant streamwise momentum responsible for drag. Consequently, lift is inherently more sensitive to subtle variations in particle orientation and particle Reynolds number, which in turn modifies the wake topology. This increased sensitivity makes the lift coefficient more challenging to represent using compact yet accurate analytical correlations. The selected correlation forms are sufficiently flexible to describe the complex variations of the lift coefficient arising from the combined effects of particle shape, orientation, and flow inertia. The quality of the fits is evaluated using the coefficient of determination, $R^2$, computed for each orientation and for all considered particle shapes. 
The minimum $R^2$ value obtained among all fitting cases is $0.42$, while substantially higher values are observed for most configurations. Although an $R^2$ value of $0.42$ means that the most challenging cases exhibit a larger degree of scatter, it still corresponds to a model that captures the dominant trends and a significant fraction of the variation given in the PR-DNS data. The lower values are primarily associated with orientations for which the lift coefficient experiences rapid changes with Reynolds number, specially for the hexahedron, resulting in a more complex response that is difficult to represent using a compact analytical expression. 
Considering the broad range of Reynolds numbers investigated, from $\mathrm{Re_p} = 0.1$ to $\mathrm{Re_p} = 300$, the diversity of particle shapes, and the highly nonlinear dependence of the lift force on particle orientation, the overall performance of the proposed correlation can be regarded as satisfactory. The results also show that the correlation provides a practical compromise between accuracy and simplicity, making it suitable for incorporation into reduced-order particle-force models and Euler--Lagrange simulations where a computationally efficient representation of orientation-dependent lift effects is required.

\subsection{Torque coefficient}

The hydrodynamic torque acting on a particle immersed in a fluid flow is naturally fundamental to determine its rotational dynamics and orientation. Unlike spherical particles, which experience no preferential orientation in uniform flow, non-spherical particles are subjected to orientation-dependent pressure and viscous stresses that generate a net torque. This torque tends to rotate the particle toward stable equilibrium orientations while destabilizing others, thereby governing the particle's rotational motion and, ultimately, its translational behaviour through the coupling between orientation and hydrodynamic forces. Similarly to the lift coefficient, due to the rotational symmetry of the hexahedron, octahedron, dodecahedron, and icosahedron, the torque coefficient presents a periodicity around $180^\circ$, and the proposed correlations are therefore valid over the angular range $0^\circ \leq \theta \leq 180^\circ$. In contrast, the tetrahedron does not possess this rotational symmetry, requiring the correlation to be defined over the full range $0^\circ \leq \theta \leq 360^\circ$ of orientation. 

\begin{figure}[htbp!]
\centering
\includegraphics[width=0.95\textwidth]{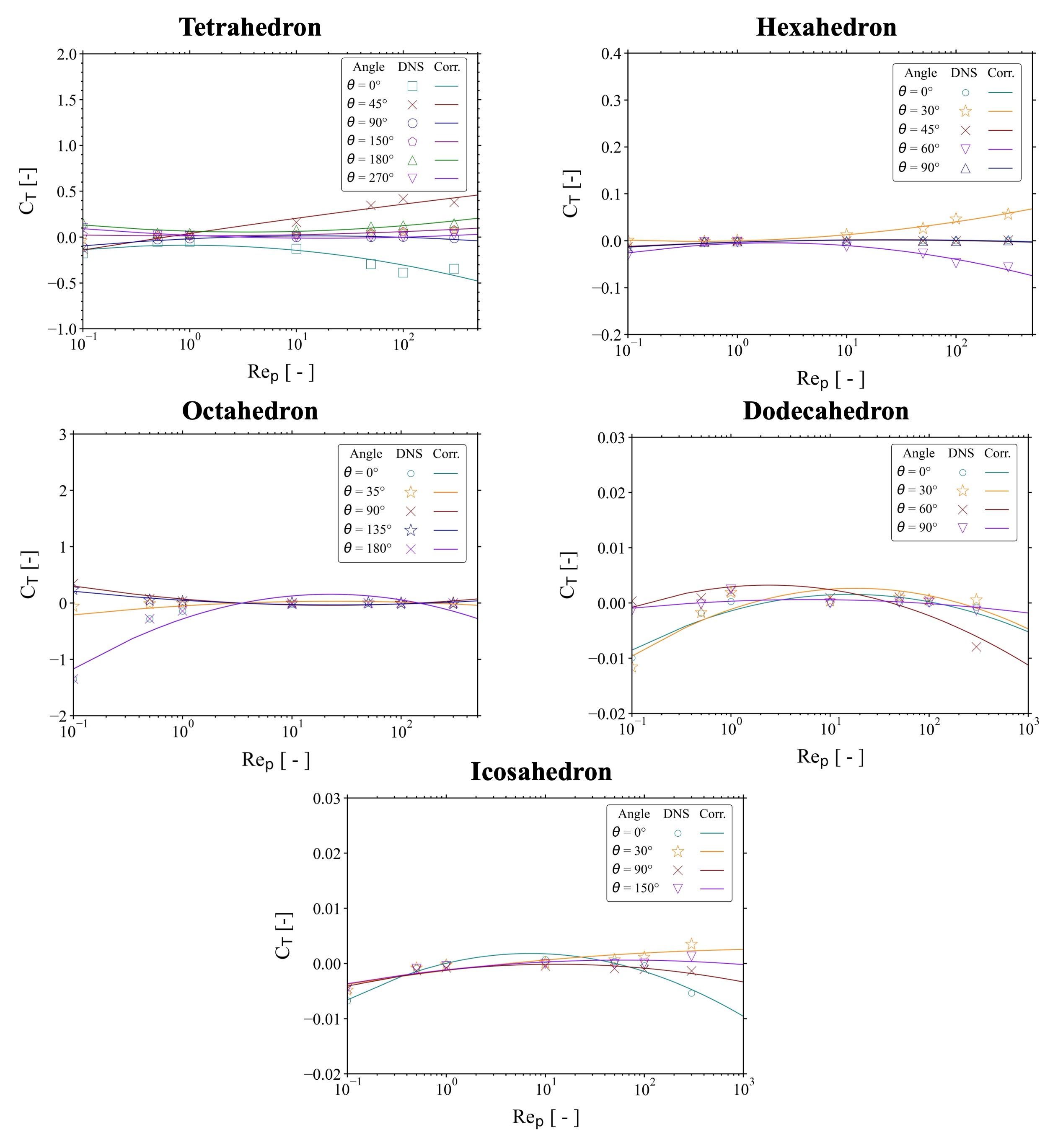}
\caption{Torque coefficient as a function of the Reynolds number and particle orientation for the considered Platonic particles. Symbols correspond to the PR-DNS data and the solid lines correspond to the proposed correlation.}
\label{fig:torqueCorrelation}
\end{figure}

The PR-DNS simulations performed in the present work show that the torque coefficient, $C_\mathrm{T}$, depends strongly on both the particle Reynolds number and orientation, see figure~\ref{fig:torqueCorrelation}. Although the five Platonic solids exhibit different rotational symmetries and torque responses, all cases can be represented using a common modelling strategy in which Reynolds-number-dependent amplitudes multiply a compact combination of trigonometric and exponential functions selected according to the particle symmetry. This approach provides sufficient flexibility to reproduce the different orientation-dependent behaviour while maintaining a compact analytical formulation. The torque coefficient is therefore expressed in the general form

\begin{equation}
C_\mathrm{T}=
A
+B\,\sin(\theta)
+C\,\cos(\theta)
+D\,\sin(k\theta)
+E\,\cos(k\theta)
+F\,e^{\theta}
+G\,e^{2\theta},
\label{eq:CT_correlation}
\end{equation}

where $k=3$ for the tetrahedron and $k=4$ for the hexahedron. The remaining particle shapes use reduced functional forms in which specific angular contribution is omitted according to their symmetry. Similar to the lift coefficient correlation, the Reynolds-number dependence of the amplitude functions $A$--$G$ is represented by quadratic polynomials:

\begin{equation}
\Phi(x)=\phi_0+\phi_1x+\phi_2x^2,
\qquad
\Phi\in\{A,B,C,D,E,F,G\},
\label{eq:CT_poly}
\end{equation}

where $x=\log_{10}(\mathrm{Re_p})$. The corresponding polynomial coefficients are provided in table~\ref{tab:torque_coefficients}. For example, the coefficients associated with the function $C$ are denoted as $c_0$, $c_1$, and $c_2$. The angular contributions depend on the rotational symmetry of each Platonic solid. For the tetrahedron, all terms $A$--$G$ are maintained, resulting in the angular dependencies $\sin(\theta)$, $\cos(\theta)$, $\sin(3\theta)$, $\cos(3\theta)$, $e^\theta$, and $e^{2\theta}$. For the hexahedron, all terms are also maintained, with the higher-order harmonic terms replaced by $\sin(4\theta)$ and $\cos(4\theta)$. For the octahedron, the third and fourth harmonic contributions are omitted, resulting in $D=E=0$. For the dodecahedron and icosahedron, the contribution associated with $G$ is omitted. The function $A$ represents the particle Reynolds-number-dependent baseline contribution to the torque coefficient, whereas the remaining terms describe the orientation-dependent variations arising from the interaction between the incoming flow and particle shape. The trigonometric terms capture the periodic behaviour associated with the rotational symmetry of the Platonic solids, while the exponential terms provide additional flexibility to represent the asymmetric evolution of the torque observed in the PR-DNS data. The specific combination of functions differs among particle shapes, reflecting their distinct symmetry properties and the response to the incoming flow. The Reynolds-number dependence of the amplitude functions shows the increasing influence of inertial effects on the rotational dynamics of the particles. At low particle Reynolds numbers, the flow remains largely attached, and the torque is primarily governed by viscous stresses. As the particle Reynolds number increases, flow separation and wake development generate increasingly asymmetric surface stress distributions, modifying both the magnitude and angular dependence of the hydrodynamic torque, as previously discussed. Expressing the amplitudes as quadratic functions of $x=\log_{10}(\mathrm{Re_p})$ provides a smooth representation over the investigated particle Reynolds-number range while avoiding separate correlations for individual flow conditions.

\begin{table}
\centering
\small
\begin{tabular}{cccccc}
\toprule
\textbf{Coefficient} &
\textbf{Tetrahedron} &
\textbf{Hexahedron} &
\textbf{Octahedron} &
\textbf{Dodecahedron} &
\textbf{Icosahedron} \\
\midrule

$a_0$ & -1.5009E-2 & -2.5688E-3 & -3.7917E-1 & 2.1163E-2 & 4.7275E-3 \\
$a_1$ & -2.5919E-2 & 1.8976E-4 & 8.6237E-1 & -1.8539E-2 & 3.7563E-2 \\
$a_2$ & -1.3021E-2 & 1.4826E-3 & -3.2282E-1 & -2.8043E-3 & -2.3669E-2 \\
\midrule

$b_0$ & 7.8962E-3 & -6.0312E-5 & 2.9087E-1 & -9.5054E-3 & -2.2959E-3 \\
$b_1$ & 1.0958E-1 & 1.2832E-3 & -6.6441E-1 & 1.3652E-2 & -6.4076E-3 \\
$b_2$ & -3.5461E-3 & -7.7077E-4 & 2.4837E-1 & -1.4550E-3 & 4.9920E-3 \\
\midrule

$c_0$ & -8.4413E-3 & -2.5881E-4 & 5.8336E-2 & -1.5507E-2 & -3.8136E-3 \\
$c_1$ & 7.8982E-2 & 5.0085E-3 & -1.2555E-1 & 1.8510E-2 & -2.7047E-2 \\
$c_2$ & -1.7343E-2 & -2.9922E-3 & 5.0268E-2 & -4.9001E-4 & 1.7251E-2 \\
\midrule

$d_0$ & 1.2448E-2 & 3.0769E-3 & -- & -4.6281E-3 & -1.0028E-4 \\
$d_1$ & 3.3135E-2 & -1.1489E-3 & -- & 4.0263E-3 & 1.7708E-3 \\
$d_2$ & 7.3429E-3 & 1.1319E-2 & -- & 2.6679E-4 & -6.3934E-4 \\
\midrule

$e_0$ & -6.4113E-2 & -8.4349E-5 & -- & -2.4896E-3 & -8.8446E-4 \\
$e_1$ & -5.7128E-2 & 7.7728E-4 & -- & 9.6081E-4 & -6.5098E-3 \\
$e_2$ & -2.2390E-2 & -2.9618E-4 & -- & 9.8315E-4 & 4.0634E-3 \\
\midrule

$f_0$ & 3.4553E-4 & -6.6153E-5 & 4.0544E-2 & 2.7803E-5 & 2.2367E-5 \\
$f_1$ & 5.7887E-4 & 1.2728E-3 & -9.2082E-2 & 4.4537E-5 & 1.6845E-4 \\
$f_2$ & 2.5719E-4 & -7.6015E-4 & 3.4350E-2 & -4.1572E-5 & -1.0377E-4 \\
\midrule

$g_0$ & -6.3845E-7 & 1.7839E-6 & -1.4614E-3 & -- & -- \\
$g_1$ & -1.0842E-6 & -3.3981E-5 & 3.3445E-3 & -- & -- \\
$g_2$ & -4.8166E-7 & 2.0281E-5 & -1.2346E-3 & -- & -- \\
\bottomrule

\end{tabular}
\caption{Polynomial coefficients for the torque coefficient correlation. Coefficients corresponding to angular contributions excluded by particle symmetry are omitted from the respective functional forms.}
\label{tab:torque_coefficients}
\end{table}

Figure~\ref{fig:torqueCorrelation} compares the torque coefficient predicted by the proposed correlations with the PR-DNS results for all considered Platonic particles. In general, the proposed correlations reproduce the dependence of the torque coefficient on both particle Reynolds number and orientation with good agreement. Despite the use of different functional forms for the individual particle shapes, the models capture the dominant orientation-dependent trends observed in the numerical data, including the main locations of maxima, minima, and zero-crossings of the torque coefficient. These features are particularly relevant because they determine the stable and unstable equilibrium orientations of the particles and consequently influence their rotational dynamics. The use of particle Reynolds-number-dependent amplitudes allows the correlations to maintain accuracy over the investigated particle Reynolds-number range while preserving a compact analytical form suitable for reduced-order particle-force models and Euler--Lagrange simulations.

The rotational equilibrium orientations of the particle can be identified directly from the torque coefficient curves. An equilibrium orientation is defined by the condition $C_{\mathrm{T}}=0$, which indicates that the net hydrodynamic torque acting on the particle vanishes. However, the existence of a zero-torque configuration alone does not determine whether the orientation is physically stable. 
The stability of each equilibrium is established by evaluating the local slope of the torque curve, $\left.\frac{\mathrm{d}C_{\mathrm{T}}}{\mathrm{d}\theta}\right|_{\theta=\theta_0}$, where $\theta_0$ denotes a zero-crossing of the torque coefficient. If $\left.\frac{\mathrm{d}C_{\mathrm{T}}}{\mathrm{d}\theta}\right|_{\theta=\theta_0}<0$, the equilibrium is stable. In this case, a small positive angular perturbation produces a negative torque, while a small negative perturbation generates a positive torque. Consequently, the hydrodynamic torque acts to restore the particle towards its original orientation. Conversely, if $\left.\frac{\mathrm{d}C_{\mathrm{T}}}{\mathrm{d}\theta}\right|_{\theta=\theta_0}>0$, the equilibrium is unstable, since a small perturbation gives rise to a torque that further increases the angular displacement, causing the particle to rotate away from the equilibrium orientation. 
Here, the torque coefficient serves as the rotational counterpart of the force, allowing stable and unstable particle orientations to be identified directly from the analytical correlation. By locating the zero-crossings of $C_{\mathrm{T}}$ and evaluating the corresponding derivative, the hydrodynamically stable equilibrium orientations of each Platonic particle can therefore be determined over the entire range of particle Reynolds numbers considered in the present study. Based on the computed torque curves, the stable and unstable equilibrium orientations for each Platonic solid and two representative particle Reynolds numbers are summarized in table~\ref{tab:stable_unstable}. The angles listed in table~\ref{tab:stable_unstable} are expressed in degrees and correspond to the complete angular range considered in the correlation. Equivalent orientations arising from particle symmetry are not repeated. The results show that the number and location of equilibrium positions depend strongly on the particle geometry and its associated rotational symmetry. 

The tetrahedron exhibits three stable and three unstable orientations over the investigated angular range at both particle Reynolds numbers. The equilibrium positions remain approximately separated by $120^\circ$, reflecting the rotational symmetry of the particle, although their exact locations shift with increasing particle Reynolds number. The displacement of the equilibrium angles from $\mathrm{Re_p}=1$ to $\mathrm{Re_p}=10$ indicates that inertial effects modify the balance between the viscous and pressure-induced contributions to the hydrodynamic torque. The hexahedron shows two stable and two unstable orientations at both Reynolds numbers. However, the equilibrium positions undergo a noticeable shift as $\mathrm{Re_p}$ increases, with the stable orientations moving from $27.4^\circ$ and $117.7^\circ$ to $46.3^\circ$ and $136.3^\circ$. The larger displacement of the unstable orientations, particularly the transition from $18.3^\circ$ and $108.1^\circ$ to $88.0^\circ$ and $178.1^\circ$, shows that wake development significantly modifies the angular dependence of the torque at higher particle Reynolds numbers. The octahedron presents a single stable and unstable equilibrium orientation for both Reynolds numbers considered. Nevertheless, the stable orientation shifts from $149.3^\circ$ at $\mathrm{Re_p}=1$ to $59.9^\circ$ at $\mathrm{Re_p}=10$, being relatively symmetrical position around $90^\circ$. This shift shows the increasing influence of separated flow structures on the torque balance. The dodecahedron exhibits two stable orientations at $\mathrm{Re_p}=1$, whereas only one stable orientation remains at $\mathrm{Re_p}=10$. This change means that increasing inertia can alter the torque by modifying the number of zero-crossings and therefore the number of possible equilibrium states. A similar modification is observed for the unstable orientations, where the two equilibrium positions at low particle Reynolds number become concentrated around a single orientation at higher particle Reynolds number. The icosahedron shows the strongest change in equilibrium structure among the considered particles. At $\mathrm{Re_p}=1$, only one stable and one unstable orientation are identified, whereas at $\mathrm{Re_p}=10$, two stable and two unstable orientations appear. This additional pair of equilibria shows that inertial effects introduce new variations into the torque distribution, likely associated with increased flow separation and wake-induced asymmetries despite the high rotational symmetry of the particle.

In general, the stable orientations correspond to angles where small perturbations generate restoring torques, whereas the unstable orientations represent angular positions from which perturbations are amplified. The results therefore provide a direct link between the hydrodynamic torque correlations and the rotational dynamics of non-spherical particles, enabling the prediction of particle Reynolds-number-dependent equilibrium orientations without requiring additional PR-DNS simulations.

\begin{table}
\centering
\begin{tabular}{lccc}
\toprule
\textbf{Shape} & 
\textbf{$\mathrm{Re_p}$} & 
\textbf{Stable ($^\circ$)} & 
\textbf{Unstable ($^\circ$)} \\
\midrule

Tetrahedron & 1  & 84.4, 206.6, 331.8 & 31.5, 146.1, 264.5 \\
Tetrahedron & 10 & 91.5, 190.4, 335.4 & 18.9, 145.6, 271.1 \\
\midrule

Hexahedron & 1  & 27.4, 117.7 & 18.3, 108.1 \\
Hexahedron & 10 & 46.3, 136.3 & 88.0, 178.1 \\
\midrule

Octahedron & 1  & 149.3 & 58.3 \\
Octahedron & 10 & 59.9 & 150.6 \\
\midrule

Dodecahedron & 1  & 92.1, 174.8 & 36.6, 142.6 \\
Dodecahedron & 10 & 97.1 & 174.4 \\
\midrule

Icosahedron & 1  & 0.9 & 179.5 \\
Icosahedron & 10 & 52.6, 155.8 & 94.2, 169.4 \\
\bottomrule

\end{tabular}
\caption{Stable and unstable configurations for all Platonic shapes at two representative particle Reynolds numbers, $\mathrm{Re_p}=1$ and $\mathrm{Re_p}=10$.}
\label{tab:stable_unstable}
\end{table}

The good agreement observed between the correlations and the PR-DNS results demonstrates that the selected trigonometric and exponential representation provides sufficient flexibility to describe the complex dependence of the hydrodynamic torque on particle shape, orientation, and flow inertia. In addition to reproducing the magnitude of the torque coefficient, the correlations preserve the main characteristics of the torque, including the orientation-dependent variations, zero-crossings, and local extrema that determine the rotational behaviour of the particles. The quality of the correlations is evaluated using the coefficient of determination, $R^2$, computed over all particle orientations and for each Platonic shape considered in this study. The minimum $R^2$ value obtained among all fitting cases is approximately $0.72$, while higher values are obtained for most configurations. Therefore, even for the most challenging cases, the proposed correlations reproduce the dominant trends obtained in the simulation data. Although the torque coefficient exhibits a considerably more complex dependence on particle Reynolds number and orientation compared with the translational force coefficients, the proposed formulation is capable of representing its main characteristics with satisfactory accuracy. The lower $R^2$ values observed for certain configurations arise from the strong sensitivity of hydrodynamic torque to small changes in the surrounding flow field. In particular, the torque coefficient can experience rapid sign changes and develop multiple local extrema over relatively narrow orientation intervals, making its representation using a compact analytical expression very challenging. This behaviour is particularly evident for the hexahedron, for which the lowest $R^2$ value is obtained, reflecting the more complex orientation-dependent variation of the torque. Unlike force coefficients, which depend primarily on the integrated magnitude of the hydrodynamic forces acting on the particle, the torque coefficient additionally depends on their spatial distribution relative to the particle centre. Consequently, small modifications in flow separation, and wake structure can generate significant changes in the resulting torque. This sensitivity increases the complexity of the fitting problem, particularly at higher particle Reynolds numbers where inertial effects introduce stronger flow asymmetries.

Despite all these challenges, the proposed correlations reproduce the dominant orientation-dependent tendencies observed in the PR-DNS results while keeping the essential physical behaviour governing the rotational dynamics of Platonic-shaped particles. Considering the broad particle Reynolds-number range investigated ($0.1 \leq \mathrm{Re_p} \leq 300$), the diversity of particle shapes, and the nonlinear dependence of torque on orientation, the obtained agreement is considered satisfactory. More importantly, the correlations preserve the key characteristics required for particle-dynamics modelling, including the prediction of hydrodynamically stable and unstable orientations. The resulting formulation therefore provides an effective compromise between accuracy and computational efficiency, making it suitable for Euler--Lagrange approaches, reduced-order models, and other numerical frameworks requiring a compact representation of orientation-dependent hydrodynamic torque.

\section{Conclusions}
\label{sec:Conclusions}

PR-DNS simulations have been performed to investigate the drag force, lift force, and hydrodynamic torque acting on Platonic particles over a broad range of particle Reynolds numbers and orientations. The results show that particle shape and orientation strongly influence the hydrodynamic response, which stress the importance of orientation-dependent fluid--particle interactions in systems where rotational dynamics affect particle transport. The simulations, performed using an immersed boundary method, cover particle Reynolds numbers in the range $0.1 \leq \mathrm{Re_p} \leq 300$ and a wide range of particle orientations. The results show that the magnitude and orientation dependence of drag, lift, and torque are strongly governed by the particle shape and the wake evolution. The tetrahedron, hexahedron, and octahedron exhibit pronounced orientation-dependent behaviour, whereas the high-sphericity dodecahedron and icosahedron display reduced sensitivity to orientation and increasingly sphere-like hydrodynamic response.

Based on the PR-DNS data, new orientation-dependent correlations are developed for the drag, lift, and torque coefficients of the considered Platonic solids. Unlike classical isotropic correlations for spherical particles, such as those of~\citet{Schiller1933}, and shape-dependent orientation-averaged models such as those of~\citet{Haider1989}, the proposed formulations explicitly account for the coupled effects of particle Reynolds number and orientation. The drag correlations achieve high predictive accuracy with coefficients of determination exceeding $R^2=0.97$, while the lift and torque correlations reproduce the dominant orientation-dependent trends observed in the PR-DNS data despite the increased complexity of their angular dependence.

The developed correlations provide compact representations of the anisotropic hydrodynamic response of Platonic-shaped particles. The drag formulation extends classical spherical-particle correlations by introducing orientation-dependent corrections associated with changes in projected geometry, flow separation, and wake structure. The lift and torque coefficients are represented through Reynolds-number-dependent trigonometric and exponential functions, which capture the periodic variations associated with the rotational symmetries of the particle shape. In addition, the torque correlations allow the identification of Reynolds-number-dependent stable and unstable equilibrium orientations, providing direct insight into the rotational states that govern particle reorientation.

The results provide a unified framework for incorporating faceted particle shapes into Euler--Lagrange simulations. By including orientation-dependent force and torque coefficients, the developed correlations enable more realistic predictions of the behaviour of particle transport and rotation in multiphase flows where particle shape effects cannot be neglected.





\section*{Acknowledgements}
\noindent 
This research was funded by the Deutsche Forschungsgemeinschaft (DFG, German Research Foundation), Project number 448292913. This funding is gratefully acknowledged.

\section*{Data Availability Statement}
The data supporting the findings of this study, together with the files required to reproduce the results, are openly available in two Zenodo repositories: DOI 10.5281/zenodo.22007912, available at \url{https://doi.org/10.5281/zenodo.22007912}, and DOI 10.5281/zenodo.22007964, available at \url{https://doi.org/10.5281/zenodo.22007964}.

\bibliographystyle{jfm}


\end{document}